\documentclass[12pt]{article}

\usepackage{bm}
\usepackage{float}
\restylefloat{table}
\restylefloat{figure}

\usepackage{upquote}
\usepackage{graphicx,amssymb,amsbsy,amsfonts,amssymb,amsmath}
\usepackage[usenames, dvipsnames]{xcolor}
\usepackage{calc}
\usepackage{url}

\usepackage{tikz}
\usetikzlibrary{calc} \usetikzlibrary{patterns} \usetikzlibrary{decorations.pathreplacing} \usetikzlibrary{decorations.markings} \usetikzlibrary{decorations.pathmorphing} \usetikzlibrary{positioning}

\allowdisplaybreaks
\usepackage{array}
\usepackage{bbm}
\usepackage{slashed}

\numberwithin{equation}{section}

\newif\ifdraft
\drafttrue
\newif\ifpreprint
\preprinttrue

\def\eqn#1{eq.~(\ref{#1})}

\def\eqns#1#2{eqs.~(\ref{#1}) and~(\ref{#2})}

\def\beq{\begin{equation}}
\def\eeq{\end{equation}}
\def\bea{\begin{eqnarray}}
\def\eea{\end{eqnarray}}
\newcommand{\ba}{\begin{array}}
\newcommand{\ea}{\end{array}}
\def\spa#1.#2{\left\langle#1\,#2\right\rangle}
\def\spb#1.#2{\left[#1\,#2\right]}
\def\spash#1.#2{\spa{\smash{#1}}.{\smash{#2}}}
\def\spbsh#1.#2{\spb{\smash{#1}}.{\smash{#2}}}
\def\sand#1.#2.#3{%
\left\langle\smash{#1}{\vphantom1}^{-}\right|{#2}%
\left|\smash{#3}{\vphantom1}^{-}\right\rangle}
\def\sandpp#1.#2.#3{%
\left\langle\smash{#1}{\vphantom1}^{+}\right|{#2}%
\left|\smash{#3}{\vphantom1}^{+}\right\rangle}
\def\sandpm#1.#2.#3{%
\left\langle\smash{#1}{\vphantom1}^{+}\right|{#2}%
\left|\smash{#3}{\vphantom1}^{-}\right\rangle}
\def\sandmp#1.#2.#3{%
\left\langle\smash{#1}{\vphantom1}^{-}\right|{#2}%
\left|\smash{#3}{\vphantom1}^{+}\right\rangle}
\def\sandmm#1.#2.#3{%
\left\langle\smash{#1}{\vphantom1}^{-}\right|{#2}%
\left|\smash{#3}{\vphantom1}^{-}\right\rangle}
\def\spab#1.#2.#3{\sandmm#1.#2.#3}
\def\spaa#1.#2.#3.#4{\sandmp#1.{#2#3}.#4}
\def\spbb#1.#2.#3.#4{\sandpm#1.{#2#3}.#4}

\def\ln{{\rm ln}}

\def\fs#1{{      
        \setbox\charbox=\hbox{$#1$}
        \setbox\slabox=\hbox{$/$}
        \dimen\charbox=\ht\slabox
        \advance\dimen\charbox by -\dp\slabox
        \advance\dimen\charbox by -\ht\charbox
        \advance\dimen\charbox by \dp\charbox
        \divide\dimen\charbox by 2
        \raise-\dimen\charbox\hbox to \wd\charbox{\hss/\hss}
        \llap{$#1$} }}

\def\subtractfour#1{\ifthenelse{#1=5}{1}{\ifthenelse{#1=6}{2}
{\ifthenelse{#1=7}{3}{\ifthenelse{#1=8}{4}{\ifthenelse{#1=9}{5}
{\ifthenelse{#1=10}{6}{\ifthenelse{#1=11}{7}{\ifthenelse{#1=12}{8}
{\ifthenelse{#1=13}{9}{\ifthenelse{#1=14}{10}{}}}}}}}}}}}

\def\b{\beta}

\def\ve{\varepsilon}

\def\s{\sigma}

\def\z{\zeta}

\newcommand{\ZR}{\mathbb R}
\newcommand{\ZZ}{\mathbb Z}

\newcommand{\ZN}{\mathbb N}
\newcommand{\ZQ}{\mathbb Q}

\newcommand{\CH}{\mathcal{H}}

\newcommand{\CN}{\mathcal{N}}      
\newcommand{\CM}{\mathcal{M}}       
\newcommand{\CO}{\mathcal{O}}      

\newcommand{\CU}{\mathcal{U}}       
\newcommand{\CV}{\mathcal{V}}

\newcommand{\CZ}{\mathcal{Z}}       

\newcommand{\BA}{\mathbf{A}}       
\newcommand{\fm}{m}

\def\YM{\textrm{YM}}

\def\open{\textrm{open}}

\def\reg{\textrm{reg}}

\def\ap{\alpha'}

\newbox\charbox
\newbox\slabox

\newcommand{\nnl}{\nonumber\\}

\newcommand{\eq}{\earel{=}}

\def\_{\;\;}
\def\^{\wedge}

\def\dd{\mbox{d}}
\def\pa{\mbox{$\partial$}}
\def\pd{\mbox{$\partial$}}
\def\>{\rangle}
\def\<{\langle}
\def\+{\dagger}
\def\={\ =\ }
\def\eq{\ \ = \ \ }
\newcommand\mand{\qquad\textrm{and}\qquad}
\DeclareMathOperator{\shuffle}{\sqcup\!\!\,\sqcup}

\newcommand{\te}{\textrm}
\newcommand{\si}{\sigma}
\newcommand{\co}{\ , \ \ \ \ \ \ }
\newcommand{\vecb}{\left(\begin{array}{c}}
\newcommand{\vece}{\end{array}\right)}
\newcommand{\ccb}{\left(\begin{array}{cc}}
\newcommand{\cce}{\end{array}\right)}
\newcommand{\cccb}{\left(\begin{array}{ccc}}
\newcommand{\ccce}{\end{array}\right)}
\newcommand{\ccccb}{\left(\begin{array}{cccc}}
\newcommand{\cccce}{\end{array}\right)}
\newcommand{\cccccb}{\left(\begin{array}{ccccc}}
\newcommand{\ccccce}{\end{array}\right)}

\newcommand{\al}{\alpha}
\newcommand{\be}{\beta}

\newcommand{\de}{\delta}

\newsavebox{\matchbox}

\newsavebox{\clashbox}

\newsavebox{\formulabox}

\begin{document}
\hfuzz 20pt

\title{\textbf{Polylogarithms, Multiple Zeta Values} \\
\textbf{and Superstring Amplitudes}\\ }
\author{\large Johannes Broedel$^{\te{a}}$, Oliver Schlotterer$^{\te{b,c}}$, and 
Stephan Stieberger$^{\te{d}}$\\[2cm]}
\date{}
\maketitle
\vskip-2cm
\centerline{\it $^{\te{a}}$ Institut f\"ur theoretische Physik, ETH Z\"urich,}
\centerline{\it 8093 Z\"urich, Switzerland,}
\vskip0.15cm
\centerline{\it $^{\te{b}}$ Department of Applied Mathematics and Theoretical Physics,}
\centerline{\it Cambridge CB3 0WA, United Kingdom,}
\vskip0.15cm
\centerline{\it $^{\te{c}}$ Max-Planck-Institut f\"ur Gravitationsphysik,} 
\centerline{\it Albert-Einstein-Institut, 14476 Potsdam, Germany,}
\vskip0.15cm
\centerline{\it $^{\te{d}}$ Max-Planck-Institut f\"ur Physik,}
\centerline{\it Werner-Heisenberg-Institut, 80805 M\"unchen, Germany}

\medskip\bigskip
\begin{abstract}
  A formalism is provided to calculate tree amplitudes in open
  superstring theory for any multiplicity at any order in the inverse string
  tension. We point out that the underlying world-sheet disk integrals 
  share substantial properties with color-ordered tree amplitudes in
  Yang-Mills field theories. In particular, we closely relate world-sheet integrands
  of open-string tree amplitudes to the Kawai-Lewellen-Tye representation of
  supergravity amplitudes. This correspondence helps to reduce the singular parts of
  world-sheet disk integrals---including their string corrections---to
  lower-point results. The remaining regular parts are systematically addressed
  by polylogarithm manipulations.
\end{abstract}
\bigskip\bigskip
\medskip
\begin{flushright}
{\small DAMTP--2013--22\\
AEI--2013--194\\
MPP--2013--119}
\end{flushright}

\thispagestyle{empty}

\newpage
\setcounter{tocdepth}{2}
\tableofcontents


\newpage
\section{Introduction}

Perturbative string theory has undergone important conceptual and technical
advances, such as the analysis of world-sheet and target-space descriptions,
the use of modular properties  and more importantly of the pure-spinor
formalism.  The application of those methods has considerably 
simplified the computation of both
tree-level and multi-loop superstring scattering amplitudes.

The structure of superstring scattering amplitudes remarkably resembles the
form and organization of field-theory amplitudes.  Moreover, some features of
the latter are encoded in and derived from the properties of the string
world-sheet.  Striking field-theory relations such as Kleiss-Kuijf (KK)
\cite{KK}, Bern-Carrasco-Johansson (BCJ) \cite{BCJ} or Kawai-Lewellen-Tye (KLT)
\cite{Kawai:1985xq} relations can be easily derived from and understood in
string theory by tracing these identities back to the monodromy properties of
the string world-sheet \cite{BjerrumBohr:2009rd,Stieberger:2009hq}.
Furthermore, recent mathematical concepts such as the appearance of twistor
space, motivic aspects, symbols and the coproduct structure for Feynman
integrals have radically changed our viewpoint and strategy of how to quickly
obtain, reorganize and express results for amplitudes in field-theory. It
appears fruitful to further investigate and understand the natural appearance
of those concepts in string theory.  One particular feature of quantum field
theories---the appearance of transcendental functions at loop-level---already
occurs at tree-level scattering in string theory as a consequence of the
underlying string world-sheet. Since generically the complete perturbative
tree-level contributions to string amplitudes are easier to access than
fully-fledged loop corrections, they provide a convenient testing ground for
transcendental structures. 

This article is devoted to tree-level open superstring amplitudes (or disk
amplitudes). Tree-level interactions of open strings are described by integrals
over the boundary of world-sheets with disk topology.  The entire polarization
 dependence of open superstring tree amplitudes was shown in \cite{Mafra:2011nv,
Mafra:2011nw} to reside in tree amplitudes of the underlying Yang-Mills (YM)
field theory, which emerges in the point-particle limit $\ap \rightarrow 0$.
String corrections to the YM amplitude resemble formfactors promoting the
kinematic factors of field theory into those of string theory. They enter
through generalized Euler or Selberg integrals which involve multiple zeta
values (MZVs) in their power-series expansion w.r.t. $\ap$ as was discussed in
 e.g. Refs. \cite{Terasoma, Francis, Oprisa:2005wu,Stieberger:2009rr}.
We shall show that generalized Selberg integrals involving different cyclic
combinations of world-sheet Green's functions share essential symmetry
properties of color-stripped YM amplitudes. In addition to cyclic symmetry in
the arguments of the Green's functions, the KK and BCJ relations between field
theory amplitudes literally carry over to disk integrals. This paves our way
towards casting the world-sheet integrand of disk amplitudes into a form
resembling the KLT representation of supergravity tree amplitudes. One copy of
YM subamplitudes therein is replaced by its world-sheet correspondent -- a
Selberg integral involving an appropriate cyclic product of Green's functions.
This provides a field theoretic intuition for the structure of the disk
integrand explicitly computed in \cite{Mafra:2011nv, Mafra:2011nw}.

The systematics of the appearance of MZVs in disk amplitudes has been analyzed
in \cite{SSMZV}. In this work, the $\ap$-expansion was shown to take a compact
and elegant form once the contributions from different classes of MZVs are
disentangled. The idea is to lift MZVs to their motivic versions endowed with a
Hopf-algebra structure. The latter induces an isomorphism which allows to cast
the amplitudes into a very symmetric form. In analogy with the symbol of a
transcendental function, this isomorphism automatically builds in all relations
between MZVs.  However, its invertibility overcomes the loss of information
inherent to the symbol approach.  In spite of this beautiful organization of
disk amplitudes, the explicit calculation of the $\ap$-dependent ``seeds''
accompanying the  single zeta values has not yet been addressed systematically.
Their exact momentum dependence is hidden within world-sheet integrals whose
complexity increases drastically with multiplicity $N$ and the order in $\ap$.
While the world-sheet integrals at five points can still be reduced to a set of
single (variable) Gaussian hypergeometric functions $_3F_2$, which has  been thoroughly
exploited in \cite{SSMZV} to probe weights $w \leq 16$, their six- and
higher-point versions comprise  multiple Gaussian hypergeometric functions
\cite{Oprisa:2005wu}, whose expansions in $\ap$ are much more involved.  Though
computing some of these expansions has been accomplished at the six-
\cite{Oprisa:2005wu,Stieberger:2006bh,Stieberger:2006te,Stieberger:2007am} and
seven-point level \cite{Stieberger:2007jv,Mafra:2011nw} a systematic approach
is still lacking.

In this article, we will provide a method to completely evaluate the
world-sheet integrals, which is---in principle---applicable at any multiplicity
and to any order in $\ap$. Once a suitable set of basis integrals is
identified, their singular parts can be rewritten in a form recycling regular
parts from lower-point superstring amplitudes. That is, the complete $N$-point
superstring amplitude including poles can be written in terms of regular parts
of the world-sheet integrals at multiplicities smaller or equal to $N$. The
only remaining task is the calculation of the regular parts. Employing the
language of multiple polylogarithms, we provide a method to iteratively solve
all integrals occurring. 

The $\ap$--expansion of the integral bases for the
multiplicity $N=5$ has been computed up to the weight $w=16$ in \cite{SSMZV}.
The complete set up to the order $w=21$ is now public at \cite{WWW}, while 
the order $w=22$ will be made available therein soon.
Nevertheless, these matrices have also recently been determined up to $w=21$ in 
\cite{Boels:2013jua}. For the multiplicities $N=6$ and $N=7$ the 
$\ap$--expansions of the integral bases have been determined up to weights $w=9$ and $w=7$, respectively and are explicitly given at \cite{WWW}.
An alternative expansion method for the
four-point and five-point integrals has been recently developed in \cite{Boels:2013jua}.

The article is organized as follows: after reviewing the structure of the
$\ap$-expansion in open-string tree-level amplitudes in section \ref{secap}, we
will point out a close relation between world-sheet integrals and color-ordered
subamplitudes in YM theories in section \ref{ftpatterns}. This correspondence
inspires the calculation of world-sheet integrals: Section \ref{secKinPoles} is
devoted to exploring the pole structure of the integrals as well as the
recursive nature of the residues, while the integration of the regular parts is
taken care of in section \ref{secPolylogs} using polylogarithm manipulations.
In section \ref{sec3}, the symmetry of disk amplitudes under cyclic shifts and
world-sheet parity is exploited in order to increase the efficiency of the
setup.

Several appendices contain more detailed information on various aspects of this
work: In appendix \ref{sec:zeta} we recapitulate the Hopf algebra structure of
motivic MZVs. In order to support and extend the discussion in section
\ref{ftpatterns} on field-theory structures in world-sheet integrals, we supply
derivations for some of the identities in appendix \ref{appA}. Examples at
multiplicities $4 \leq N\leq 7$ on how the massless poles of disk integrals can
be quickly read off from their integrand are gathered in appendix
\ref{exppole}. In addition, we provide further examples for six-point residues
and their recursively organized $\ap$-expansion in appendix \ref{app6pt}. The
general discussion of polylogarithm identities in section \ref{secPolylogs} is
supplemented by the examples in appendix \ref{polyexpl}. Finally, some of the
intermediate steps for seven-point $\ap$-expansions are outsourced to appendix
\ref{app7pt}.


\section{String tree-level amplitudes and their $\bm{\ap}$-expansion}
\label{secap}

The $N$-point tree-level open superstring amplitude was computed in references
\cite{Mafra:2011nv, Mafra:2011nw} based on pure-spinor cohomology methods
\cite{Mafra:2010jq}. Remarkably, the entire polarization dependence of the
amplitude is carried by color-ordered tree amplitudes $A_{\YM}$ of the
underlying YM field theory. In terms of the $(N-3)!$-element basis\footnote{Throughout this work, expressions of the form $\si(2,3,\ldots,N-2)$ with permutation $\si$ have to be understood as $\si(2),\si(3),\ldots,\si(N-2)$.}
$A_{\YM}(1,\si(2,\ldots,N-2),N-1,N)$ with $\si \in S_{N-3}$ \cite{BCJ},
the $N$-point superstring amplitude reads 
\beq
A_{\open}(\Pi,\ap) = \sum_{\s \in S_{N-3}} F_\Pi{}^\s(\ap) \, A_{\YM}(1,\s(2,\ldots,N-2),N-1,N) \ .
\label{00,1}
\eeq
In the above equation, $A_\open(\Pi)$ with $\Pi\in S_N$ denotes the string
amplitude associated with an ordering of vertex operator positions $z_i$ on the
disk boundary according to $z_{\Pi(i)}<z_{\Pi(i+1)}$. This amounts to
calculating the partial string amplitude for the particular color-ordering
$\Pi$. The objects $F_\Pi{}^\s(\ap)$ originate in the string world-sheet
integrals and encode the string-theory modifications to the field theory
amplitude\footnote{As a consequence of the manifestly supersymmetric
derivation in \cite{Mafra:2011nv} the labels $1,2,\ldots,N$ in the subamplitude \eqn{00,1}
may denote any state from the $\CN=1$  super YM multiplet  in $D=10$
space--time dimensions without any particular reference to its polarization. 
Moreover, pure gluon tree-level amplitudes are not affected by
dimensional reduction and supersymmetry breaking. Thus, \eqn{00,1} remains
valid for external gluons in any $D<10$ superstring compactification which
allows for a CFT description with the field theory subamplitudes
$A_\YM$  adapted to the appropriate spacetime dimension.}.

In analogy to the YM situation, not all color-orderings $\Pi$ of the string
amplitude $A_\open(\Pi)$ are independent
\cite{Stieberger:2009hq,BjerrumBohr:2009rd}. Choosing a basis amounts to
single out three legs: we will choose to fix legs $1$, $N-1$ and $N$ leading
to the basis $A_\open(1,\Pi(2,\ldots,N-2),N-1,N)$ of string amplitudes, with 
$\Pi\in S_{N-3}$. The common $(N-3)!$-structure of the basis for open-string
amplitudes $A_\open$ and YM amplitudes $A_\YM$ suggests to
rewrite \eqn{00,1} in matrix notation, which relates the two vectors comprising the  basis amplitudes to a
$(N-3)!\times(N-3)!$-matrix $F$ \cite{Mafra:2011nw}:
\begin{equation}
  \BA_\open=F\,\BA_\YM\,.
  \label{00,1a}
\end{equation} 
Explicitly, one finds 
\beq
\begin{small}  \begin{pmatrix}
    A_{\open}(1,\Pi_1,N-1,N)\\
    \vdots\\    
    A_{\open}(1,\Pi_{(N-3)!},N-1,N)
  \end{pmatrix} = 
  \begin{pmatrix} 
    F_{\Pi_1}{}^{\s_1}&\cdots&F_{\Pi_1}{}^{\s_{(N-3)!}}\\
    \vdots&\ddots&\vdots\\
    F_{\Pi_{(N-3)!}}{}^{\s_1}&\cdots&F_{\Pi_{(N-3)!}}{}^{\s_{(N-3)!}}
  \end{pmatrix}
  \begin{pmatrix}
    A_{\YM}(1,\s_1,N-1,N)\\
    \vdots\\
    A_{\YM}(1,\s_{(N-3)!},N-1,N)
  \end{pmatrix},
  \label{00,1b}
\end{small}\eeq
where $\si_i$ and $\Pi_i$ with $i\in\{1,\ldots,(N-3)!\}$ denote the
permutations of $(2,3,\ldots,N-2)$, respectively\footnote{We enumerate
$S_{N-3}$ permutations in canonical order.}.  The matrix $F$, however, contains
redundant information: Knowing the functions 
$F_{\Pi_1}{}^{\si_1},\ldots,F_{\Pi_1}{}^{\si_{(N-3)!}}$ in the first line of
$F$ allows to obtain all other entries by a suitable relabeling.

After having organized the open-string amplitude in the form as \eqn{00,1a}, let
us now turn to the entries in the first line of the matrix $F$. The dependence of the disk amplitude on the inverse string tension $\ap$ is incorporated in the set of
$(N-3)!$ functions $\{F_{\Pi_1}{}^{\si}(\ap), \ \si \in S_{N-3} \}$. The latter
represent multiple Gaussian hypergeometric functions originating from generalized Euler or Selberg integrals \cite{Oprisa:2005wu,Stieberger:2006te}. The origin of their $(N-3)$-fold integrations 
stems from vertex operator positions along the
boundary of the disk world-sheet. Taylor expanding the string amplitudes (\ref{00,1}) 
w.r.t. to small $\ap$ reproduces their low-energy behavior. Each order in $\ap$ 
appears with a rational function in the dimensionless Mandelstam variables 
\beq
s_{i_1 i_2 \ldots i_p} \ \= \ \ \ap\, (k_{i_1}+k_{i_2}+\ldots +k_{i_p})^2\,
\label{01,2}
\eeq
supplemented by  some MZVs 
\begin{equation}
  \zeta_{n_1,\ldots,n_r}\ \ := \ \ \zeta(n_1,\ldots,n_r) \ \ =\ \ \sum_{0<k_1<\ldots<k_r}k_1^{-n_1}\ldots k_r^{-n_r}\co n_l\in\ZN^+\co n_r\geq2 \ .
  \label{00,4}
\end{equation}
or products thereof.
In (\ref{00,4}) the number $r$ of the arguments\footnote{If there is only one argument $n_1$,
the above definition reduces to the usual definition of the single Riemann zeta
function $\zeta(n_1)$.} and their sum $w=\sum_{i=1}^r n_i$ are called depth and
weight of the MZV, respectively. The overall weight of MZV products matches the
corresponding power in $s_{i_1 i_2 \ldots i_p}$. The $\ap$-expansion of these iterated
integrals in terms of MZVs has been discussed in both mathematics \cite{Terasoma, Francis,
Aomoto} and physics literature \cite{Stieberger:2009rr, Mafra:2011nw, SSMZV} extensively. 

There are numerous relations over rational numbers $\ZQ$ between different
MZVs, all of which preserve the weight $w$. A convenient way to automatically
take all these relations into account in the $\ap$-expansion of $F$, is to
endow the MZVs (\ref{00,4}) with a Hopf-algebra structure and map $F$ to a
non-commutative algebra comodule\footnote{A discussion of the mathematical
concepts is provided in appendix \ref{sec:zeta}.} with a more transparent basis
\cite{BrownDecomp, SSMZV}. The result is a sum over all
non-commutative words\footnote{
As pointed out by Don Zagier, one can rewrite (\ref{00,13}) as a geometric series
\begin{equation}
  \phi(\BA^\fm_\open) \eq\left(\sum_{k=0}^{\infty} f_2^k\ P_{2k}  \right)
\left\{\left(1- \sum\limits_{k=1}^{\infty} f_{2k+1} M_{2k+1}  \right)^{-1}\right\}^t \BA_\YM
  \label{zagier}
\end{equation}
where $\{\ldots\}^t$ reverses the order of the $f_{i_k}$'s enclosed.
} in cogenerators $f_3,f_5,\ldots$, supplemented by a commutative element $f_2$ \cite{SSMZV},
\begin{equation}
  \phi(\BA^\fm_\open) \eq\left(\sum_{k=0}^{\infty} f_2^k\ P_{2k}  \right)
\left\lbrace \sum_{p=0}^{\infty} 
\ \sum_{ i_1,\ldots, i_p \atop\in 2 \ZN^+ + 1}
f_{i_1} f_{i_2}\ldots f_{i_p}\ M_{i_p} \ldots M_{i_2} M_{i_1}\  \right\rbrace \BA_\YM \ .
  \label{00,13}
\end{equation}
The $(N-3)! \times (N-3)!$-matrices $P_{2k}$ and $M_{2k+1}$ are defined to be
the coefficients of Riemann zeta values,
\beq
M_{2k+1} \ \ := \ \  F \, \Big|_{\zeta_{2k+1}} \co  P_{2k} \ \ := \ \ F \, \Big|_{(\zeta_{2})^k}
\label{MsPs}
\eeq
with respect to a particular $\ZQ$ basis of MZVs, see e.g. table
\ref{zetaBasis} in appendix \ref{sec:zeta1}. The non-commutative monomial
$f_2^k f_{i_1} f_{i_2}\ldots f_{i_p}$ is accompanied by a matrix product
$P_{2k} M_{i_p} \ldots M_{i_2} M_{i_1}$. Appendix \ref{sec:zeta} gives a more
detailed account on the aforementioned Hopf algebra structure, the isomorphism
$\phi$ and the motivic version $\zeta^\fm$ of MZVs (\ref{00,4}) to which the
superscript in (\ref{00,13}) alludes.

Thus, for our choice of basis expansion (\ref{MsPs}), the whole information about the open
superstring amplitude, which is not fixed by the structure in \eqn{00,13}, is contained in
the matrices $P_w$ and $M_w$ associated with the single zeta values $\zeta_w$ as (\ref{MsPs}). For the multiplicities $N=5,6$ and $N=7$ these matrices are available  up to weights $w=21,9$ and $w=7$, respectively at \cite{WWW}. In the following we shall be concerned with the computation of these matrices, 
composed from the set of the $(N-3)!$ functions $F^\s:=
F_{\Pi_1}{}^\si$ and relabelings thereof. Prior to this, let us consider some of their
properties yielding a simplification of their
world-sheet integrals.


\section{Field-theory patterns in world-sheet integrals}
\label{ftpatterns}

In this section, we establish a connection between color-ordered subamplitudes
$A_\YM$ in YM theories and world-sheet disk integrals. 
In particular, we 
show that in the open-string amplitude (\ref{00,1}) the role of  the
YM subamplitudes and world-sheet integrals can be swapped. This feature is reminiscent of the
observation of Bern et al. \cite{BCJ, Bern:2010ue, Bern:2012uf,Bern:2011ia} (and references
therein) that color and kinematic contributions to gauge-theory amplitudes
share their algebraic properties and are freely interchangeable. We refer to
the string-theory property as a correspondence between disk integrals and YM
subamplitudes $A_\YM$.

This correspondence is manifested by employing the KLT relations \cite{Kawai:1985xq} and
supplemented by a world-sheet analogues of KK \cite{KK,
DelDuca:1999rs} and BCJ relations \cite{BCJ}. Furthermore, different ways of writing the KLT
relations \cite{BjerrumBohr:2010hn} translate into integration-by-part
identities between world-sheet integrals. The properties and relations derived
in this section will set the stage to get a convenient handle on
the $(N-3)!$ basis of disk integrals\footnote{The 
structure of the $(N-3)!$ dimensional basis of $N$-point disk integrals 
and the formal similarity between the
underlying partial-fraction and integration-by-parts identities and the KK and BCJ
relations, respectively has already been investigated to some extent in \cite{Mafra:2011nw}.}
and to correctly identify their singular behavior in section~\ref{secKinPoles}. 

\subsection{Basic definitions}

According to the calculation in \cite{Mafra:2011nv, Mafra:2011nw}, the set of
functions $F_\Pi{}^\s(\ap)$ introduced in \eqn{00,1} reads
\begin{align}
F_{\Pi}{}^{\si} \eq &(-1)^{N-3} \prod_{i=2}^{N-2} \int_{D(\Pi)} \dd z_i \ \prod_{i<j}^{N-1} |z_{ij}|^{s_{ij}} \\
&\times\si \left\{ \, \frac{s_{12}}{z_{12}} \, \left( \frac{s_{13}}{z_{13}}+\frac{s_{23}}{z_{23}} \right) \, \ldots \, \left(\frac{s_{1,N-2}}{z_{1,N-2}}+\ldots+\frac{s_{N-3,N-2}}{z_{N-3,N-2}} \right) \, \right\} \notag \\[3mm]
\eq  &(-1)^{N-3} \prod_{i=2}^{N-2} \int_{D(\Pi)} \dd z_i \ \prod_{i<j}^{N-1} |z_{ij}|^{s_{ij}} \ \si \left\{ \, \prod_{k=2}^{N-2} \sum_{m=1}^{k-1} \frac{ s_{mk}}{z_{mk}} \, \right\} \ .
\label{01,1}
\end{align}
We are always working in coordinates $z$ mapping the boundary of the disk
world-sheet to the real axis such that $z \in \ZR$. The integration domain $D(\Pi)$ in
(\ref{01,1}) is then defined by $z_{\Pi(i)}<z_{\Pi(i+1)}$ corresponding to a
cyclic ordering of vertex operators along the world-sheet boundary. Conformal
invariance guarantees that world-sheet positions only enter through the differences $z_{ij}
:= z_{i,j} := z_i-z_j$, and three of them are fixed
as\footnote{Note that once we set $z_N=\infty$, the $|z_{iN}|^{s_{iN}}$
contributions to the Koba-Nielsen factor will converge to $1$ by virtue of
momentum conservation.} 
\beq
z_1 \eq 0\co z_{N-1} \eq 1 \co z_N \ \ \rightarrow \ \ \infty
\label{01,fix}
\eeq
in order to mod out the redundancy of the conformal Killing group (CKG) $SL(2,\ZR)$ of
the disk topology. The momentum dependence of the functions $F_\Pi{}^\si$ is
carried by the dimensionless Mandelstam variables (\ref{01,2}). 
 
The functions (\ref{01,1}) can be expressed as linear combinations of certain disk
integrals
\beq
Z_\Pi(1,2,3,\ldots,N-1,N) \ \ := \ \ \frac{1}{{\CV}_{\te{CKG}}} \prod_{i=1}^{N} \int_{D(\Pi)} \dd z_i \  \frac{\prod^{N}_{i<j} |z_{ij}|^{s_{ij}}}{  z_{12} z_{23} \ldots z_{N-1,N} z_{N,1} }\,,
\label{01,3}
\eeq 
whose integrand  is characterized by a cycle of $N$ world-sheet propagators
\makebox{$(z_i-z_j)^{-1} := z_{ij}^{-1}$}, which result
from the superstring CFT computation\footnote{In contrast to bosonic string
theory, there are no closed subcycles of $z_{ij}^{-1}$ in the world-sheet integrand of the superstring which reflects the absence of tachyonic propagators \cite{Mafra:2011nw,Stieberger:2012rq}.} on the disk.
In order to keep manifest the cyclic symmetry $Z_\Pi(1,2,3,\ldots,N-1,N)=
Z_\Pi(2,3,\ldots,N-1,N,1)$, we do not fix the insertion points of
vertex operators as in \eqn{01,fix}. At any rate, the inverse volume $\CV_\te{CKG}$ of
$SL(2,\ZR)$ can be respected anytime by fixing three positions $z_i,
z_j,z_k$ and inserting the Jacobian $|z_{ij} z_{ik} z_{jk}|$. 
In eq. (\ref{01,3}) the cyclic products of world-sheet Green's functions
$z_{ij}^{-1}$ on the disk share several properties with the corresponding field-theory
subamplitudes $A_\YM$. This fact, which
will become clear below, explains the reason for 
their appearance in the $F_\Pi{}^\si$.

\subsection{Gravity tree amplitudes versus superstring disk amplitudes}
\label{KLTsec}

In \eqn{01,1} the integrand of the function $F_\Pi{}^\s$ , in particular the expression in
parenthesis, looks pretty complicated and requires a more intuitive
understanding. For this purpose, one notes the resemblance to the structure of
the KLT relations \cite{Kawai:1985xq}, which allow to write the $N$-graviton
tree amplitudes $\CM$ in perturbative gravity in terms of bilinears in gauge
theory subamplitudes $A_\YM,\tilde A_\YM$:
\begin{align}
\CM&(1,2,\ldots,N) \eq (-1)^{N-3} \sum_{\si \in S_{N-3}} A_\YM(1,\si(2,3,\ldots,N-2),N-1,N) \notag \\
&\! \! \times \sum_{\rho \in S_{N-3}} \! S[\, \rho(2,\ldots,N-2) \, | \, \si(2,\ldots,N-2) \, ]_1 \, \tilde A_\YM(1,\rho(2,3,\ldots,N-2),N,N-1)\,.
\label{01,4}
\end{align}
In the above equation, $S[\rho|\si]_1$ is the field-theory limit of the
momentum kernel\footnote{For convenience we reversed the first permutation in
comparison to the original reference \cite{BjerrumBohr:2010hn}, which turns
$S[\rho|\si]_1$ into a symmetric matrix.}  \cite{BjerrumBohr:2010hn}, a $(N-3)!
\times (N-3)!$-matrix\footnote{In a string theory context, the momentum kernel
gathers monodromy phases due to complex contour deformations on a genus-zero
world-sheet which are used to derive closed string tree amplitudes
\cite{Kawai:1985xq, BjerrumBohr:2010hn} and open-string subamplitudes relations
\cite{BjerrumBohr:2009rd, Stieberger:2009hq}. The $\ap \rightarrow 0$ limit
relevant to our discussion amounts to replacing factors $\sin (\pi s_{ij})$ by
the arguments of the $\sin$ functions.} homogeneous of degree $(N-3)$ in the
Mandelstam variables $s_{ij}$. The subscript $1$ refers to the reference
momentum $k_1$ which shows up in the entries
\beq
S[\, 2_\rho,\ldots,(N-2)_\rho \, | \, 2_\si,\ldots,(N-2)_\si \, ]_1\ \ := \ \ \prod_{j=2}^{N-3} \Big( \, s_{1,j_\rho} \ + \ \sum_{k=2}^{j-1} \theta(j_\rho,k_\rho) \, s_{j_\rho,k_\rho} \, \Big)\,.
\label{01,4a}
\eeq
with shorthand $i_\rho := \rho(i)$. The object $\theta(j_\rho,k_\rho)$
equals 1 if the ordering of the legs $j_\rho,k_\rho$ is the same in the ordered sets
$\rho(2,\ldots,N-2)$ and $\si(2,\ldots,N-2)$, and zero if the ordering is
opposite. In other words, it keeps track of labels which swap their relative
positions in the two permutations $\rho$ and $\si$.

After substituting \eqn{01,4a} into \eqn{01,4} and applying partial-fraction
identities in the integrand of $Z_\Pi$ as 
\begin{equation}
\frac{1}{z_{ij} z_{ik}} +\frac{1}{z_{ji} z_{jk}} +\frac{1}{z_{ki} z_{kj}} =0\,,
  \label{partialfraction}
\end{equation}
one finds the second line of \eqn{01,4} to match the formula (\ref{01,1}) for
the functions\footnote{Note that the extended set of $(N-2)!$ functions
$F_\Pi{}^{\si(23\ldots N-1)}$ considered in subsection 2.5 of
\cite{Mafra:2011nw} accordingly follows by the action of $\si \in S_{N-2}$ on
$\sum_{\rho \in S_{N-3}}  S[2,\ldots,N-2| \rho ]_1
Z_{\Pi}(1,\rho(2,3,\ldots,N-2),N,N-1)$ or any integration-by-parts equivalent
representation thereof.} $F_\Pi{}^\si$,
\beq
F_\Pi{}^\si \eq (-1)^{N-3} \sum_{\rho \in S_{N-3}} S[\, \rho(2,\ldots,N-2) \, | \, \si(2,\ldots,N-2) \, ]_1 \, Z_{\Pi}(1,\rho(2,3,\ldots,N-2),N,N-1)\ ,
\label{01,5}
\eeq
once we perform the $SL(2,\ZR)$ fixing  (\ref{01,3})
\beq
Z_\Pi(1,\rho(2,3,\ldots,N-2),N,N-1) \eq \prod_{i=2}^{N-2} \int_{D(\Pi)} \dd z_i \  \frac{\prod\limits_{i<j}^{N-1} |z_{ij}|^{s_{ij}}}{  z_{1\rho(2)} z_{\rho(2),\rho(3)} \ldots z_{\rho(N-3),\rho(N-2)} } 
\label{01,6}
\eeq
and identify the functions $Z_\Pi$ with $\tilde{A}_\YM$. 

Note, that the computation of the $D=4$ maximally helicity-violating (MHV) superstring disk 
amplitude quite naturally arrives at the basis (\ref{01,6}) after choosing appropriate reference 
spinors and performing partial-fraction decompositions \cite{Stieberger:2012rq}.
 According to this reference, a tree-graph can be associated 
  to the rational-function part of the integrand (\ref{01,3}) and partial fractions  
(\ref{partialfraction}) can be graphically described yielding the 
functions (\ref{01,6}) as a basis. In the language of graphs the set of $(N-3)!$ integrals (\ref{01,6}) represent the same Hamilton 
basis as introduced in \cite{Stieberger:2013hza}.


In the KLT formula (\ref{01,4}), the transposition of legs $N-1$ and $N$
between $A_\YM$ and $\tilde A_\YM$ makes sure that these bilinears exhaust all
pole channels present in the corresponding field theory amplitude. At the level
of the functions $Z_\Pi(1,\ldots,N,N-1)$, the order of legs $N,N-1,1,\ldots$
(with $N-1$ and $1$ adjacent) implies that the Green's function
$z_{N-1,1}^{-1}$ in the cyclic denominator cancels the factor $z_{1,N-1}$ from
the $SL(2,\ZR)$ Jacobian due to \eqn{01,fix}. The remaining rational function is then more suitable
to the methods of section \ref{secPolylogs}.

The  representation (\ref{01,5}) of the functions $F_\Pi{}^\si$ in terms of $Z_\Pi$ casts
the open-string amplitude (\ref{00,1}) into the same form as (\ref{01,4})
\begin{align}
&A_\open(1,\Pi(2,\ldots,N-2),N-1,N) \eq (-1)^{N-3} \sum_{\si \in S_{N-3}} A_\YM(1,\si(2,\ldots,N-2),N-1,N) \notag \\
&\ \ \ \ \times \, \sum_{\rho \in S_{N-3}} S[\, \rho(2,\ldots,N-2) \, | \, \si(2,\ldots,N-2) \, ]_1 \, Z_\Pi(1,\rho(2,3,\ldots,N-2),N,N-1)\;,
\label{01,7}
\end{align}
where the replacement $\tilde A_\YM(\rho) \rightarrow Z_\Pi(\rho)$ builds up the
functions (\ref{01,5}) in the second line. The result (\ref{01,7}) does not depend on which YM sector $\tilde A_\YM$ or 
$A_\YM$ in (\ref{01,4}) is replaced by the integral $Z_\Pi$. As a
consequence, the open-string amplitude (\ref{01,7}) is 
symmetric under the exchange of the YM subamplitude $A_{\YM}$ and 
the world-sheet integral $Z_\Pi$. Note also that color ordering $\Pi$ of the string amplitude is a spectator in the $S_{N-3}$ summation over $\si$ and
$\rho$. That is, the open superstring amplitude $A_\open$ is totally symmetric
in all indices as long as the color ordering $\Pi$ remains unspecified. 
The formal equivalence between (\ref{01,4}) and (\ref{01,7}) as
\beq\label{CORR}
\tilde A_\YM(\rho) \ \ \simeq \ \ Z_\Pi(\rho)
\eeq
makes the exchange symmetry between YM subamplitudes and disk integrals manifest. We shall point
out further faces of this correspondence in subsection \ref{sec1,12}. 

Let us remark, 
that in the four-dimensional spinor helicity formalism, MHV disk
amplitudes \cite{Stieberger:2012rq} allow to establish the correspondence (\ref{CORR}) even at the level of individual
Green's functions $z_{ij}^{-1}$  \cite{Stieberger:2013hza}.

\subsection{Examples}

Let us illustrate the statements above by some examples at multiplicities $N=4,5$ and $N=6$. 

\subsubsection{Four points}
The four-point amplitude 
\beq
A_\open( \Pi(1,2,3,4) ) \eq - \, A_\YM(1,2,3,4) \, s_{12} \, Z_\Pi(1,2,4,3)
\eeq
is governed by the integral
\beq
Z_\Pi(1,2,4,3) \eq \frac{1}{\CV_{\te{CKG}}} \prod_{i=1}^4 \int_{D(\Pi)} \dd z_i \ \frac{ \prod_{i<j}^4 |z_{ij}|^{s_{ij}}}{  z_{12} z_{24} z_{43} z_{31} } \eq  \int_{D(\Pi)} \dd z_2 \ \frac{ |z_{12}|^{s_{12}} \, |z_{23}|^{s_{23}} }{  z_{12}  }  \ .
\label{01,8}
\eeq
In the limit $z_4 \rightarrow \infty$, the factors of $z_{24} z_{43} z_{31}$
are cancelled by the Jacobian $z_{13} z_{14} z_{34}$. Given the $1\times 1$
momentum kernel $S[2|2]_1=s_{12}$, we can rewrite the function $F_\Pi{}^{(2)}$
as
\beq
F_\Pi{}^{(2)} \eq - \, S[\, 2 \,| \, 2 \, ]_1 \, Z_\Pi(1,2,4,3) \eq - \int_{D(\Pi)} \dd z_2 \  |z_{12}|^{s_{12}} \, |z_{23}|^{s_{23}}  \ \frac{ s_{12} }{  z_{12}  }  \ ,
\label{01,9}
\eeq
which is in agreement with (\ref{01,1}).

\subsubsection{Five points}
The five-point integrand involves two permutations $2_\rho = \rho(2), \
3_\rho=\rho(3)$ of the labels $2,3$, \beq
Z_\Pi(1,2_\rho,3_\rho,5,4) \eq \frac{1}{\CV_{\te{CKG}}} \prod_{i=1}^5 \int_{D(\Pi)} \dd z_i \ \frac{ \prod_{i<j}^5 |z_{ij}|^{s_{ij}}}{  z_{12_\rho} z_{2_\rho 3_\rho} z_{3_\rho 5} z_{54} z_{41} } \eq  \int_{D(\Pi)} \dd z_2 \, \dd z_3 \ \frac{ \prod_{i<j}^4 |z_{ij}|^{s_{ij}}}{  z_{12_\rho} z_{2_\rho3_\rho}  } 
\label{01,10}
\eeq
which are tied together by the $2\times 2$ momentum kernel
\beq
S[ \, \rho(2,3) \, | \, \si(2,3) \, ]_1 \eq \ccb S[\,23\, |\, 23\, ]_1 & S[\, 32\, |\,23\, ]_1 \\S[\, 23\,|\,32\,]_1 & S[\, 32\, |\,32\,]_1 \cce \eq \ccb s_{12} (s_{13}+s_{23}) &s_{12} s_{13} \\ s_{12} s_{13} &s_{13}(s_{12}+s_{23}) \,  \cce 
\label{01,11}
\eeq
Inserting these expressions into (\ref{01,5}) reproduces the form (\ref{01,1})
of the $F_{\Pi}{}^{\si}$ after using partial-fraction identities: 
\begin{align}
F_{\Pi}{}^{(23)} \ = \ &\sum_{\rho \in S_2} S[\, \rho(2,3) \, | \, 2,3\,]_1 \, Z_\Pi(1,2_\rho,3_\rho,5,4) \ = \  \int_{D(\Pi)} \dd z_2 \, \dd z_3 \ \prod_{i<j}^4 |z_{ij}|^{s_{ij}} \ \frac{s_{12}}{z_{12}} \, \left( \,\frac{s_{13}}{z_{13}} + \frac{ s_{23}}{z_{23}} \,\right) 
\label{01,12} \\
F_{\Pi}{}^{(32)}\ = \ &\sum_{\rho \in S_2} S[\, \rho(2,3) \, | \, 3,2\, ]_1 \, Z_\Pi(1,2_\rho,3_\rho,5,4) \ = \  \int_{D(\Pi)} \dd z_2 \, \dd z_3 \ \prod_{i<j}^4 |z_{ij}|^{s_{ij}} \ \frac{s_{13}}{z_{13}} \, \left( \,\frac{s_{12}}{z_{12}} + \frac{ s_{32}}{z_{32}}\, \right) \notag
\end{align}


\subsubsection{Six points}

The six-point string corrections are governed by the $\rho \in S_{3}$ basis of $(6-3)!=6$ functions
\beq
Z_\Pi(1,2_\rho,3_\rho,4_\rho,6,5) 
\eq  \int_{D(\Pi)} \dd z_2 \, \dd z_3 \, \dd z_4 \ \frac{ \prod_{i<j}^5 |z_{ij}|^{s_{ij}}}{  z_{12_\rho} z_{2_\rho3_\rho} z_{3_\rho 4_\rho  }} 
\label{01,13}
\eeq
and the $(6\times 6)$-momentum kernel with the following entries in its first row:
\beq
\begin{array}{ll}
S[ \, 234\, | \, 234\, ]_{1} \eq s_{12} (s_{13}+s_{23}) (s_{14}+s_{24}+s_{34})  \ , \ \ \
&S[ \, 342\, | \, 234\, ]_{1} \eq s_{12} s_{13} (s_{14}+s_{34}) \\
S[ \, 243\, | \, 234\, ]_{1} \eq s_{12} (s_{13}+s_{23}) (s_{14}+s_{24})  \ , \ \ \
&S[ \, 423\, | \, 234\, ]_{1} \eq s_{12} (s_{13}+s_{23}) s_{14}\\
S[ \, 324\, | \, 234\, ]_{1} \eq s_{12} s_{13} (s_{14}+s_{24}+s_{34}) \ , \ \ \
&S[ \, 432 \, | \, 234\, ]_{1} \eq s_{12} s_{13} s_{14}
\end{array}
\label{01,14}
\eeq
Repeated use of partial-fraction identities casts the result
of (\ref{01,5}) into the form (\ref{01,1})
\begin{align}
F_{\Pi}{}^{\si} &\ \, = \, \ - \sum_{\rho \in S_3} S[ \, \rho(2,3,4)\, |\, \sigma(2,3,4) \,]_1\, Z_\Pi(1,2_\rho,3_\rho,4_\rho,6,5) \notag \\
&\ \, = \, \ \int_{D(\Pi)} \dd z_2 \, \dd z_3 \, \dd z_4 \ \prod_{i<j}^5 |z_{ij}|^{s_{ij}} \ \frac{s_{12_\si}}{z_{2_\si 1}} \, \left( \,\frac{s_{13_\si}}{z_{13_\si}} + \frac{ s_{2_\si3_\si}}{z_{2_\si3_\si}} \, \right)  \, \left( \,\frac{s_{14_\si}}{z_{14_\si}} + \frac{ s_{2_\si4_\si}}{z_{2_\si4_\si}}+ \frac{ s_{3_\si4_\si}}{z_{3_\si4_\si}} \, \right)   \ .
\label{01,15}
\end{align}
Higher--point analogues of the functions consist of more and more factors
$\sum_{m=1}^{k-1} \frac{ s_{mk}}{z_{mk}}$ with an increasing number of terms
each.

\subsection{World-Sheet analogues of KK and BCJ relations}
\label{sec1,12}

The gravity amplitude (\ref{01,4}) does not have any notion of color ordering,
it is a totally symmetric function w.r.t. its labels $1,2,\ldots,N$. Although
this symmetry is obscured on the right hand side of the KLT formula, it can be
verified to hold through the KK relations (discovered in \cite{KK}
and proven in \cite{DelDuca:1999rs}) and BCJ relations \cite{BCJ} obeyed by the
amplitudes $A_\YM(\si)$ and $\tilde A_\YM(\rho)$. The connection between
gravity amplitudes (\ref{01,4}) and open-string subamplitudes (\ref{01,7})
motivates to investigate whether the underlying disk integrals $Z_\Pi(\rho)$
(\ref{01,6}) taking the role of $\tilde A_\YM(\rho)$ satisfy the same KK and
BCJ relations.

As demonstrated in appendix \ref{appA}, the $Z_\Pi(\rho)$ at fixed
color ordering $\Pi$ share all the algebraic properties of $A_\YM(\rho)$ when
the permutation $\rho$ determining the integrand is varied. Firstly, they
satisfy a world-sheet analogue of KK relations
\beq
Z_\Pi(1, \al , N-1, \be ) \eq (-1)^{|\be|} \sum_{\si \in \al \shuffle \be^t} Z_\Pi(1,\si,N-1) 
\label{KKrelations}
\eeq
where $\al$ and $\b$ denote disjoint ordered subsets of $\{2,\ldots,N-2,N\}$
such that $\al \,\cup\, \be = \{2,\ldots,N-2,N\}$, the $^t$ operation reverses the order of the elements in $\be$, and $|\be|$ is given by the number of
elements in $\beta$. The summation range $\al \shuffle \be^t$ includes those
permutations of $\al \,\cup\, \be$ which preserve the order of elements in
$\al$ and $\b^t$.
Secondly, the BCJ relations among $A_{\te{YM}}$ \cite{BCJ} literally
translate into
\begin{align}
s_{1,N-2} \, &Z_\Pi(1,N-2,2,3,\ldots,N,N-1) \notag \\
&+\sum_{j=2}^{N-3} \sum_{k=1}^j s_{k,N-2}\, Z_\Pi(1,2,\ldots,j,N-2,j+1,\ldots,N,N-1) \notag \\
& - \ s_{N-2,N-1} \, Z_{\Pi}(1,2,\ldots,N-3,N,N-2,N-1) \eq 0 \ . 
\label{01,19new}
\end{align}
A detailed discussion and derivation of (\ref{KKrelations}) and
(\ref{01,19new}) based on world-sheet manipulations can be found in appendix
\ref{appA}.

 These analogues of KK and BCJ relations fulfilled by the integrals $Z_\Pi(\rho)$ (regardless of
$\Pi$) imply that the open-string CFT correlator enjoys the same total exchange
symmetry in $1,2,\ldots,N$ as the gravity amplitude. Individual disk
subamplitudes $A_\open(\Pi)$ then stem from integrating a totally symmetric
correlator over a domain specified by $z_{\Pi(i)} < z_{\Pi(i+1)}$, that is, the
ordering of the vertex operators along the disk boundary. The integration
region determined by $\Pi$ breaks the complete permutation symmetry $S_N$ of
the integrand down to a cyclic subgroup $\ZZ_N$. Hence, establishing an
analogue of KK and BCJ relations among the $Z_\Pi(\rho)$ ultimately guarantees
cyclic symmetry of $A_\open(\Pi)$ given by (\ref{01,7}).

\subsection{Different KLT representations and integration by parts}
\label{sec1,13}

An important feature to conveniently determine the $\ap$-expansion of the functions
$F_\Pi{}^\si$ in the string amplitude (\ref{00,1}) is the choice of 
a convenient basis of integrals $Z_\Pi$'s. In this subsection, we will show that there are
$N-2$ different $S_{N-3}$-families of such $Z_\Pi(\ldots)$. One is free to
focus on the $S_{N-3}$ slice of $Z_\Pi(\ldots)$ whose low-energy behavior is
easiest to access (the criterion will follow in the later section
\ref{secKinPoles}). The freedom in constructing the $F_{\Pi}{}^\si$ can be
explained both by world-sheet integration by parts and by the analogy of the
disk amplitude (\ref{01,7}) with the KLT formula (\ref{01,4}) for gravity tree
amplitudes.

Contour deformation and the 
underlying monodromy relations \cite{BjerrumBohr:2009rd,Stieberger:2009hq} 
allow to rewrite the KLT formula in a variety of ways
such that the number of terms in (\ref{01,4}) or
(\ref{01,7}) can be reduced from $(N-3)!\times (N-3)!$ down to $(N-3)!\times
(\nu-1)! \times (N-2-\nu)!$ for some $\nu=1,2,\ldots,N-2$ 
\cite{Kawai:1985xq,Bern:1998sv}. In the language of
the momentum kernel \cite{BjerrumBohr:2010hn}, this gives rise to equivalent
representations\footnote{Similar to (\ref{KKrelations}), the permutation
$\tau^t$ is related to $\tau$ by reversal of its elements.} 
\begin{align}
\CM(1,2,\ldots,N) &\eq (-1)^{N-3} \sum_{\si \in S_{N-3}} A_\YM(1,\si(2,3,\ldots,N-2),N-1,N)   \notag \\
&\times \sum_{\rho \in S_{\nu-1}} \sum_{\tau \in S_{N-2-\nu}}  S[\, \rho(2_\si,\ldots,\nu_\si) \, | \, 2_\si,\ldots,\nu_\si \, ]_1  \notag \\
&\times S[\, \tau((N-2)_\si,\ldots,(\nu+1)_\si) \, | \, (N-2)_\si ,\ldots,(\nu+1)_\si \, ]_{N-1} \notag \\
&\times\tilde A_\YM(1,\rho(2_\si,\ldots,\nu_\si),N, \tau^t((N-2)_\si,\ldots, (\nu+1)_\si), N-1)\ ,
\label{01,20}
\end{align}
for all values $\nu=1,\ldots,N-2$. Eq. (\ref{01,4}) corresponds to taking $\nu=N-2$ in (\ref{01,20}). At odd multiplicity $N$, the most economic
choice $\nu = \frac{1}{2}(N-1)$ w.r.t. the number of terms makes
reflection symmetry manifest: At five points, $\nu=2$ yields a two-term
representation
\beq
\CM(1,2,\ldots,5) \eq \sum_{\si \in S_{2}} A_\YM(1,\si(2,3),4,5) \, S[\, 2_\si \, | \, 2_\si \, ]_1 \, S[\, 3_\si \, | \, 3_\si \, ]_{4} \, \tilde A_\YM(1,2_\si,5, 3_\si, 4)\ ,
\label{01,21}
\eeq
with $S[2_\si  | 2_\si  ]_1=s_{12_\si}$ and $S[ 3_\si  | 3_\si  ]_{4} =
s_{3_\si 4}$, and the seven-point amplitude at $\nu=3$ takes the form
\begin{align}
\CM(1,2,&\ldots,7) \eq  \sum_{\si \in S_{4}} A_\YM(1,\si(2,3,4,5),6,7)   \sum_{\rho \in S_{2}}  S[\, \rho(2_\si,3_\si) \, | \, 2_\si,3_\si \, ]_1 \notag \\
&\times \  \sum_{\tau \in S_{2}}  S[\, \tau(5_\si,4_\si) \, | \, 5_\si ,4_\si \, ]_{6} \,\tilde A_\YM(1,\rho(2_\si,3_\si),7, \tau^t(5_\si,4_\si), 6)\ ,
\label{01,22}
\end{align}
with $24 \times 4$ terms rather than $24\times 24$. Even multiplicities, on the other hand, leave
two values $\nu = \frac{1}{2}N - 1, \tfrac{1}{2}N$ for the minimal number of
terms in (\ref{01,20}).

The  dictionary (\ref{CORR}) between the supergravity  amplitude $\CM$ and
the open superstring amplitude $A_\open(\Pi)$ maps the freedom
in the choice of the parameter $\nu$ in (\ref{01,20}) to world-sheet integrations by parts. By discarding a total $z_{N-2}$ derivative,
we can reduce the number of terms in the representation (\ref{01,1})  of
$F_{\Pi}{}^\si$ by $N-3$
\beq
 \int_{D(\Pi)}\dd z_{N-2} \  \prod_{i<j}^{N-1}  |z_{ij}|^{s_{ij}} \, \left( \, \frac{ s_{1,N-2}}{ z_{1,N-2} } + \ldots + \frac{ s_{N-3,N-2}}{ z_{N-3,N-2} }  \, \right) \eq  \int_{D(\Pi)} \dd z_{N-2}\  \prod_{i<j}^{N-1}  |z_{ij}|^{s_{ij}} \, \frac{ s_{N-2,N-1}}{ z_{N-2,N-1} }  \ . \label{01,23}
\eeq 
This results in moving from case $\nu = N-2$ to case $\nu= N-3$. Generically, for any
$\nu=1,\ldots,N-2$ we have:
\begin{align}
A_\open(1,\Pi(2,\ldots,N-2)&,N-1,N) \eq (-1)^{N-3} \sum_{\si \in S_{N-3}} A_\YM(1,\si(2,3,\ldots,N-2),N-1,N)   \notag \\
& \times \sum_{\rho \in S_{\nu-1}} \sum_{\tau \in S_{N-2-\nu}}
 S[\, \rho(2_\si,\ldots,\nu_\si) \, | \, 2_\si,\ldots,\nu_\si \, ]_1   \notag \\  
& \times S[\, \tau((N-2)_\si,\ldots,(\nu+1)_\si) \, | \, (N-2)_\si ,\ldots,(\nu+1)_\si \, ]_{N-1} \notag \\
& \times  Z_\Pi(1,\rho(2_\si,\ldots,\nu_\si),N, \tau^t((N-2)_\si,\ldots, (\nu+1)_\si), N-1) \ .
\label{01,24}
\end{align}
Note, that the Green's functions in
$Z_\Pi(1,\rho(2_\si,\ldots,\nu_\si),N, \tau^t((N-2)_\si,\ldots, (\nu+1)_\si), N-1)$ factorize into $\rho(z_{12_\si} \ldots z_{(\nu-1)_\si,\nu_\si} )^{-1}$ and $\tau(z_{(\nu+1)_\si,(\nu+2)_\si} \ldots z_{(N-2)_\si,N-1})^{-1}$ after $SL(2,\ZR)$ fixing. Carrying out the $\rho$-sum in (\ref{01,24}) yields
\beq
\sum_{\rho \in S_{\nu-1}} \frac{S[\, \rho(2_\si,\ldots,\nu_\si) \, | \, 2_\si,\ldots,\nu_\si \, ]_1}{\rho(z_{12_\si} \ldots z_{(\nu-1)_\si,\nu_\si}) } 
\eq \si \left( \, \prod_{k=2}^{\nu} \sum_{m=1}^{k-1} \frac{ s_{mk}}{z_{mk}} \, \right) \ .
\label{facto}
\eeq
Eventually, after repeating the manipulation (\ref{01,23}) for derivatives in $z_{N-3},
z_{N-4},\ldots,z_{\lfloor N/2 \rfloor+1}$ casts the function $F_\Pi{}^\si$ 
into a form with a minimal number of terms at $\nu = \lfloor N/2 \rfloor$:
\begin{align}
F_\Pi{}^\si \eq &(-1)^{N-3} \, \sum_{\rho \in S_{\lfloor N/2 \rfloor-1} } \, S[\, \rho(2_\si,\ldots,\lfloor N/2 \rfloor_\si) \, | \, 2_\si,\ldots,\lfloor N/2 \rfloor_\si \, ]_1 \notag \\
& \ \times \ \sum_{\tau \in S_{\lceil N/2 \rceil-2}} S[\, \tau((N-2)_\si,\ldots,(\lfloor N/2 \rfloor+1)_\si) \, | \, (N-2)_\si ,\ldots,(\lfloor N/2 \rfloor+1)_\si \, ]_{N-1}  \notag \\
& \ \times \ Z_\Pi(1,\rho(2_\si,\ldots,\lfloor N/2 \rfloor_\si),N, \tau^t((N-2)_\si,\ldots, (\lfloor N/2 \rfloor+1)_\si), N-1) \label{01,25} \\
\eq &(-1)^{N-3} \prod_{i=2}^{N-2} \int_{D(\Pi)} \dd z_i \ \prod_{i<j}^{N-1} |z_{ij}|^{s_{ij}} \ \si \left\{ \, \left( \, \prod_{k=2}^{\lfloor N/2 \rfloor} \sum_{m=1}^{k-1} \frac{ s_{mk}}{z_{mk}} \, \right) \, \left( \, \prod_{k=\lfloor N/2 \rfloor+1}^{N-2} \sum_{n=k+1}^{N-1} \frac{ s_{kn}}{z_{kn}} \, \right) \, \right\} \ .
\notag
\end{align}
At five points, e.g. the  representation for $\nu=2$
\beq
F_\Pi{}^\si \eq \int_{D(\Pi)} \dd z_2 \, \dd z_3 \ \prod_{i<j}^4 |z_{ij}|^{s_{ij}} \ \frac{s_{12_\si}}{z_{12_\si}} \,
\frac{s_{3_\si 4}}{z_{3_\si 4}}\  \ \ : \ \nu =2
\label{01,26}
\eeq
involves fewer terms than the  representations   of the basis functions for 
$\nu=1,3$:
\beq
F_\Pi{}^{\si} \eq \left\{ \ \begin{array}{ll}\displaystyle \int_{D(\Pi)} \dd z_2 \, \dd z_3 \ \prod_{i<j}^4 |z_{ij}|^{s_{ij}} \ \frac{s_{12_\si}}{z_{12_\si}} \, \left( \,\frac{s_{13_\si}}{z_{13_\si}} + \frac{ s_{2_\si3_\si}}{z_{2_\si3_\si}} \,\right) &: \ \nu =3\ ,\\
\displaystyle \int_{D(\Pi)} \dd z_2 \, \dd z_3 \ \prod_{i<j}^4 |z_{ij}|^{s_{ij}} \   \left( \,  \frac{ s_{2_\si3_\si}}{z_{2_\si3_\si}} +\frac{ s_{2_\si4}}{z_{2_\si 4}}  \,\right) \, \frac{s_{3_\si 4}}{z_{3_\si 4}}&: \ \nu =1\ . \end{array} \right.
\label{01,27}
\eeq
As we can see in (\ref{01,25}), the freedom in rewriting the KLT-like formula for $A_\open$ enables to reconstruct the $F_\Pi{}^\si$ from any set of $(N-3)!$ functions $Z_{\Pi}(1,2_\si,\ldots,\nu_\si,N,(\nu+1)_\si,\ldots,(N-2)_\si,N-1)$ with $\nu=1,2,\ldots,N-2$. The situation is summarized in figure \ref{fig1}.
\begin{center}
\begin{figure}[H]
\begin{tikzpicture} [scale = 1, line width=0.30mm, xshift=-1cm]
\draw (0,0.5) -- (0,-0.5) ;
\draw (0,0.5) -- (4,0.5) ;
\draw (0,-0.5) -- (4,-0.5) ;
\draw (4,0.5) -- (4,-0.5) ;
\scope[xshift=6.5cm]
\draw (-0.7,0.5) -- (-0.7,-0.5) ;
\draw (-0.7,0.5) -- (4.7,0.5) ;
\draw (-0.7,-0.5) -- (4.7,-0.5) ;
\draw (4.7,0.5) -- (4.7,-0.5) ;
\endscope
\scope[xshift=13cm]
\draw (0,0.5) -- (0,-0.5) ;
\draw (0,0.5) -- (4,0.5) ;
\draw (0,-0.5) -- (4,-0.5) ;
\draw (4,0.5) -- (4,-0.5) ;
\endscope
\draw (2,0) node{$Z_\Pi(1,\rho_{N-3},N,N-1)$};
\draw (4.5,-1.7) node{$Z_\Pi(1,\rho_{N-4},N,\tau_1,N-1)$};
\draw (8.5,0) node{$Z_\Pi(1,\rho_{\nu-1},N,\tau_{N-2-\nu},N-1)$};
\draw (12.5,-1.7) node{$Z_\Pi(1,\rho_1,N,\tau_{N-4},N-1)$};
\draw (15,0) node{$Z_\Pi(1,N,\tau_{N-3},N-1)$};

\scope[xshift=2.5cm,yshift=-1.7cm]
\draw (-0.3,0.5) -- (-0.3,-0.5) ;
\draw (-0.3,0.5) -- (4.3,0.5) ;
\draw (-0.3,-0.5) -- (4.3,-0.5) ;
\draw (4.3,0.5) -- (4.3,-0.5) ;
\endscope
\scope[xshift=10.5cm,yshift=-1.7cm]
\draw (-0.3,0.5) -- (-0.3,-0.5) ;
\draw (-0.3,0.5) -- (4.3,0.5) ;
\draw (-0.3,-0.5) -- (4.3,-0.5) ;
\draw (4.3,0.5) -- (4.3,-0.5) ;
\endscope
\draw (0,3.5) -- (0,2.5);
\draw (0,3.5) -- (17,3.5) ;
\draw (0,2.5) -- (17,2.5) ;
\draw (17,3.5) -- (17,2.5);
\draw (8.5,3) node{$F_\Pi{}^{\si}$};
\draw[dashed] (7.5,3) -- (5.5,3);
\draw (4,3) node{int. by parts};
\draw[dashed,->] (2.5,3) -- (0.5,3);
\draw[dashed] (9.5,3) -- (11.5,3);
\draw (13,3) node{int. by parts};
\draw[dashed,->] (14.5,3) -- (16.5,3);
\draw[<->] (2,0.5) -- (2,2.5);
\draw[<->] (4.5,-1.2) -- (4.5,2.5);
\draw[<->] (8.5,0.5) -- (8.5,2.5);
\draw[<->] (12.5,-1.2) -- (12.5,2.5);
\draw[<->] (15,0.5) -- (15,2.5);
\draw (1.1,1.5)node{$S[\rho_{N-3}]$};
\draw (3.6,1.5)node{$S[\rho_{N-4}]$};
\draw (5.1,1.5)node{$S[\tau_{1}]$};
\draw (7.65,1.5)node{$S[\rho_{\nu-1}]$};
\draw (9.4,1.5)node{$S[\tau_{N-2-\nu}]$};
\draw (11.9,1.5)node{$S[\rho_{1}]$};
\draw (13.3,1.5)node{$S[\tau_{N-4}]$};
\draw (15.9,1.5)node{$S[\tau_{N-3}]$};
\draw (5.15,0)node{$\hdots$};
\draw (11.85,0)node{$\hdots$};
\draw (6.5,1.5)node{$\hdots$};
\draw (10.5,1.5)node{$\hdots$};
\end{tikzpicture}
\caption{Each of the $N-2$ integration-by-parts equivalent representations of
the basis functions $F_\Pi{}^\si$ can be mapped to another $(N-3)!$ basis of KK
integrals $Z_\Pi(1,\ldots,N-1)$. Depending on the position $\nu+1$ of leg $N$
in the $Z_\Pi$, the transformation matrix is given by a product of $(\nu-1)$-
and $(N-2-\nu)$-particle momentum kernels.}
\label{fig1}
\end{figure}
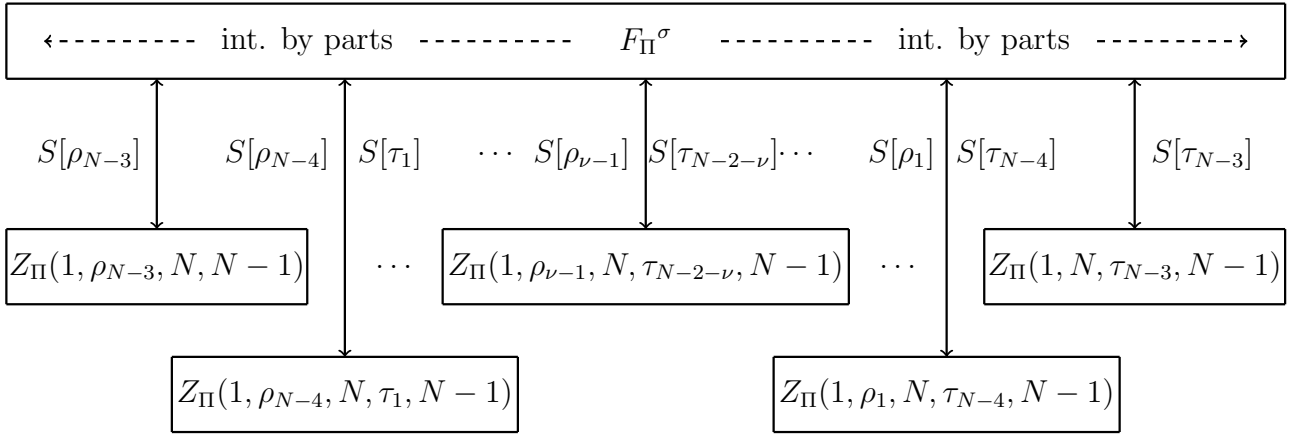
\end{center}

\subsection{Basis expansion of individual integrals}
\label{sec:inv}

Even though the general aim of this article is the construction of basis
functions $F_\Pi{}^\si$ from disk integrals $Z_\Pi(1,\ldots,N-1)$ as shown in
\eqn{01,25}, we shall now invert these relations to facilitate the translation
between the two types of objects. The expansion of KK basis integrals in terms
of $F_\Pi{}^\si$ introduces the inverse of the momentum kernel whose entries
involve $N-3$ simultaneous poles in Mandelstam invariants. The four-point case
$S^{-1}[2|2]_1=s_{12}^{-1}$ extends as follows to five and six points:
\beq
\ccb S^{-1}[\,23\, |\, 23\, ]_1 & S^{-1}[\, 32\, |\,23\, ]_1 \\S^{-1}[\, 23\,|\,32\,]_1 & S^{-1}[\, 32\, |\,32\,]_1 \cce \eq \frac{1}{s_{123} }\,\ccb \frac{1}{s_{12}}+\frac{1}{s_{23}} &-\frac{1}{s_{23}} \\ -\frac{1}{s_{23}} &\frac{1}{s_{13}}+\frac{1}{s_{23}} \cce
\eeq
\begin{align}
S^{-1}[ \, 234 \, | \, 234 \, ]_1 &\eq \frac{1}{s_{1234}} \, \Big( \, \frac{1}{s_{234}}  \Big(\frac{1}{s_{23}}+\frac{1}{s_{34}} \Big) + \frac{1}{s_{12}s_{34}}+ \frac{1}{s_{123}}  \Big(\frac{1}{s_{12}}+\frac{1}{s_{23}} \Big) \, \Big)   \notag \\
S^{-1}[ \, 243 \, | \, 234 \, ]_1 &\eq - \frac{1}{s_{1234}s_{34}} \, \Big(\frac{1}{s_{12}}+\frac{1}{s_{234}} \Big) \co S^{-1}[ \, 342 \, | \, 234 \, ]_1 \eq  -\frac{1}{s_{1234}s_{34}s_{234}}
 \notag \\
 S^{-1}[ \, 324 \, | \, 234 \, ]_1 &\eq - \frac{1}{s_{1234}s_{23}} \, \Big(\frac{1}{s_{123}}+\frac{1}{s_{234}} \Big)
  \ ,\ \ \ \ \ S^{-1}[ \, 423 \, | \, 234 \, ]_1 \eq -\frac{1}{s_{1234}s_{23}s_{234}}  \notag\\
 S^{-1}[ \, 432 \, | \, 234 \, ]_1 &\eq \frac{1}{s_{1234}s_{234}} \, \Big(\frac{1}{s_{23}}+\frac{1}{s_{34}} \Big) \,.
\label{invmom1}
\end{align}
These combinations of $(N-3)$-fold poles govern the inverse of (\ref{01,25}),
\begin{align}
&Z_\Pi(1,2_\si,\ldots,\nu_\si,N,(\nu+1)_\si ,\ldots ,(N-2)_\si,N-1) \eq (-1)^{N-3} \sum_{\rho \in S_{N-3}} \, F_\Pi{}^{\rho(2_\si,3_\si,\ldots,(N-2)_\si)} \notag \\
&\times \, S^{-1}[\, 2_\si ,\ldots, \nu_\si \,|\, \rho(2_\si ,\ldots, \nu_\si) \, ]_1 \, S^{-1}[ \, (N\!-\!2)_\si,\ldots,(\nu\!+\!1)_\si \, | \,  \rho ( (N\!-\!2)_\si  ,\ldots, (\nu\!+\!1)_\si) \, ]_{N-1} \ ,
\label{invmom2}
\end{align}
where $S^{-1}[\al|\be]_{j}=0$ unless $\al$ is a permutation of $\be$, that
is, the inverse momentum kernel can only take nonzero values if the same set of
labels appears in its two slots.

Let us illustrate (\ref{invmom2}) by two examples: 
\begin{itemize}
\item At $N=5$ and $\nu=2$, we have
\begin{align}
Z_\Pi(1,2_\si,5,3_\si,4) \eq \sum_{\rho \in S_2} F_{\Pi}{}^{\rho(2_\si,3_\si)} \, S^{-1}[\, 2_\si  \,|\, \rho(2_\si) \, ]_1\, S^{-1}[\, 3_\si  \,|\,   \rho(3_\si)  \, ]_4  \eq \frac{ F_{\Pi}{}^{\si(23)} }{s_{1 2_\si} s_{3_\si 4}}
\end{align}
since the second term with $\rho(2_\si,3_\si)=(3_\si,2_\si)$ yields $S^{-1}[\, 2_\si  \,|\, 3_\si \, ]_1= S^{-1}[\, 3_\si  \,|\, 2_\si  \, ]_4=0$. 
\item At $N=6$ and $\nu=1$, on the other hand,
\begin{align}
Z_\Pi&(1,6,2_\si,3_\si,4_\si,5) \eq - \sum_{\rho \in S_3} F_{\Pi}{}^{\rho(2_\si,3_\si,4_\si)} \, S^{-1}[ \, 4_\si3_\si2_\si \, |\, \rho(4_\si,3_\si,2_\si)  \,]_5  \notag \\
&\eq - \, \frac{ F_\Pi{}^{\si(234)} }{ s_{2345} } \, \Big( \, \frac{ 1 }{s_{3_\si 4_\si} s_{3_\si 4_\si 5}} + \frac{ 1 }{s_{4_\si 5} s_{3_\si 4_\si 5}} + \frac{ 1 }{s_{2_\si 3_\si} s_{4_\si 5}} +  \frac{ 1 }{s_{3_\si 4_\si} s_{234}} + \frac{ 1 }{s_{2_\si 3_\si} s_{234}} \, \Big) \notag \\
& \ \ \ \ \ \ + \ \frac{ F_\Pi{}^{\si(243)} }{ s_{2345} s_{3_\si 4_\si} } \, \Big( \, \frac{ 1 }{s_{234}}+\frac{ 1 }{s_{3_\si 4_\si 5} } \, \Big)
\ + \ \frac{ F_\Pi{}^{\si(324)} }{ s_{2345} s_{2_\si 3_\si} } \, \Big( \, \frac{ 1 }{s_{234}}+\frac{ 1 }{s_{4_\si 5} } \, \Big)
\ + \ \frac{ F_\Pi{}^{\si(342)} }{ s_{2345}s_{234} s_{3_\si 4_\si} } 
\notag \\
& \ \ \ \ \ \ + \ \frac{ F_\Pi{}^{\si(423)} }{ s_{2345}s_{234} s_{2_\si 3_\si} } \ - \ \frac{ F_\Pi{}^{\si(432)} }{ s_{2345}s_{234}  } \, \Big( \, \frac{ 1 }{s_{2_\si 3_\si}} +\frac{ 1 }{s_{3_\si 4_\si}} \, \Big)\,.
 \end{align}
\end{itemize}
We will make use of the relation (\ref{invmom2}) in assembling the final
results in section \ref{sec3} below.


\section{Kinematic poles}
\label{secKinPoles}

In the previous section we have seen  how the string
corrections $F_\Pi{}^\si(\ap)$ to disk amplitudes are built from generalized
Euler integrals $Z_{\Pi}(\rho)$ with $\rho \in S_{N}$. Their integrands are
characterized by the totally symmetric Koba-Nielsen factor $\prod_{i<j}^N
|z_{ij}|^{s_{ij}}$ decorated by a cyclic product $\rho(z_{12} z_{23}\ldots
z_{N-1,N} z_{N,1})^{-1}$ of Green's functions (on the sphere). The latter  endows the
functions $Z_{\Pi}(\rho)$ with the same algebraic properties as the YM amplitudes $A_\YM(\rho)$. 
In particular, the freedom due to world-sheet integration by parts allows to
assemble the complete basis of $F_\Pi{}^\si(\ap)$ from each of the $N-2$
different sectors of $Z_{\Pi}(\rho)$, classified by $\nu=1,2,\ldots,N-2$:
\beq
\bigl\{ \, Z_{\Pi}(1,2_\rho,\ldots,\nu_\rho,N,(\nu+1)_\rho,\ldots,(N-2)_\rho,N-1) \ , \ \ \rho \in S_{N-3} \, \big\} 
\ , \ \ \ \nu=1,2,\ldots,N-2\;.
\label{02,1}
\eeq

The multiple resonance exchange in an $N$-point scattering process is reflected
in up to $N-3$ simultaneous poles in the Mandelstam variables. They are the
source of field theory propagators within disk amplitudes (or, equivalently,
non-vanishing field theory limits for some of the $F_\Pi{}^\si$) and pose a major
complication in performing an $\ap$-expansion of the $Z_{\Pi}$. The
identification and classification of pole channels has been thoroughly
explained in \cite{Mafra:2011nw}. In the following, we will review selected
aspects thereof and formulate the resulting prescriptions to directly read off
the pole structure from the permutations $\Pi,\rho$ characterizing
$Z_\Pi(\rho)$. Moreover, we will investigate the $\ap$-expansion of pole
residues and reveal their recursive structure. 
Massless poles occur in integration regions where several neighboring vertex
operators collide. So far, we have left the disk ordering $\Pi$ unspecified
since it has no impact on the structure and possible manipulations of the
integrand. Kinematic poles of the integrated amplitude, however, are sensitive
to the color ordering. Hence, we choose the canonical disk ordering $\Pi
=(1,2,\ldots,N)$ in the following (keeping the usual $SL(2,\ZR)$ fixing $z_1 =
0, \ z_{N-1}=1, \ z_N \rightarrow \infty$) and omit the $\Pi$ subscripts of
$Z_\Pi(\rho)$ and $F_\Pi{}^\si$. 
The integration domain $0 = z_1 \leq z_2 \leq z_3 \leq \ldots \leq
z_{N-2} \leq z_{N-1}=1$ allows for singularities in the multiparticle
Mandelstam variables $s_{i,i+1\ldots,j-1,j}$ defined by (\ref{01,2}) from the
regime $z_{i+1},z_{i+2},\ldots,z_{j-1},z_{j} \rightarrow z_i$ for $1\leq i < j
\leq N-1$. Momentum conservation
$s_{j+1,\ldots,N,1,\ldots,i-1}=s_{i,i+1\ldots,j-1,j}$ guarantees that we can
avoid explicit reference to the momentum $k_N$ from the vertex operator at $z_N
\rightarrow \infty$ without omitting any pole channel. 

  The first subsection \ref{sec1,21} provides a criterion to find the
complete set of pole channels present in the individual $Z(\rho)$ of the form
(\ref{02,1}). Subsection \ref{sec1,22} then explains how to reduce such
$Z(\rho)$ to integrals whose low-energy limit
reproduces the propagators of a {\it single} diagram from the underlying YM field
theory comprising cubic vertices only (we shall refer to such diagrams as
`cubic' below). Factorization properties imply that the residues of $N$-point
amplitudes on a $k$-fold massless pole are given by (off-shell) $(N-k)$-point
amplitudes (or products of $p$ lower-point amplitudes with overall multiplicity
$N-k+3(p-1)$). The manifestation of factorization at the level of integrals and
the systematics of the residues' recursive structure is presented in subsection
\ref{sec1,23}.

\subsection{Identifying pole channels}
\label{sec1,21}

The kinematic poles present in a given $Z(\rho)$ are not immediately obvious
from the integrand. Thus we will give a criterion (\ref{pole}) below which
relates the pole structure of an integral $Z(\rho)$ with the $\rho$-dependent
integrand, see (\ref{01,6}). We address this problem for the complete KK
basis\footnote{We want to emphasize that any other integral beyond the KK basis
$\{ Z(1,\rho(2,3,\ldots,N-2,N),N) , \ \rho \in S_{N-2} \}$ can be brought into
the form \eqn{02,2} by means of KK relations \eqn{KKrelations}. So there is no
loss of generality in restricting the discussion to integrals of the form
(\ref{02,2}) with a degree $N-3$ polynomial $R^{\nu,\rho}$ in $z_{ij}$.},
firstly for the sake of generality, and secondly to gather evidence in favor of
the $\nu = \lfloor N/2\rfloor$ slice of (\ref{02,1}) which leads to a complete
$F^\si$ basis in the most economic way. Once we have stripped off the
Koba-Nielsen factor, one can refer to the Green's functions in the integrand by
a polynomial $R^{\nu,\rho}(z_{ij})$ of degree $(N-3)$: 
\begin{align}
Z(1,2_\rho,&\ldots,\nu_\rho,N,(\nu+1)_\rho,\ldots,(N-2)_\rho,N-1) \eq \prod_{i=2}^{N-2} \int^{z_{i+1}}_0 \dd z_i \  \prod_{i<j}^{N-1} |z_{ij}|^{s_{ij}} \, \frac{1}{R^{\nu,\rho}(z_{ij})}\notag \\
&R^{\nu,\rho}(z_{ij}) \ \ := \ \  ( z_{12_\rho} z_{2_\rho 3_\rho}\ldots z_{(\nu-1)_\rho,\nu_\rho}) \ \times \ (z_{(\nu+1)_\rho,(\nu+2)_\rho} \ldots z_{(N-2)_\rho,N-1})  \ .
\label{02,2}
\end{align}
Note that the world-sheet parity transformation $(z_{i},k_i \mapsto
z_{N-i},k_{N-i})$ for $i=1,2,\ldots,N-1$ preserves the form of (\ref{02,2})
including the $SL(2,\ZR)$ fixing and exchanges $S_{N-3}$ sectors of the KK
basis associated with $\nu$ and $N-1-\nu$.

The following change of integration variables allows to conveniently probe
several pole channels which will turn out to be sufficient for finding the
general criterion (\ref{pole}):
\beq
z_{2} \eq x_1\, x_2 \ldots x_{N-3} \co z_3 \eq x_2\, x_3 \ldots x_{N-3} \ , \ \ \ \ \ldots \ \ \  \co z_{N-2} \eq x_{N-3} \ .
\label{02,3}
\eeq
This transformation maps the integration region from a simplex $0 \leq z_2 \leq
z_3 \leq \ldots \leq z_{N-2} \leq 1$ to the unit cube $0 \leq x_i \leq 1$, and
the integration measure becomes
\beq
\prod_{i=2}^{N-2} \int^{z_{i+1}}_0 \dd z_i \  \prod_{i<j}^{N-1}  |z_{ij}|^{s_{ij}}  \eq \prod_{i=1}^{N-3} \int^1_0 \dd x_i \ \prod_{j=1}^{N-3} x_j^{j-1+s_{12\ldots j+1}} \, (1-x_j)^{s_{j+1,j+2}} \, \prod_{l=j+1}^{N-3} \left( \, 1 - \prod_{k=j}^{l} x_k \, \right)^{s_{j+1,l+2}} 
\label{02,4}
\eeq
with $\prod_{j=1}^{N-3} x_j^{j-1}$ emerging from the Jacobian $\big| \frac{ \pa
z_i }{\pa x_j} \big|$. Including the polynomials $R^{\nu,\rho}(z_{ij})$ of
(\ref{02,2}) into (\ref{02,4}) effectively shifts $s_{ij}\mapsto s_{ij} - 1$
for each $z_{ij}$-factor present in $R^{\nu,\rho}$, see e.g.
\cite{Mafra:2011nw}. This is crucial for identifying kinematic poles: For
functions $f,g$ which are regular at $x=0$ and $x=1$, respectively, we find the
following behavior\footnote{This is a consequence of the $\de$-function
representation $\de(x) = \lim_{s\rightarrow0} s x^{s-1}$.}
\beq
\int^a_0 \dd x \ x^{s-1} \, f(x) \eq \frac{f(0)}{s} \  + \ {\cal O}(s^0) \co \int^1_{1-b} \dd x \ (1-x)^{s-1} \, g(x) \ \ = \ \ \frac{g(1)}{s} \  + \ {\cal O}(s^0)
\label{asymp}
\eeq
due to the limits of the integration range. Hence, the exponents
$j-1+s_{12\ldots j+1}$ and $s_{j+1,j+2}$ of the $x_j$ and $(1-x_j)$ factors in
the integrand (\ref{02,4}) trigger propagators $s_{12\ldots j+1}^{-1}$ and
$s_{j+1,j+2}^{-1}$, whenever presence of $R^{\nu,\rho}$ shifts them to
$x_j^{s_{12\ldots j+1}-1}$ and $(1-x_j)^{s_{j+1,j+2}-1}$, respectively:
\begin{itemize}
  \item $(j+1)$-particle pole channels $s_{12\ldots j+1}^{-1}$ encompassing legs $1,2,\ldots,j+1 \leq N-2$ emerge from the $x_j \rightarrow 0$ regime of the integral (\ref{02,4}) whenever $R^{\nu,\rho}$ covers $j$ factors of $z_{pq}$ with $1\leq p<q\leq j+1$
\item two-particle channels $s_{j+1,j+2}^{-1}$ emerge as $x_j \rightarrow 1$ whenever $R^{\nu,\rho}\sim z_{j+1,j+2}$
\end{itemize}
The factors of $(1-x_j x_{j+1} \ldots x_{l})^{s_{j+1,l+2}}$ in (\ref{02,4})
with $l > j$ or potential $z_{j+1,l+2}^{-1}$ admixtures from $R^{\nu,\rho}$ do
not affect these poles. They take the role of the functions $f,g$ in
(\ref{asymp}) with regular behavior in the critical range of integration.

This criterion to find the pole channels can neither depend on the particular
$j+1$-particle channel $s_{12\ldots j+1},s_{23\ldots j+2},\ldots,s_{N-j-1
\ldots N-1}$ in question nor on the parametrization of the disk boundary.
Hence, we can lift the above correlation between $s_{12\ldots j+1}$ poles and
$R^{\nu,\rho}$ to cyclic images $s_{k,k+1,\ldots j+k}$ with $k \leq N-1$ and
formulate the following general criterion\footnote{Although the rule
(\ref{pole}) is formulated for a particular $SL(2,\ZR)$ fixed version of the
integrals it does not depend on this choice: 
in the picture $\frac{1}{\CV_{\te{CKG}}} \prod_{i=1}^N \int \dd z_i\ldots$ with
unspecified $SL(2,\ZR)$ fixing, the $S_{N-3}$ basis of $Z(1,\ldots,N,N-1)$ at
$\nu = N-2$ exhibits a pole in $s_{N-1,N}$ from the $(z_{N-1,N})^{s_{N-1,N}-1}$
part of the integrand. Fixing $z_{N-1}=1$ and $z_{N}\rightarrow \infty$ appears
to prevent these two positions from colliding, but the rule (\ref{pole}) reveals
$s_{N-1,N} = s_{12\ldots N-2}$ as an $(N-2)$-particle channel caused by the
$N-3$ factors of $z_{pq}$ with $1\leq p < q \leq N-2$ in the associated
polynomial $R^{\nu=N-2,\rho}$. A similar argument applies to the
$s_{N,1}=s_{23\ldots N-1}$ pole present in any $Z(1,N,\ldots,N-1)$.} for
massless poles:

\beq\label{pole}
\ba{lcl}
\te{\it Integrals $Z$ of type (\ref{02,2}) contribute to the $(j-i+1)$-particle pole channel $s_{i,i+1\ldots j-1,j}$} \\
 \te{\it with $1 \leq i < j \leq N-1$ if $R^{\nu,\rho}$ contains $(j-i)$ factors of $z_{pq}$ in the range $i \leq p < q \leq j$.}
  \ea
  \eeq
\vskip0.3cm

The pole channels accessible in \eqn{02,4} additionally point out
\textit{incompatible} (or dual) channels\footnote{Compatibility conditions for
two channels involving $\geq 3$ particles are not directly obvious from
\eqn{02,4}. One can however show by other means that they are compatible if
they can appear as simultaneous propagators in a cubic diagram.}
\cite{Mafra:2011nw, Frampton}: The two pole channels $s_{12\ldots j+1}$ and
$s_{j+1,j+2}$ are caused by complementary integration regions $x_{j}
\rightarrow 0$ and $x_j \rightarrow 1$, respectively, this is why a
simultaneous pole in $s_{12\ldots j+1}$ and $s_{j+1,j+2}$ cannot appear. Again,
this statement must be independent on the label 1 bounding the multi-particle
channel, so we conclude that the pairs $(s_{i,i+1} ,s_{i+1,i+2\ldots j-1,j})$
and $(s_{i,i+1\ldots j-1,j}, s_{j,j+1})$ are incompatible pole channel for any
$1 \leq i<j \leq N-1$ and always appear in separate denominators (see
\eqn{02,12} below for an example).

Examples for the pole criterion (\ref{pole}) are given in appendix \ref{exppole}.

\subsubsection{Momentum kernel representation of field theory limits}

We can cast the results of the pole analysis (\ref{pole}) into a closed form by
demanding a consistent field theory limit for the disk amplitude (\ref{00,1}):
The pole structure of the functions (\ref{02,2}) must be compatible with
\beq
F_\Pi{}^\si  \, \Big|_{\ap \rightarrow 0} \eq \de_\Pi^\si \ .
\eeq
The representation (\ref{invmom2}) for $F_\Pi{}^\si$ in terms of inverse momentum kernels yields
\begin{align}
&Z_\Pi(1,2_\si,\ldots,\nu_\si,N,(\nu+1)_\si,\ldots,(N-2)_\si,N-1) \, \Big|_{\ap \rightarrow 0} \eq (-1)^{N-3} \notag \\
&\ \times \, S^{-1}[ \, 2_\si,\dots, \nu_\si \, | \, \Pi(2,\ldots,\nu) \,]_1 \, S^{-1}[ \, (N\!-\!2)_\si,\dots, (\nu\!+\!1)_\si \, | \, \Pi(N\!-\!2,\ldots,\nu\!+\!1) \,]_{N-1} \ .
\label{consFT}
\end{align}
Like in subsection \ref{sec:inv}, the entry $S^{-1}[ \al |\be]$ is bound to vanish unless $\al$ is a permutation of $\be$.

\subsection{From a KK basis to a pole channel basis}
\label{sec1,22}

We would like to take advantage of the rule (\ref{pole}) identifying pole
channels to efficiently construct the $\ap$-expansion of an integral basis. The
examples of appendix \ref{exppole} show that generic KK integrals contribute
to several cubic diagrams of YM field theory. In the later subsection
\ref{sec1,23}, we aim to identify the residues for less singular string
corrections to YM diagrams where only a subset of the $N-3$ poles remain (the
others are said to be collapsed in the following). Since there is a total of
$2^{N-3}$ such subsets per cubic diagram, it is preferable to rearrange the
integrals such that they generate only one cubic diagram each, even on the
expense of leaving the KK basis (\ref{02,2}). So we will start with the KK
basis and rewrite its elements by means of partial-fraction relations in order to obtain
a pole channel basis---another $(N-2)!$ set of integrals which by construction
involve at most one field theory diagram. Subsectors of different
$\nu=1,2,\ldots,N-2$ do not mix under these partial-fraction manipulations.

In order to determine the function associated with a particular field theory
diagram which is closest to the KK basis (\ref{02,2}), one is faced with the
inverse problem to the pole detection rule (\ref{pole}): For the given cubic
diagram characterized by $N-3$ compatible pole channels
$s_{a_i,a_i+1,\ldots,b_i-1,b_i}$ at $1 \leq a_i<b_i\leq N-1$ for
$i=1,2,\ldots,N-3$, we want to construct a polynomial\footnote{We have to
demand polynomial form $\sim (z_{pq})^{N-3}$ for the $R[a_i,b_i]$ because
rational functions $\sim (z_{pq})^{N-2}/z_{rs}$ of the same degree are not
accessible to the integration methods of section \ref{secPolylogs}.}
$R[a_i,b_i] \sim (z_{pq})^{N-3}$ such that the integral 
\beq
Z[s_{a_i,a_i+1,\ldots,b_i-1,b_i}] \ \ := \ \ \prod_{i=2}^{N-2} \int^{z_{i+1}}_0 \dd z_i \ \prod_{i<j}^{N-1} |z_{ij}|^{s_{ij}} \ \frac{1}{R[a_i,b_i]}
\label{02,71}
\eeq
contributes to those $N-3$ channels exclusively. Indeed, we can verify using
(\ref{pole}) that the desired polynomial is given by
\beq
R[a_i,b_i] \eq \prod_{i=1}^{N-3} z_{b_i,a_i}\,,
\label{02,71a}
\eeq 
where each multiparticle Mandelstam variable is associated with a factor
$z_{pq}$ according to its bounding momenta $k_{a_i}$ and $k_{b_i}$, for example
$s_{a_i,a_i+1,\ldots,b_i} \mapsto z_{b_i,a_i}$. In other words, the diagram
with poles $\prod_{i=1}^{N-3} s_{a_i,a_i+1,\ldots,b_i}$ can be generated by the
function
\beq
Z[s_{a_i,a_i+1,\ldots,b_i-1,b_i}]  \ \ = 
\ \ \prod_{i=2}^{N-2} \int^{z_{i+1}}_0 \dd z_i \ \prod_{i<j}^{N-1} |z_{ij}|^{s_{ij}} \, \prod_{i=1}^{N-3} \frac{1}{z_{b_i,a_i}}
\ \ \sim \ \  \prod_{i=1}^{N-3} \frac{1}{s_{a_i,a_i+1,\ldots,b_i-1,b_i}}  \ .
\label{02,72}
\eeq
In our setup, an integral $Z(\ldots)$ involved in the construction of the
$F^\si$-basis must be written as a linear combinations of functions
(\ref{02,72}). We shall see in the following examples that this can always be
achieved through repeated use of partial-fraction identities.

As mentioned in appendix \ref{sec1,213}, the KK basis of $Z(1,\ldots,N-1)$
also incorporates functions with $k\leq N-5$ simultaneous poles whose
$\ap$-expansion starts at transcendentality $\zeta_{N-3-k}$. In this case, the
mapping $s_{a_i,a_i+1,\ldots,b_i} \mapsto z_{b_i,a_i}$ for $k \leq N-5$ pole
channels leaves $N-3-k$ factors of $z_{pq}$ undetermined.  It is desirable to
complete the pole basis in a way such that the minimal number of poles need to
be considered in performing the $\ap$-expansion. On the other hand, this
guideline can conflict with the form of the integrand \eqn{constraint} required
by the integration methods of section \ref{secPolylogs}, we will come back to this point later on.

We should point out that higher $\ap$-corrections of the functions
constructed from (\ref{02,72}) do not transform into each other under a
shift $i \mapsto i+1$ subject to cyclic identifications $i := i+1$. This is a consequence of the fact that neither the set
of $R[a_i,b_i] $ nor the KK basis (\ref{02,2}) is preserved under cyclic
shifts.

\subsubsection{Five-point examples}

The four-point KK basis functions \eqns{z4pt1}{z4pt2} each generate no more
than one cubic diagram, $Z[s_{12}] = -Z(1,2,4,3)$ and $Z[s_{23}] =
-Z(1,4,2,3)$, so there is no need to modify the basis. Starting from five
points, the map (\ref{02,72}) from cubic diagrams to disk integrals leads to
functions beyond the KK basis elements. According to (\ref{02,11}) and
(\ref{02,12}), we find
\begin{align}
Z[s_{23}s_{123}] &\eq \int^1_0 \dd z_3 \int^{z_3}_0 \dd z_2 \ \frac{ \prod_{i<j}^4 |z_{ij}|^{s_{ij}} }{z_{31} z_{32}} \eq -\, Z(1,3,2,5,4) \notag \\
Z[s_{23} s_{234}] &\eq \int^1_0 \dd z_3 \int^{z_3}_0 \dd z_2 \ \frac{ \prod_{i<j}^4 |z_{ij}|^{s_{ij}} }{z_{32} z_{42}} \eq - \,Z(1,5,3,2,4) 
\label{02,73}
\end{align}
whereas the diagrams with poles in $s_{12}s_{123}$ and $s_{23}s_{234}$ require
a linear combination of KK functions,
\begin{align}
Z[s_{12}s_{123}] \eq & \int^1_0 \dd z_3 \int^{z_3}_0 \dd z_2 \ \frac{ \prod_{i<j}^4 |z_{ij}|^{s_{ij}} }{z_{21} z_{31}} \eq Z(1,3,2,5,4) \ + \ Z(1,2,3,5,4) \notag \\
 Z[s_{34} s_{234}] \eq &\int^1_0 \dd z_3 \int^{z_3}_0 \dd z_2 \ \frac{ \prod_{i<j}^4 |z_{ij}|^{s_{ij}} }{z_{42} z_{43}} \eq Z(1,5,3,2,4) \ + \ Z(1,5,2,3,4) \ .
\label{02,74}
\end{align}
Partial fraction manipulations mediate between the KK basis and the pole basis
containing all the functions generated by (\ref{02,72}). This change of basis
acts block-diagonally on the $S_{2}$ sectors labeled by $\nu = 1,2,3$ (with
self-inverse $2\times 2$ blocks):
\beq
\vecb Z(1,2,3,5,4) \\ Z(1,3,2,5,4) \\ Z(1,2,5,3,4) \\ Z(1,3,5,2,4) \\ Z(1,5,2,3,4) \\ Z(1,5,3,2,4) \vece \eq
\left(\begin{array}{cccccc}
1 &1 &&&&\\
0 &-1 &&&& \\
& &1 &0 && \\
& &0 &1 && \\
&&& &1 &1 \\
&&& &0 &-1
\end{array} \right) \, \vecb Z[s_{12} s_{123}] \\ Z[s_{23} s_{123}] \\ Z[s_{12} s_{34}] \\ Z[\zeta_2] \\ Z[s_{34} s_{234}] \\ Z[s_{23} s_{234}] \vece \,.
\label{02,75}
\eeq
The $\nu =2$ elements of the KK basis are suitable to enter the pole basis, 
\begin{align}
Z[s_{12} s_{34}] \eq & \int^1_0 \dd z_3 \int^{z_3}_0 \dd z_2 \ \frac{ \prod_{i<j}^4 |z_{ij}|^{s_{ij}} }{z_{21} z_{43}} \eq Z(1,2,5,3,4)
\notag \\
Z[\zeta_2]  \eq &\int^1_0 \dd z_3 \int^{z_3}_0 \dd z_2 \ \frac{ \prod_{i<j}^4 |z_{ij}|^{s_{ij}} }{z_{31} z_{42}} \eq Z(1,3,5,2,4) \ .
\label{02,75a}
\end{align}
Here we introduced the notation $Z[\zeta_2]$ for the only regular function
$Z(1,3,5,2,4)$ in the KK basis. Given the absence of poles in $Z[\z_2]$, it is
most economic to construct the basis functions $F^{(23)}$ and $F^{(32)}$ from
the $\nu=2$ sector formed by $Z[s_{12} s_{34}]$ and $Z[\z_2]$.

\subsubsection{Six-point examples}
\label{subsec}

Let us discuss the $\nu =4,3$ sectors of the six-point KK separately because
the partial-fraction transformation towards a pole channel basis acts block
diagonally. As already mentioned in subsection \ref{sec1,21}, the $\nu=1,2$
sectors follow from world-sheet parity $i\mapsto 6-i$ and are thus not addressed
explicitly. The functions therein complete the set of fourteen field-theory
pole-channels (and six subleading pole channels $\z_2/s_{i,i+1}$ in integrals
with a vanishing field theory limit).

\begin{itemize}
\item
At $\nu=4$, a series of partial-fraction operations leads to the transformation matrix
\beq
\vecb Z(1,2,3,4,6,5) \\ Z(1,2,4,3,6,5) \\ Z(1,3,2,4,6,5) \\ Z(1,3,4,2,6,5) \\ Z(1,4,2,3,6,5) \\ Z(1,4,3,2,6,5) \vece \eq
\left(\begin{array}{cccccc}
-1 &-1 &-1&0&-1&-1\\
0 &0 &1&0&1&0 \\
0 &1 &0 &-1 &0 &1 \\
0 &0 &0 &1 &1 &0 \\
0 &0 &0 &0 &0 &1 \\
0 &0 &0 &0 &-1 &-1
\end{array} \right) \, \vecb Z[s_{12} s_{123} s_{1234}] \\ Z[s_{23} s_{123}s_{1234}] \\ Z[s_{12} s_{34}s_{1234}] \\ Z[\zeta_2 s_{1234}] \\ Z[s_{34} s_{234} s_{1234}] \\ Z[s_{23} s_{234} s_{1234}] \vece \,, 
\label{02,76}
\eeq
which is in agreement with equations (\ref{02,13}),(\ref{02,14}) and (\ref{02,15}). In addition to the triple pole functions defined by (\ref{02,72}), we have introduced a function
\beq
Z[\zeta_2 s_{1234}] \eq  \int^{1}_0 \dd z_4 \int^{z_{4}}_0 \dd z_3 \int^{z_{3}}_0 \dd z_2 \  \prod_{i<j}^5 |z_{ij}|^{s_{ij}}  \, \frac{ 1}{z_{31} z_{41} z_{42}} \eq \frac{ \zeta_2 }{s_{1234}} \ + \ {\cal O}(s^0)
\label{02,77}
\eeq
which generalizes the five-point integral $Z(1,3,5,2,4) = Z[\zeta_2]$ to a six-point setting with a single pole in $s_{1234}$.
\item In the $\nu=3$ sector of the six-point KK basis, the factorization (\ref{facto}) of the $F^\si$ integrands' rational part leads to a $2\times 2$ block-diagonal transformation to the desired pole channel basis:
\beq
\vecb Z(1,2,3,6,4,5) \\ Z(1,3,2,6,4,5) \\ Z(1,2,4,6,3,5) \\ Z(1,4,2,6,3,5) \\ Z(1,3,4,6,2,5) \\ Z(1,4,3,6,2,5) \vece \eq
\left(\begin{array}{cccccc}
-1 &-1 &&&&\\
0 &1 &&&& \\
& &-1 &-1 &&  \\
& &0 &1 && \\
&&& &-1 &-1 \\
&&& &0 &1
\end{array} \right) \, \vecb Z[s_{12} s_{123} s_{45}] \\ Z[s_{23} s_{123}s_{45}] \\ Z[\zeta_2 s_{12} ] \\ \tilde Z[\zeta_3] \\ Z[\zeta_3] \\ Z[\zeta_2s_{34}] \vece 
\label{02,79}
\eeq
The $\ap$-expansion for the following pole basis elements starts at subleading order:
\begin{align}
Z[\zeta_2 s_{12}] \eq& \prod_{i=2}^4 \int^{z_{i+1}}_0 \dd z_i \  \frac{ \prod_{i<j}^5 |z_{ij}|^{s_{ij}} }{z_{21} z_{41} z_{53}}
\co \tilde Z[\zeta_3] \eq  \prod_{i=2}^4 \int^{z_{i+1}}_0 \dd z_i \    \frac{ \prod_{i<j}^5 |z_{ij}|^{s_{ij}}}{z_{41} z_{42} z_{53}}
\label{02,81} \\
Z[\zeta_2 s_{34}] \eq&  \prod_{i=2}^4 \int^{z_{i+1}}_0 \dd z_i  \   \frac{ \prod_{i<j}^5 |z_{ij}|^{s_{ij}} }{z_{41} z_{43} z_{52}}
\co
Z[\zeta_3] \eq \prod_{i=2}^4 \int^{z_{i+1}}_0 \dd z_i  \  \frac{  \prod_{i<j}^5 |z_{ij}|^{s_{ij}}  }{z_{31} z_{41} z_{52}}
\label{02,82} 
\end{align}
The functions (\ref{02,81}) and (\ref{02,82}) are not the unique choice of completing the pole basis. For the sake of computing the $F^{\si}$ economically, it is preferable to include as many functions as possible with leading low-energy behavior $\zeta_3$ rather than $\zeta_2/s_{i,i+1}$. It has been already noticed in \cite{Stieberger:2006te} that a six-point basis requires four linearly independent functions whose $\ap$-expansion starts with $\zeta_3$. Those functions can for instance be taken as $Z[\zeta_3], \tilde Z[\zeta_3]$ and parity or cyclicity images thereof.
\end{itemize}
Comparing (\ref{02,76}) with (\ref{02,79}), one can identify the $\nu=3$ sector
to be the most suitable starting point towards an $\ap$-expansion of the
$F^{\si(234)}$: apart from the block-diagonal partial-fraction transformations,
it contains only two instead of five functions with triple poles.

\subsubsection{Seven-point examples}
\label{sec:block}

In view of the six-point example above, both the pole channel analysis and the
number of terms in (\ref{01,24}) suggests to compute the low-energy expansion
of the $F^{\si}$ from the $\nu = \lfloor N/2 \rfloor$ sector of the KK basis
(\ref{02,1}). At seven points, this amounts to considering the $\nu=3$ sector
$\{  Z(1,2_\rho,3_\rho,7,4_\rho,5_\rho,6) , \ \rho \in S_{4} \}$ associated
with polynomials $R^{\nu=3,\rho}= (z_{12_\rho} z_{2_\rho 3_\rho}) \times
(z_{4_\rho 5_\rho} z_{5_\rho 6})$. A suitable pole channel basis can be picked
separately in the six subsectors of $Z(1,2_\rho,3_\rho,7,4_\rho,5_\rho,6)$
according to $ \{ 2_\rho,3_\rho\} = \{2,3\},\,\{2,4\}, \, \{2,5\}, \,
\{3,4\},\,\{3,5\}, \,\{4,5\}$, with four elements each. This generalizes the
block diagonal form of (\ref{02,79}) for the six-point $\nu=3$ functions due to
the factorization (\ref{facto}) of the rational functions in the $F^\si$
integrand.

In subsection \ref{7basis} we will construct the seven-point $\ap$-expansion in blocks of four functions
each, the $\{ 2_\rho,3_\rho \}$ block encompassing
$F^{\rho(2345)},F^{\rho(2354)},F^{\rho(3245)}$ and $F^{\rho(3254)}$.

\subsection{Identifying pole residues}
\label{sec1,23}

This subsection is devoted to determining the full $\ap$-expansion of kinematic
pole residues. The factorization of $N$-point amplitudes into (products of)
lower-point amplitudes on kinematic poles extends to all orders in $\ap$. That
is why residues on $N-3-p$ simultaneous poles carry the fingerprints of the
full $(p+3)$-point amplitudes including string corrections. In order to make
these statements precise, we shall focus on integrals of type (\ref{02,72})
generating one cubic field-theory diagram and recall the association of each
propagator $s_{a_i , a_i+1\ldots b_i}^{-1}$ with the rational factor of
$z_{b_i,a_i}^{-1}$ in the integrand. At the residue of the pole, the
integration over the associated $z_{b_i,a_i}$ variable collapses to the
singular region due to (\ref{asymp}). The same correspondence $s_{a_i , a_i+1\ldots b_i}
\leftrightarrow z_{b_i,a_i}$ holds for functions with less than $N-3$
simultaneous poles, so their residues can be determined by the same methods.

Singularities in Mandelstam variables pose the obstruction to perform a Taylor
expansion of the Koba-Nielsen factor
\beq
\prod_{i<j}^{N-1}|z_{ij}|^{s_{ij}}
\eq \prod_{i<j}^{N-1} \sum_{n_{ij}=0}^{\infty} \frac{ (s_{ij} \, \ln |z_{ij}| )^{n_{ij}} }{n_{ij}!}
\label{02,201}
\eeq
before carrying out the world-sheet integrals. Putting it differently,
integrating the product of the Taylor series (\ref{02,201}) with the rational
function $\prod_{l=1}^{N-3} z_{u_l v_l}^{-1}$ yields a regular object without
any poles in $s_{u_i , u_i+1\ldots v_i}$,
\beq
I_{u_1v_1,u_2 v_2,\ldots,u_{N-3} v_{N-3}}^\reg
 \ \ := \ \ \prod_{i<j}^{N-1} \sum_{n_{ij}=0}^{\infty} \frac{ (s_{ij})^{n_{ij}} }{n_{ij}!} \prod_{l=1}^{N-3} \int^{z_{l+2}}_0 \frac{\dd z_{l+1}}{ z_{u_l v_l}} \; (\ln |z_{ij}|)^{n_{ij}} \ .
 \label{02,202}
\eeq
These regularized world-sheet integrals are the natural objects in order to
describe the residues on kinematic poles. The polylogarithmic integrals therein will
be evaluated in section \ref{secPolylogs}. Of course, there are many
alternative ways to obtain pole residues of the $Z(\rho)$, examples for five
and six points can be found in \cite{Stieberger:2006te}. In this section we
want to cast both the residues and the regular parts of relevant world-sheet
integrals into the form (\ref{02,202}). The integration techniques of section
\ref{secPolylogs} then render the $\ap$-expansion suitable for computer
automatization.

In what follows, we promote the regular part  $I_{u_1v_1,u_2 v_2,\ldots,u_{N-3}
v_{N-3}}^\reg$ of $N$-point integrals to functions $I_{u_1v_1,u_2
v_2,\ldots,u_{N-3} v_{N-3}}^\reg(k_1,k_2,\ldots,k_{N-1})$ of $N-1$ on-shell
momenta. (By convention, $k_1,\ldots,k_{N-1}$ are understood to be the
arguments for any $I_{\ldots}^\reg$ unless we specify them differently.) This
allows for a diagrammatic and intuitive method to determine the behavior of the
Koba-Nielsen factor at pole residues. The following procedure systematically
resolves the singularity structure of an integral with poles in $s_{a_i ,
a_i+1\ldots b_i}$ due to $z_{b_i,a_i}$ and leaves residues and regular parts of
the form (\ref{02,202}) which can be treated with the methods of section
\ref{secPolylogs}:
\begin{itemize}
\item Sum over all ways to collapse a subset of the $(s_{a_i , a_i+1\ldots
  b_i}^{-1})$-propagators in the cubic YM diagram associated with the integral
  in question. It is convenient to start with the maximally singular term
  (leaving the cubic YM diagram untouched) and to gradually increase the number
  of relaxed poles. Draw a diagram for each set of collapsed propagators
  $s_{a_i , a_i+1\ldots b_i}^{-1}$ and replace them by a contact vertex (drawn
  as a bubble) associated with the rational factor $z^{-1}_{b_i,a_i}$ in the
  integrand, see the five-point example in figure \ref{fig:expl5}.
\begin{figure}[htbp]
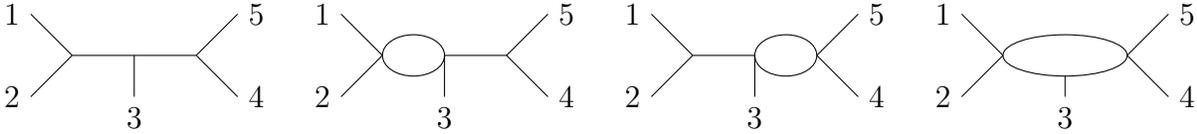
 
\begin{center}
\tikzpicture[scale=0.55]
%
\draw (0,0) -- (-1,1) node[left]{$1$};
\draw (0,0) -- (-1,-1) node[left]{$2$};
\draw (0,0) -- (3,0);
\draw (1.5,0) -- (1.5,-1) node[below]{$3$};
\draw (3,0) -- (4,1) node[right]{$5$};
\draw (3,0) -- (4,-1) node[right]{$4$};
%
%
\scope[xshift=7.5cm]
\draw (0,0) -- (-1,1) node[left]{$1$};
\draw (0,0) -- (-1,-1) node[left]{$2$};
\draw (1.5,0) -- (3,0);
\draw (1.5,0) -- (1.5,-1) node[below]{$3$};
\draw (3,0) -- (4,1) node[right]{$5$};
\draw (3,0) -- (4,-1) node[right]{$4$};
\draw (0.75,0) ellipse (0.75cm and 0.5cm);
%
\endscope
\scope[xshift=15cm]
\draw (0,0) -- (-1,1) node[left]{$1$};
\draw (0,0) -- (-1,-1) node[left]{$2$};
\draw (0,0) -- (1.5,0);
\draw (1.5,0) -- (1.5,-1) node[below]{$3$};
\draw (3,0) -- (4,1) node[right]{$5$};
\draw (3,0) -- (4,-1) node[right]{$4$};
\draw (2.25,0) ellipse (0.75cm and 0.5cm);
\endscope
\scope[xshift=22.5cm]
\draw (0,0) -- (-1,1) node[left]{$1$};
\draw (0,0) -- (-1,-1) node[left]{$2$};
\draw (1.5,-0.5) -- (1.5,-1) node[below]{$3$};
\draw (3,0) -- (4,1) node[right]{$5$};
\draw (3,0) -- (4,-1) node[right]{$4$};
\draw (1.5,0) ellipse (1.5cm and 0.5cm);
\endscope
\endtikzpicture
\caption{Propagator collapses contributing to the $Z[s_{12}s_{123}]$ integral.}
\label{fig:expl5}
\end{center}
\end{figure}

\item Each connected set of $p$ contact vertices represents a $p$-fold
  world-sheet integral as it appears in a $(p+3)$-point amplitude. The
  propagator structure determines the $p+3$ inflowing composite (or region-)
  momenta $k_A$ of type
\beq
k_{12\ldots p} \ \ := \ \ k_1+ k_2+\ldots+k_p
\label{multimom}
\eeq
with $A = \{1,2,\ldots,p\}$ in (\ref{multimom}) which are generically off-shell
with $k_A^2 \neq 0$. The $SL(2,\ZR)$ fixing $z_N \rightarrow \infty$ decouples
the composite momentum encompassing $k_N$ at infinity, i.e. we don't list it as
a separate argument. The $p+2$ region momenta
$k_{A_1},k_{A_2},\ldots,k_{A_{p+2}}$ remain as independent arguments of the
$(p+3)$-point integral, examples are shown in figure \ref{figexpl}.

\begin{center}
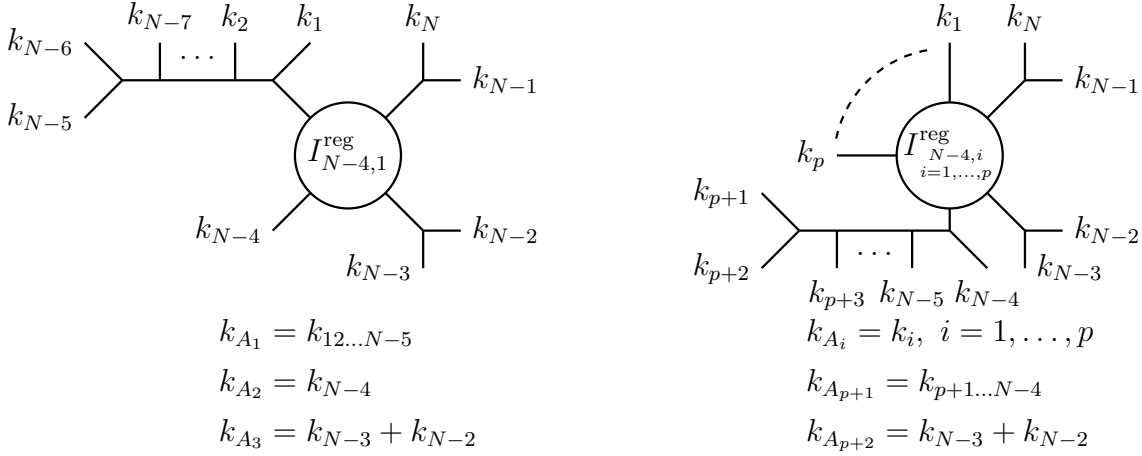
\begin{figure}
\begin{tikzpicture} [scale = 1, line width=0.30mm, xshift=-1cm]
\draw (0,0) circle (0.705cm);
\draw (0,0) node{$I_{N-4,1}^\reg$};
\draw (-0.5,0.5) -- (-1,1);
\draw (-1,1) -- (-0.5,1.5)node[above]{$k_1$};
\draw (-1,1) -- (-3,1);
\draw (-1.5,1) -- (-1.5,1.5)node[above]{$k_2$};
\draw (-2.5,1) -- (-2.5,1.5)node[above]{$k_{N-7}$};
\draw (-3,1) -- (-3.5,1.5)node[left]{$k_{N-6}$};
\draw (-3,1) -- (-3.5,0.5)node[left]{$k_{N-5}$};
\draw (-2,1.3) node{$\ldots$};
\draw (-0.5,-0.5) -- (-1,-1) node[left]{$k_{N-4}$};
\draw (0.5,-0.5) -- (1,-1) ;
\draw (1,-1) -- (1.5,-1) node[right]{$k_{N-2}$};
\draw (1,-1) -- (1,-1.5) node[left]{$k_{N-3}$};
\draw (0.5,0.5) -- (1,1) ;
\draw (1,1) -- (1.5,1) node[right]{$k_{N-1}$};
\draw (1,1) -- (1,1.5) node[above]{$k_{N}$};
\draw (0,-3) node{$\begin{array}{l}  k_{A_1}=k_{12\ldots N-5} \\ k_{A_2}=k_{ N-4} \\ k_{A_3}=k_{N-3}+k_{N-2} \end{array}$};
\begin{scope}[xshift=8cm]
\draw (0,0) circle (0.705cm);
\draw (0,0) node{$I_{N-4,i \atop {i=1,\ldots,p}}^\reg$};
%
\draw (0,0.705) -- (0,1.5) node[above]{$k_1$};
\draw (-0.705,0) -- (-1.5,0) node[left]{$k_p$};
\draw (0,-0.705) -- (0,-1);
\draw (0,-1) -- (0.5,-1.5) node[below]{$k_{N-4}$};
\draw (0,-1) -- (-2,-1);
\draw (-0.5,-1) -- (-0.5,-1.5) node[below]{$k_{N-5}$};
\draw (-1,-1.3) node{$\ldots$};
\draw (-1.5,-1) -- (-1.5,-1.5) node[below]{$k_{p+3}$};
\draw (-2,-1)--(-2.5,-0.5) node[left]{$k_{p+1}$};
\draw (-2,-1)--(-2.5,-1.5) node[left]{$k_{p+2}$};
\draw (0.5,-0.5)--(1,-1);
\draw (1,-1) -- (1,-1.5) node[right]{$k_{N-3}$};
\draw (1,-1) -- (1.5,-1) node[right]{$k_{N-2}$};
\draw (0.5,0.5)--(1,1);
\draw (1,1) -- (1,1.5) node[above]{$k_{N}$};
\draw (1,1) -- (1.5,1) node[right]{$k_{N-1}$};
\draw[dashed] (-0.3,1.4) arc (100:170:1.5cm);
\draw (0,-3) node{$\begin{array}{l}  k_{A_i}=k_{i} , \ i=1,\ldots,p \\ k_{A_{p+1}}=k_{p+1 \ldots N-4} \\ k_{A_{p+2}}=k_{N-3}+k_{N-2} \end{array}$};
\end{scope}
\end{tikzpicture}
\caption{Examples of a four-point and a $(p+3)$-point contact vertex together with their region momenta $k_{A_i}$.}
\label{figexpl}
\end{figure}
\end{center}

\item Relabel the $z_{b_i ,a_i}$ from the collapsed propagators according to
  the composite momenta $k_A$ of the associated contact vertices where legs
  $b_i$ and $a_i$ are attached:
\beq
a_{i} \in A_{v_i}, \ b_i \in A_{u_i} \ \ \Rightarrow \ \ z_{b_i,a_i} \ \mapsto \ z_{u_i,v_i}
\label{02,203}
\eeq
Figure \ref{figexpl} gives two examples with a four-point and a $(p+3)$-point contact vertex.
\item The momentum expansion of a $(p+3)$-point contact vertex at momenta $k_{A_1},k_{A_2},\ldots,k_{A_{p+2}}$ is given by the regular part (\ref{02,202}) of a $(p+3)$-point integral
\begin{align}
I_{u_1v_1,u_2 v_2,\ldots,u_p w_p}^\reg(k_{A_1},k_{A_2},\ldots,&k_{A_{p+2}}) \ \ = \ \ \prod_{i<j}^{p+2} \sum_{n_{ij}=0}^{\infty} \frac{ (2\ap k_{A_i}\cdot k_{A_j} )^{n_{ij}} }{n_{ij}!} \notag \\
\times \
&\prod_{l=1}^{p} \int^{z_{l+2}}_0  \frac{ \dd z_{l+1} }{z_{u_l,v_l}}
\; ( \ln |z_{ij}| )^{n_{ij}} \, \Big|^{z_{p+2}=1}_{z_1 = 0} \ + \ \CO( k_{A_i}^2 )
\label{02,204}
\end{align}
\item In contrast to the regular $N$-point integral (\ref{02,202}), generic
  contact diagrams (\ref{02,204}) in presence of propagators $s_{a_i ,
  a_i+1\ldots b_i}^{-1} \sim 1/k^2_{A_i}$ depend on off-shell momenta
  $k_{A_i}^2 \neq 0$. The dependence of on-shell integrals (\ref{02,202}) on
  the vanishing $k_i^2$ is invisible, so for each pair $1 \leq i<j \leq p+2$,
  it is not a priori clear whether their off-shell generalizations
  (\ref{02,204}) depend on $2\ap k_{A_i}\cdot k_{A_j}$, on $\ap
  (k_{A_i}+k_{A_j})^2$ or on one of $2\ap k_{A_i}\cdot k_{A_j} + \ap k_{A_{i,j}}^2$.
  Each of the $\frac{1}{2}p(p+3)$ Mandelstam invariants governing a
  $(p+3)$-point contact vertex admits shifts by one of $0,\ap
  k_{A_i}^2$, $\ap k_{A_j}^2$ or $\ap k_{A_i}^2+\ap k_{A_j}^2$, and our present
  method does not single out a preferred choice. This is a minor drawback of
  the otherwise constructive method. As we shall see in section \ref{sec3},
  there are plenty of consistency conditions on the amplitude (such as
  cyclicity) which fix the missing $p(p+3)$ ``bit'' of information, and it will
  turn out that the majority of the potential $k_{A_i}^2$-dependencies is
  absent\footnote{In fact, the unique five-point example for a nontrivial
  ``mass'' dependence occurs at the $s_{23}$ single pole residue of
  $Z[s_{23}s_{123}]$, see (\ref{02,26}) and (\ref{ktos3}): In this example, the associated $I_{21}^\reg$
  function depends on the two combinations of Mandelstam variables given by $2\ap k_1 \cdot
  k_{23}+\ap k_{23}^2$ and $2 \ap k_{23}\cdot k_4$.}. Moreover, the $k_{A_i}^2$-dependence 
  is a global property of contact vertices (\ref{02,204}), so the
  information on $(2\ap k_{A_i}\cdot k_{A_j})$-shifts only needs to be determined
  once for each set of labels $u_i,v_i$ in $I_{u_1v_1,u_2 v_2,\ldots,u_p
  w_p}^\reg$. Any dependence of the (partially off-shell) integral
  (\ref{02,204}) on the masses $k_{A_i}^2 \neq 0$ vanishes at the residue of
  the multiple pole in question but contributes to less singular parts and
  interferes with the associated $I^\reg_{b_1 a_1,b_2
  a_2,\ldots,b_{N-3}a_{N-3}}$. 
\item The same reasoning applies to integrals with low-energy behavior $
  \zeta_{N-3-k}  \prod_{i=1}^{k}s_{a_i , a_i+1\ldots b_i}^{-1}$.
\end{itemize}
In many cases, the validity of these diagrammatic rules can be verified through
the cube parametrization (\ref{02,3}) and (\ref{02,4}) introduced in subsection
\ref{sec1,21}: Sending individual cube variables to $x_i \rightarrow 0,1$
leads to poles in $s_{12\ldots i+1},s_{i+1,i+2}$ and fixes the associated
$z_{b_i,a_i}$ in the simplex parametrization. The behavior of the Koba-Nielsen
factor at the cube boundaries then determines the momentum configuration
describing the residue.

A word of caution is appropriate here: The symmetries in the graphical
arrangement of contact vertices in general do not apply to the dependence on
the external momenta. Firstly, the omnipresent choice to send $z_N \rightarrow
\infty$ treats the cubic subdiagram containing $k_N$ on special footing.
Secondly the integral representations (\ref{02,204}) are coherently modded out
by the conformal Killing group such that also $k_{A_1}$ and $k_{A_{p+2}}$
associated with $z_1=0$ and $z_{p+2}=1$ play a distinguished role. The diagrams
serve as a mnemonic to determine the labels $u_i,v_i$ and momenta
$k_{A_i}$ in (\ref{02,204}) and not to reflect any exchange symmetries in
$k_{A_i} \leftrightarrow k_{A_j}$.

\subsubsection{Four-point examples}

Four-point integrals involve a kinematic pole whose residue is determined by
the cubic YM vertex without any $\ap$-correction. Hence, the $s$-channel
integral $Z[s_{12}]=Z(1,2,4,3)$ is related to its regularized part as follows:

\begin{figure}[htbp] 
\begin{center}
\tikzpicture
\draw (-3.7,0) node{$\displaystyle Z[s_{12}] \ \ = \ \ \frac{1}{s_{12}} \ + \ I_{21}^\reg \eq$};
\draw (0,0) -- (-1,1) node[left]{$1$};
\draw (0,0) -- (-1,-1) node[left]{$2$};
\draw (0,0) -- (2,0);
\draw (1,-0.3) node{$k_{12}$};
\draw (2,0) -- (3,1) node[right]{$4$};
\draw (2,0) -- (3,-1) node[right]{$3$};
\draw (4,0) node{$+$};
\scope[xshift=6cm]
\draw (0,0) -- (-1,1) node[left]{$1$};
\draw (0,0) -- (-1,-1) node[left]{$2$};
\draw (1,0) ellipse (1cm and 0.5cm);
\draw (1,0) node{$I_{21}^\reg$};
\draw (2,0) -- (3,1) node[right]{$4$};
\draw (2,0) -- (3,-1) node[right]{$3$};
\endscope
\endtikzpicture
\caption{Pole structure of the function $Z[s_{12}]$.}
\label{fig:s12}
\end{center}
\end{figure}

The general expression (\ref{02,202}) for regular parts simplifies to
\beq
I_{21}^\reg \ \ := \ \ I_{21}^\reg(k_1,k_2,k_3) \eq \sum_{n_{12}=0}^{\infty} \sum_{n_{23}=0}^{\infty} \frac{ s_{12}^{n_{12}} \, s_{23}^{n_{23}} }{n_{12}! \, n_{23}!} \ \int^1_0 \frac{\dd z_2}{z_2} \ (\ln |z_2|)^{n_{12}} \, (\ln |1-z_2|)^{n_{23}} 
\label{02,211}
\eeq
which can be resummed in closed form: the basis function is given by the Veneziano amplitude
\beq
F^{(2)} \eq s_{12} \, Z[s_{12}] \eq 1 \ + \ s_{12} I_{21}^\reg \eq \exp \left( \, \sum_{k=2}^{\infty} \frac{ \zeta_k }{k } \,(-1)^k \, (s_{12}^k+s_{23}^k - (s_{12}+s_{23})^k) \,, \right)
\label{02,212}
\eeq
which translates into the following $\ap$-expansion for $I_{21}^\reg$:
\begin{align}
I_{21}^\reg(k_1,k_2,k_3) \eq& \frac{1}{s_{12}} \, \exp \left( \, \sum_{k=2}^{\infty} \frac{ \zeta_k }{k } \,(-1)^k \, (s_{12}^k+s_{23}^k - (s_{12}+s_{23})^k) \, \right)  \ - \ \frac{1}{s_{12}} \notag \\
\eq & -  \zeta_2 s_{23} + \zeta_3 s_{23}(s_{12}+s_{23}) - \zeta_4 s_{23}(s_{12}^2+ \tfrac{1}{4}s_{12}s_{23}+s_{23}^2) \notag \\
& \ \ + \zeta_5 s_{23} (s_{12}^3 + 2 s_{12}^2 s_{23} + 2 s_{12} s_{23}^2 + s_{23}^3) - \zeta_2 \zeta_3 s_{12} s_{23}^2 (s_{12} + s_{23}) \ + \ \ldots\,.
\label{02,213}
\end{align}
It will prove convenient to also introduce the regular part $I^\reg_{32}$ of the $u$-channel function
\beq
Z[s_{23}] \eq \frac{1}{s_{23}} \ + \ I^\reg_{32}
\label{02,214}
\eeq
which is related to its parity image via
\begin{align}
I_{32}^\reg(k_A,k_B,k_C)& \eq I_{21}^\reg(k_C,k_B,k_A) \eq  -  \zeta_2 s_{12} + \zeta_3 s_{12}(s_{12}+s_{23}) - \zeta_4 s_{12}(s_{12}^2+ \tfrac{1}{4}s_{12}s_{23}+s_{23}^2) \notag \\
& \ \ + \zeta_5 s_{12} (s_{12}^3 + 2 s_{12}^2 s_{23} + 2 s_{12} s_{23}^2 + s_{23}^3) - \zeta_2 \zeta_3 s_{12}^2 s_{23} (s_{12} + s_{23}) \ + \ \ldots
\ .
\label{02,215}
\end{align}
In the following subsections, we will not spell out the expansion of the $I_{\ldots}^\reg$ explicitly. A systematic way to obtain their expansion is explained in section \ref{secPolylogs}.

\subsubsection{Five-point examples}

At five points, the two basis functions $F^{\si}$ are proportional to the
integrals $Z[s_{12}s_{34}]=Z(1,2,5,3,4)$ and $Z[\zeta_2]= Z(1,3,5,2,4)$. The
former requires a pole treatment following the procedure of subsection
\ref{sec1,23}, this time with an infinite tower of $\ap$-corrections at the
single pole residues.

The two propagators $s_{12} ^{-1} $ and $s^{-1}_{34}$ in the cubic field-theory
diagram of $Z[s_{12}s_{34}]$ can collapse separately. This leads to quartic
contact vertices whose contributing momenta (in addition to $k_5$ from $z_5\rightarrow \infty$) can be read off to be
$(k_{1},k_2,k_{34})$ and $(k_{12},k_3,k_4)$ from the second and third diagram
in figure \ref{fig:s12s34}. The $z_{u_i v_i}$ correspondent of the collapsed
propagator determines those vertices to be of $I_{21}^\reg$ and $I_{32}^\reg$-type, respectively: the
$s_{12}^{-1}$ residue leaves the rational function $z_{43} \leftrightarrow
s_{34}$ referring to the second and third region momentum $(k_{12},k_3,k_4)
:= (k_{A_1},k_{A_2},k_{A_3})$. Applying the relabeling prescription
(\ref{02,203}) to $z_{43}$ with $a_i = 3 \in A_2$ and $b_i = 4 \in A_3$
identifies the $s_{12}^{-1}$ residue to be of $I_{32}^\reg$-type. At the
$s_{34}^{-1}$ residue, on the other hand, the rational function $z_{21}
\leftrightarrow s_{12}$ stays invariant under the relabeling prescription
(\ref{02,203}) leading to a contact vertex of $I_{21}^\reg$-type. Hence, the
pole structure of $Z[s_{12}s_{34}]$ is determined by:

\begin{figure}[htbp]
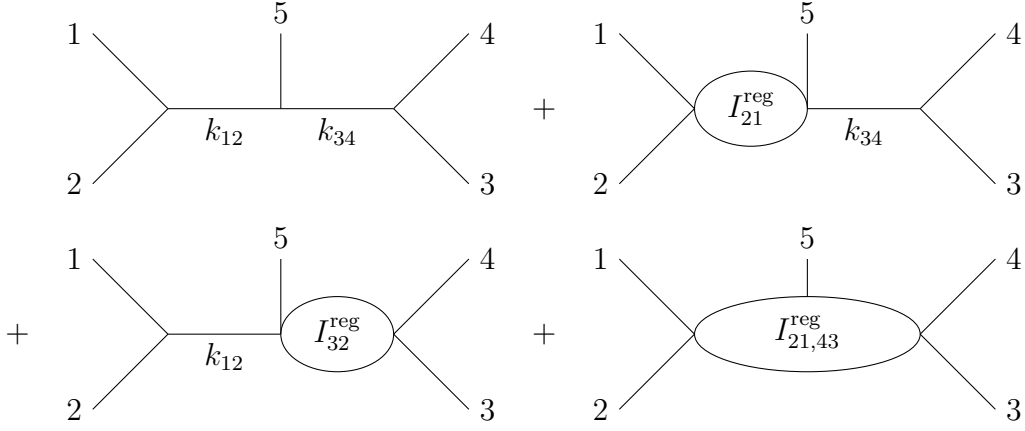
 
\begin{center}
\tikzpicture
%
\draw (0,0) -- (-1,1) node[left]{$1$};
\draw (0,0) -- (-1,-1) node[left]{$2$};
\draw (0,0) -- (3,0);
\draw (0.75,-0.3) node{$k_{12}$};
\draw (2.25,-0.3) node{$k_{34}$};
\draw (1.5,0) -- (1.5,1) node[above]{$5$};
\draw (3,0) -- (4,1) node[right]{$4$};
\draw (3,0) -- (4,-1) node[right]{$3$};
%
\draw (5,0) node{$+$};
\draw (-2,-3) node{$+$};
\draw (5,-3) node{$+$};
\scope[xshift=7cm]
\draw (0,0) -- (-1,1) node[left]{$1$};
\draw (0,0) -- (-1,-1) node[left]{$2$};
\draw (1.5,0) -- (3,0);
\draw (2.25,-0.3) node{$k_{34}$};
\draw (0.75,0) ellipse (0.75cm and 0.5cm);
\draw (0.75,0) node{$I_{21}^\reg$};
\draw (1.5,0) -- (1.5,1) node[above]{$5$};
\draw (3,0) -- (4,1) node[right]{$4$};
\draw (3,0) -- (4,-1) node[right]{$3$};
\endscope
\scope[xshift=0cm,yshift=-3cm]
\draw (0,0) -- (-1,1) node[left]{$1$};
\draw (0,0) -- (-1,-1) node[left]{$2$};
\draw (0,0) -- (1.5,0);
\draw (0.75,-0.3) node{$k_{12}$};
\draw (2.25,0) ellipse (0.75cm and 0.5cm);
\draw (2.25,0) node{$I_{32}^\reg$};
\draw (1.5,0) -- (1.5,1) node[above]{$5$};
\draw (3,0) -- (4,1) node[right]{$4$};
\draw (3,0) -- (4,-1) node[right]{$3$};
\endscope
\scope[xshift=7cm,yshift=-3cm]
\draw (0,0) -- (-1,1) node[left]{$1$};
\draw (0,0) -- (-1,-1) node[left]{$2$};
\draw (1.5,0) ellipse (1.5cm and 0.5cm);
\draw (1.5,0) node{$I_{21,43}^\reg$};
%
\draw (1.5,0.5) -- (1.5,1) node[above]{$5$};
\draw (3,0) -- (4,1) node[right]{$4$};
\draw (3,0) -- (4,-1) node[right]{$3$};
\endscope
\endtikzpicture
\caption{Pole structure of the function $Z[s_{12}s_{34}]$.}
\label{fig:s12s34}
\end{center}
\end{figure}

\beq
Z[s_{12}s_{34}] \ \ = \ \ 
\frac{1}{s_{12} s_{34}} \ + \ \frac{1}{s_{34}} \, I_{21}^\reg(k_1,k_2,k_{34}) \ + \ \frac{1}{s_{12}} \, I_{32}^\reg(k_{12},k_3,k_4) \ + \ I^\reg_{21,43}
\label{02,21}
\eeq
If we view $I_{21}^\reg$ and $I_{32}^\reg$ as a function of the four-point
Mandelstam invariants $s_{12}$ and $s_{23}$,
\beq
I_{uv}^\reg(k_1,k_2,k_3) \eq I_{uv}^\reg[s_{12},s_{23}] \ ,
\label{ktos1}
\eeq
then the composite momenta in $I_{21}^\reg(k_1,k_2,k_{34})$ and
$I_{32}^\reg(k_{12},k_3,k_4)$ replace the Mandelstam variables 
$[s_{12},s_{23}]$ as follows:
\beq
I_{21}^\reg(k_1,k_2,k_{34}) \eq I_{21}^\reg[s_{12},s_{23}+s_{24}] \co 
I_{32}^\reg(k_{12},k_3,k_4)\eq I_{32}^\reg[s_{13}+s_{23},s_{34}]\,.
\label{ktos2}
\eeq
The regular part $I^\reg_{21,43}$, on the other hand, constitutes an intrinsic
five-point vertex whose low-energy behavior is computed in the later section
\ref{secPolylogs}.

\begin{figure}[htbp]
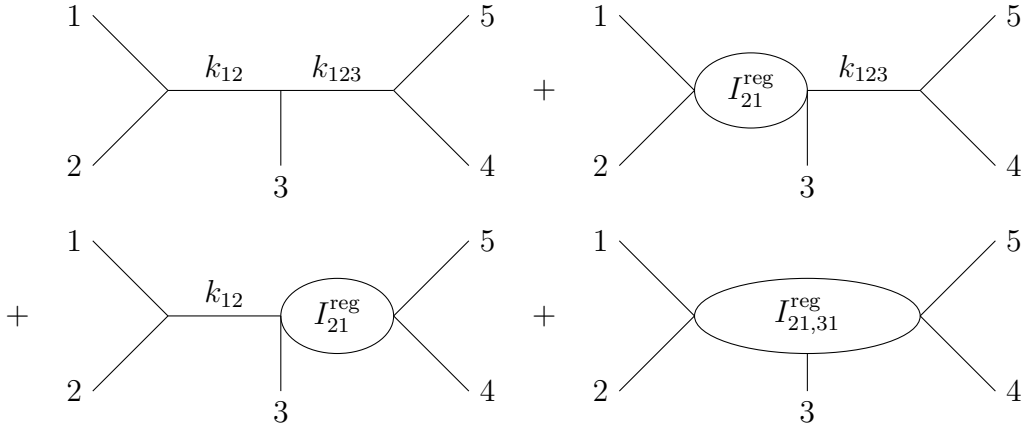
 
\begin{center}
\tikzpicture
%
\draw (0,0) -- (-1,1) node[left]{$1$};
\draw (0,0) -- (-1,-1) node[left]{$2$};
\draw (0,0) -- (3,0);
\draw (0.75,0.3) node{$k_{12}$};
\draw (2.25,0.3) node{$k_{123}$};
\draw (1.5,0) -- (1.5,-1) node[below]{$3$};
\draw (3,0) -- (4,1) node[right]{$5$};
\draw (3,0) -- (4,-1) node[right]{$4$};
%
\draw (5,0) node{$+$};
\draw (-2,-3) node{$+$};
\draw (5,-3) node{$+$};
\scope[xshift=7cm]
\draw (0,0) -- (-1,1) node[left]{$1$};
\draw (0,0) -- (-1,-1) node[left]{$2$};
\draw (1.5,0) -- (3,0);
\draw (2.25,0.3) node{$k_{123}$};
\draw (1.5,0) -- (1.5,-1) node[below]{$3$};
\draw (3,0) -- (4,1) node[right]{$5$};
\draw (3,0) -- (4,-1) node[right]{$4$};
\draw (0.75,0) ellipse (0.75cm and 0.5cm);
\draw (0.75,0) node{$I_{21}^\reg$};
\endscope
\scope[xshift=0cm,yshift=-3cm]
\draw (0,0) -- (-1,1) node[left]{$1$};
\draw (0,0) -- (-1,-1) node[left]{$2$};
\draw (0,0) -- (1.5,0);
\draw (0.75,0.3) node{$k_{12}$};
\draw (1.5,0) -- (1.5,-1) node[below]{$3$};
\draw (3,0) -- (4,1) node[right]{$5$};
\draw (3,0) -- (4,-1) node[right]{$4$};
\draw (2.25,0) ellipse (0.75cm and 0.5cm);
\draw (2.25,0) node{$I_{21}^\reg$};
\endscope
\scope[xshift=7cm,yshift=-3cm]
\draw (0,0) -- (-1,1) node[left]{$1$};
\draw (0,0) -- (-1,-1) node[left]{$2$};
\draw (1.5,-0.5) -- (1.5,-1) node[below]{$3$};
\draw (3,0) -- (4,1) node[right]{$5$};
\draw (3,0) -- (4,-1) node[right]{$4$};
\draw (1.5,0) ellipse (1.5cm and 0.5cm);
\draw (1.5,0) node{$I_{21,31}^\reg$};
\endscope
\endtikzpicture
\caption{Pole structure of the function $Z[s_{12}s_{123}]$.}
\label{fig:s12s123}
\end{center}
\end{figure}

The second integral basis element $Z[\zeta_2]$ is regular by itself,
\beq
Z[\zeta_2] \eq I^\reg_{31,42} \ .
\label{02,22}
\eeq
This introduces another five-point contact vertex $I^\reg_{31,42}$ to be
expanded in section \ref{secPolylogs}. Together with (\ref{02,21}), this
completes the $F^\si$ basis, and any other five-point integral follows from
linear combinations (\ref{invmom2}) of $F^{(23)}$ and $F^{(32)}$, e.g.
\begin{align}
Z[s_{12}s_{123}] \eq &
\frac{F^{(23)}}{s_{12} s_{123}} \ + \ \frac{F^{(32)}}{s_{13} s_{123}} 
\label{02,23}
\\
Z[s_{23}s_{123}] \eq & \frac{F^{(23)} }{s_{23} s_{123}} \ - \ \Big( \frac{1}{s_{13}} + \frac{1}{s_{23} } \Big) \, \frac{F^{(32)}} {s_{123}} \ .
\label{02,24}
\end{align}
Nevertheless, we carry out the pole analysis for the remaining parity
independent\footnote{The functions $Z[s_{34} s_{234}]$ and $Z[s_{23} s_{234}]$
associated with the remaining channels follow from parity $(1,2,3,4,5) \mapsto
(4,3,2,1,5)$.} field-theory channels to illustrate our method. Repeating the
diagrammatic method applied to $Z[s_{12}s_{34}]$ leads to
\begin{align}
Z[s_{12}s_{123}] \eq& \frac{1}{s_{12} s_{123}} \ + \  
\frac{I_{21}^\reg(k_1,k_2,k_3)}{s_{123}} \ + \  \frac{I_{21}^\reg( k_{12},k_3,k_4 )}{s_{12}}  \ + \ I_{21,31}^\reg
\label{02,25}
 \\
Z[s_{23}s_{123}] \eq & \frac{1}{s_{23} s_{123}} \ + \
\frac{I_{32}^\reg(k_1,k_2,k_3)}{s_{123}} \ + \ 
 \frac{I_{21}^\reg(k_1,k_{23},k_4)}{s_{23}} \ + \ I_{32,31}^\reg \ ,
 \label{02,26}
\end{align}
see figure \ref{fig:s12s123} for the former.

The $I^\reg_{21,31}$ and $I^\reg_{31,32}$ vertices enter various six- and
seven-point residues. Combining (\ref{02,23}) and (\ref{02,24}) with
(\ref{02,25}) and (\ref{02,26}) allows to infer their $\ap$-expansion from
$F^{(23)}$ and $F^{(32)}$, bypassing the need to compute them from the scratch.

The residues at the pole in $s_{123}$ involve three on-shell
momenta because the composite momentum $k_{123}=k_{45}$ associated with
$z_5\rightarrow \infty$ is suppressed, see figure \ref{fig:s12s123}. More
importantly, the $s_{23}^{-1}$ residue in (\ref{02,26}) is the first examples
where the norm of an off-shell momentum $k_{23}$ enters:
 \beq
I_{21}^\reg( k_{12},k_3,k_4 ) \eq I_{21}^\reg[s_{13}+s_{23},s_{34}] \co
 I_{21}^\reg(k_1,k_{23},k_4) \eq I_{21}^\reg[s_{123},s_{24} + s_{34}] \ .
 \label{ktos3}
 \eeq
The first Mandelstam argument of the latter is composed of $s_{123} = 2\ap k_1
\cdot k_{23} + \ap k_{23}^2$. Together with (\ref{ktos2}), this identifies the
off-shell completion of the $I_{21}^\reg$ and $I_{32}^\reg$ vertices:
\begin{align}
I_{21}^\reg(k_A,k_B,k_C) \eq& I_{21}^\reg[ 2\ap k_A \cdot k_B + \ap k_B^2,  \ 2\ap k_B \cdot k_C] \notag \\
I_{32}^\reg(k_A,k_B,k_C) \eq& I_{32}^\reg[ 2\ap k_A \cdot k_B ,  \ 2\ap k_B \cdot k_C+ \ap k_B^2] \ .
\label{21mass}
\end{align}
The $k_{23}^2$ dependence is invisible at the $s_{23}$ residue since
$I_{21}^\reg[s_{123},s_{24} + s_{34}] - I_{21}^\reg[s_{12}+s_{13},s_{24} +
s_{34}] = \CO(s_{23})$. However, it can be detected by comparing the regular
part $I_{32,31}^\reg$ computed in section \ref{secPolylogs} with the basis
decomposition (\ref{02,24}) of $Z[s_{23} s_{123}]$.

\subsubsection{Six-point examples}
\label{subsec:pole6pt}

According to subsection \ref{subsec}, the six-point basis $F^{\si}$ of
integrals can be most efficiently built from the six elementary integrals in
(\ref{02,79}) corresponding to the $\nu=3$ sector of the KK basis (\ref{02,2}).
Two of them introduce a field-theory channel with three simultaneous poles
$s_{12} s_{123} s_{45}$ and $s_{23} s_{123} s_{45}$, respectively. The
diagrammatic method identifies the following residues for the subleading poles:
\begin{align}
Z[s_{12}s_{123}s_{45}] \ \ = \ \ &\frac{1}{s_{12} s_{123} s_{45}} 
\ + \ \frac{I_{21}^\reg(k_1,k_2,k_3)}{s_{123} s_{45}} \ + \ \frac{ I_{21}^\reg(k_{12},k_3,k_{45}) }{s_{12} s_{45}} \ + \ \frac{ I_{32}^\reg(k_{123},k_4,k_5) }{ s_{123} s_{12}} 
\notag \\ 
&+ \ \frac{I_{21,31}^\reg(k_1,k_2,k_3,k_{45})}{ s_{45} } \ + \  \frac{ I_{21}^\reg(k_1,k_2,k_3)  I_{32}^\reg(k_{123},k_4,k_5) }{s_{123} } \notag \\
& + \ \frac{I_{21,43}^\reg(k_{12},k_3,k_4,k_5) }{s_{12}} \ + \ I^\reg_{21,31,54}
      \label{02,27} \\
Z[s_{23}s_{123}s_{45}] \eq &
\frac{1}{s_{23}s_{123}s_{45}}  \ + \ \frac{ I_{32}^\reg(k_1,k_2,k_3) }{s_{123} s_{45}} \ + \ \frac{ I_{21}^\reg(k_1,k_{23},k_{45}) }{s_{23}s_{45}} \  + \ \frac{I_{32}^\reg(k_{123},k_4,k_5) }{s_{23} s_{123} }
 \notag \\
 & + \ \frac{ I_{31,32}^\reg(k_1,k_2,k_3,k_{45}) }{s_{45}}  \ + \ \frac{ I_{32}^\reg(k_1,k_2,k_3) I_{32}^\reg(k_{123},k_4,k_5)  }{s_{123}}   \notag \\
 & + \ \frac{ I_{21,43}^\reg(k_1,k_{23},k_4,k_5) }{s_{23}}  \ + \ I_{31,32,54}^\reg
 \label{02,28}
 \end{align}
The diagrams contributing to the first case are shown in figure
\ref{fig:s12s123s45}:

\begin{figure}[h]
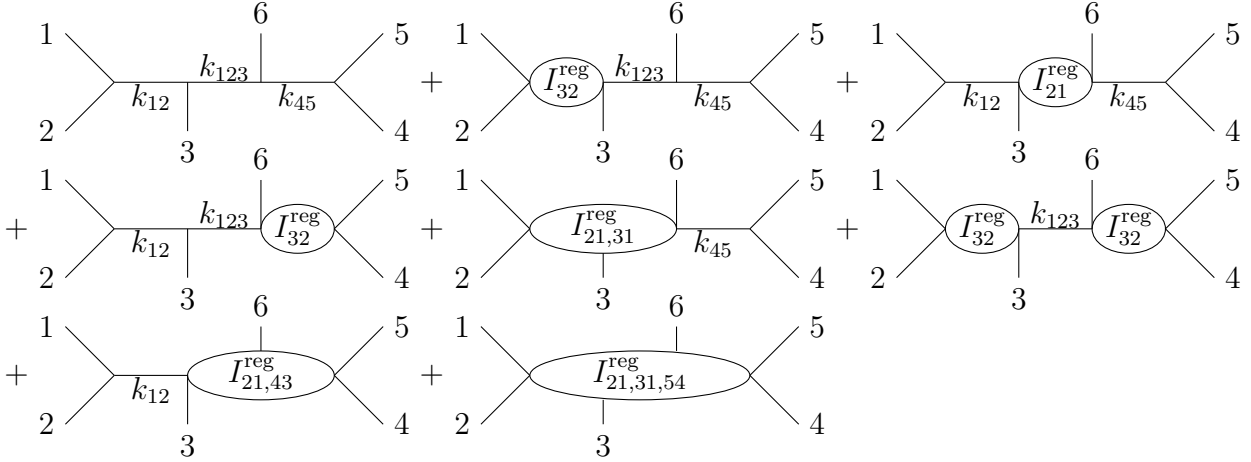
 
\begin{center}
\tikzpicture[scale=0.65]
\draw (0,0) -- (-1,1) node[left]{$1$};
\draw (0,0) -- (-1,-1) node[left]{$2$};
\draw (0,0) -- (4.5,0);
\draw (0.75,-0.3) node{$k_{12}$};
\draw (2.25,0.3) node{$k_{123}$};
\draw (3.75,-0.3) node{$k_{45}$};
\draw (1.5,0) -- (1.5,-1) node[below]{$3$};
\draw (3,0) -- (3,1) node[above]{$6$};
\draw (4.5,0) -- (5.5,1) node[right]{$5$};
\draw (4.5,0) -- (5.5,-1) node[right]{$4$};
\draw (6.5,0) node{$+$};
\draw (6.5,-3) node{$+$};
\draw (6.5,-6) node{$+$};
\draw (15,0) node{$+$};
\draw (-2,-3) node{$+$};
\draw (-2,-6) node{$+$};
\draw (15,-3) node{$+$};
\scope[xshift=8.5cm]
\draw (0,0) -- (-1,1) node[left]{$1$};
\draw (0,0) -- (-1,-1) node[left]{$2$};
\draw (1.5,0) -- (4.5,0);
%
\draw (2.25,0.3) node{$k_{123}$};
\draw (3.75,-0.3) node{$k_{45}$};
\draw (1.5,0) -- (1.5,-1) node[below]{$3$};
\draw (3,0) -- (3,1) node[above]{$6$};
\draw (4.5,0) -- (5.5,1) node[right]{$5$};
\draw (4.5,0) -- (5.5,-1) node[right]{$4$};
\draw (0.75,0) ellipse (0.75cm and 0.5cm);
\draw (0.75,0) node{$I_{32}^\reg$};
\endscope
\scope[xshift=17cm,yshift=-0cm]
\draw (0,0) -- (-1,1) node[left]{$1$};
\draw (0,0) -- (-1,-1) node[left]{$2$};
\draw (0,0) -- (1.5,0);
\draw (4.5,0) -- (3,0);
\draw (0.75,-0.3) node{$k_{12}$};
\draw (3.75,-0.3) node{$k_{45}$};
\draw (1.5,0) -- (1.5,-1) node[below]{$3$};
\draw (3,0) -- (3,1) node[above]{$6$};
\draw (4.5,0) -- (5.5,1) node[right]{$5$};
\draw (4.5,0) -- (5.5,-1) node[right]{$4$};
\draw (2.25,0) ellipse (0.75cm and 0.5cm);
\draw (2.25,0) node{$I_{21}^\reg$};
\endscope
\scope[xshift=0cm,yshift=-3cm]
\draw (0,0) -- (-1,1) node[left]{$1$};
\draw (0,0) -- (-1,-1) node[left]{$2$};
\draw (0,0) -- (3,0);
\draw (0.75,-0.3) node{$k_{12}$};
\draw (2.25,0.3) node{$k_{123}$};
\draw (1.5,0) -- (1.5,-1) node[below]{$3$};
\draw (3,0) -- (3,1) node[above]{$6$};
\draw (4.5,0) -- (5.5,1) node[right]{$5$};
\draw (4.5,0) -- (5.5,-1) node[right]{$4$};
\draw (3.75,0) ellipse (0.75cm and 0.5cm);
\draw (3.75,0) node{$I_{32}^\reg$};
\endscope
\scope[xshift=8.5cm,yshift=-3cm]
\draw (0,0) -- (-1,1) node[left]{$1$};
\draw (0,0) -- (-1,-1) node[left]{$2$};
\draw (3,0) -- (4.5,0);
%
\draw (3.75,-0.3) node{$k_{45}$};
\draw (1.5,-0.5) -- (1.5,-1) node[below]{$3$};
\draw (3,0) -- (3,1) node[above]{$6$};
\draw (4.5,0) -- (5.5,1) node[right]{$5$};
\draw (4.5,0) -- (5.5,-1) node[right]{$4$};
\draw (1.5,0) ellipse (1.5cm and 0.5cm);
\draw (1.5,0) node{$I_{21,31}^\reg$};
\endscope
\scope[xshift=17cm, yshift=-3cm]
\draw (0,0) -- (-1,1) node[left]{$1$};
\draw (0,0) -- (-1,-1) node[left]{$2$};
\draw (1.5,0) -- (3,0);
%
\draw (2.25,0.3) node{$k_{123}$};
\draw (1.5,0) -- (1.5,-1) node[below]{$3$};
\draw (3,0) -- (3,1) node[above]{$6$};
\draw (4.5,0) -- (5.5,1) node[right]{$5$};
\draw (4.5,0) -- (5.5,-1) node[right]{$4$};
\draw (0.75,0) ellipse (0.75cm and 0.5cm);
\draw (0.75,0) node{$I_{32}^\reg$};
\draw (3.75,0) ellipse (0.75cm and 0.5cm);
\draw (3.75,0) node{$I_{32}^\reg$};
\endscope
\scope[yshift=-6cm]
\draw (0,0) -- (-1,1) node[left]{$1$};
\draw (0,0) -- (-1,-1) node[left]{$2$};
\draw (0,0) -- (1.5,0);
\draw (0.75,-0.3) node{$k_{12}$};
\draw (1.5,0) -- (1.5,-1) node[below]{$3$};
\draw (3,0.5) -- (3,1) node[above]{$6$};
\draw (4.5,0) -- (5.5,1) node[right]{$5$};
\draw (4.5,0) -- (5.5,-1) node[right]{$4$};
\draw (3,0) ellipse (1.5cm and 0.5cm);
\draw (3,0) node{$I_{21,43}^\reg$};
\endscope
\scope[xshift=8.5cm,yshift=-6cm]
\draw (0,0) -- (-1,1) node[left]{$1$};
\draw (0,0) -- (-1,-1) node[left]{$2$};
%
\draw (1.5,-0.5) -- (1.5,-1) node[below]{$3$};
\draw (3,0.5) -- (3,1) node[above]{$6$};
\draw (4.5,0) -- (5.5,1) node[right]{$5$};
\draw (4.5,0) -- (5.5,-1) node[right]{$4$};
\draw (2.25,0) ellipse (2.25cm and 0.5cm);
\draw (2.25,0) node{$I_{21,31,54}^\reg$};
\endscope
\endtikzpicture
\caption{Pole structure of the function $Z[s_{12}s_{123}s_{45}]$.}
\label{fig:s12s123s45}
\end{center}
\end{figure}

Two simultaneous poles remain whenever one collapsed propagator is bypassed by
a quartic vertex $I_{21}^\reg$ or $I_{32}^\reg$, possibly at composite momenta
such as $k_{123},k_{12}$ and $k_{45}$. The nature of the three single-pole
diagrams depends on the relative positions of the collapsed propagators: If
they are connected, the resulting contact vertex is quintic and depends on
momenta through $I_{21,31}^\reg$ and $ I_{21,43}^\reg$, see (\ref{02,21}) and
(\ref{02,25}). In the disconnected situation, two independent quartic vertices
remain, and the residue factorizes into $I_{21}^\reg(k_1,k_2,k_3)
I_{32}^\reg(k_{123},k_4,k_5)$ and $I_{32}^\reg(k_1,k_2,k_3)
I_{32}^\reg(k_{123},k_4,k_5)$, respectively.

Some of the residues in (\ref{02,28}) depend on the norm of their off-shell
momenta, first of all $I_{21}^\reg(k_1,k_{23},k_{45}) =
I_{21}^\reg[s_{123},s_{24}+s_{25}+s_{34}+s_{35}]$ in agreement with the general
off-shell completion (\ref{21mass}) of the four-point contact vertices. In
addition, we have the following mass-dependent five-point vertex
\beq
I_{21,43}^\reg(k_1,k_{23},k_4,k_5)  \eq I_{21,43}^\reg[s_{123},s_{14},s_{24}+s_{34},s_{25}+s_{35},s_{45}] \ ,
\label{ktos4}
\eeq
see the explanations below (\ref{02,204}).
We observe that the norm $k_{A_1}^2$ and $k_{A_{p+2}}^2$ of the first and last
momentum drops out from any contact vertex $I_{\ldots}^\reg$
investigated up to seven points.

According to \eqn{02,79} the remaining four integrals (\ref{02,81}) and
(\ref{02,82}) contributing to $F^{\si}$ have at most single poles. The residues
within $Z[\zeta_2 s_{12}]$ and $Z[\zeta_2 s_{34}]$ are
\begin{align}
Z[\zeta_2 s_{12}] \eq &\frac{ I^\reg_{31,42}(k_{12},k_3,k_4,k_5) }{s_{12}} \ + \ I_{21,41,53}^\reg 
\label{02,29} \\
Z[\zeta_2 s_{34}] \eq &\frac{ I^\reg_{31,42}(k_1,k_2,k_{34},k_5) }{s_{34}} \ + \ I_{41,43,52}^\reg  \ ,
\label{02,30}
\end{align}
see figure \ref{fig:s12}. We can identify another explicit dependence on a
momentum norm $k_{34}^2$,
\beq
I^\reg_{31,42}(k_1,k_2,k_{34},k_5) \eq I^\reg_{31,42}[s_{12},s_{134},s_{23}+s_{24},s_{25},s_{35}+s_{45}] \ .
\label{ktos5}
\eeq

\begin{figure}[htbp]
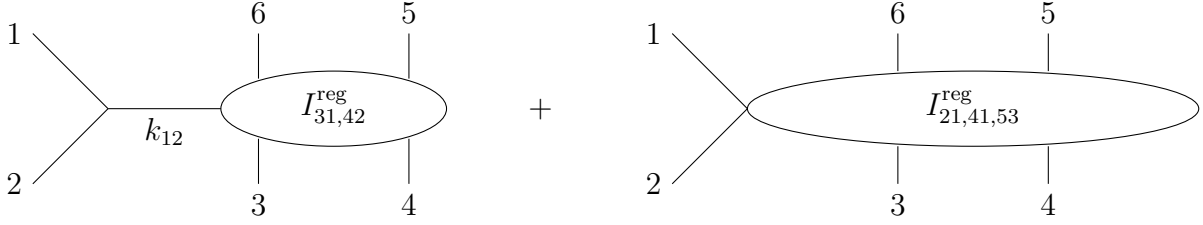
 
\begin{center}
\tikzpicture
\scope
\draw (0,0) -- (-1,1) node[left]{$1$};
\draw (0,0) -- (-1,-1) node[left]{$2$};
\draw (0,0) -- (1.5,0);
\draw (0.75,-0.3) node{$k_{12}$};
\draw (2,0.4) -- (2,1) node[above]{$6$};
\draw (2,-0.4) -- (2,-1) node[below]{$3$};
\draw (4,0.4) -- (4,1) node[above]{$5$};
\draw (4,-0.4) -- (4,-1) node[below]{$4$};
\draw (3,0) ellipse (1.5cm and 0.5cm);
\draw (3,0) node{$I_{31,42}^\reg$};
\endscope
\draw (5.75,0)node{$+$};
\scope[xshift=8.5cm]
\draw (0,0) -- (-1,1) node[left]{$1$};
\draw (0,0) -- (-1,-1) node[left]{$2$};
\draw (2,0.5) -- (2,1) node[above]{$6$};
\draw (2,-0.5) -- (2,-1) node[below]{$3$};
\draw (4,0.5) -- (4,1) node[above]{$5$};
\draw (4,-0.5) -- (4,-1) node[below]{$4$};
\draw (3,0) ellipse (3cm and 0.5cm);
\draw (3,0) node{$I_{21,41,53}^\reg$};
\endscope
\endtikzpicture
\caption{Pole structure of the function $Z[\zeta_2 s_{12}]$.}
\label{fig:s12a}
\end{center}
\end{figure}

Finally, the integrals $\tilde Z[\zeta_3]$ and $Z[\zeta_3]$ are regular by themselves,
\beq
\tilde Z[\zeta_3] \eq I^\reg_{41,42,53} \co
Z[\zeta_3] \eq I^\reg_{41,43,52} \ .
\label{02,31}
\eeq
Once we have expanded the regular parts in (\ref{02,27}) to (\ref{02,31}) using
the methods of section \ref{secPolylogs}, the information on the six-point
basis is complete. In appendix \ref{app6pt}, we discuss the treatment of the
remaining field-theory pole channels and thereby introduce further six-point
vertices $I_{ij,kl,mn}^\reg$ relevant for residues in $(N\geq 7)$-point
integrals.
            
\subsubsection{Seven-point examples}
\label{7ptexpl}

At seven points, integrals which generate one cubic YM diagram in their field
theory limit leave fourteen residues associated with subleading poles. This
cornucopia of diagrams motivates to introduce an economic setup in section
\ref{sec3} to construct the $F^\si$ basis from integrals with a minimal number
of poles. As it will turn out, it is sufficient to address one field-theory
diagram, e.g. 
\begin{align}
Z[&s_{12}s_{123}s_{1234}s_{56}] \ \, = \, \ \frac{1}{s_{12}s_{123}s_{1234}s_{56}} 
\; + \; \frac{ I_{21}^\reg(k_1,k_2,k_3) }{s_{123}s_{1234}s_{56}}
 \; + \; \frac{ I^\reg_{21}(k_{12},k_3,k_{4}) }{s_{12}s_{1234}s_{56} }
  \; + \; \frac{ I^\reg_{21}(k_{123},k_4,k_{56}) }{s_{12}s_{123}s_{56}} \notag \\
&\; + \; \frac{ I^\reg_{32}(k_{1234},k_5,k_6) }{s_{12}s_{123}s_{1234}}
\; + \; \frac{ I_{21,31}^\reg(k_1,k_2,k_3,k_{4}) }{s_{56} s_{1234}} \; + \; \frac{  I_{21}^\reg(k_1,k_2,k_3) I^\reg_{21}(k_{123},k_4,k_{56})  }{s_{123}s_{56}} \label{02,32} \\
 &\; + \; \frac{  I_{21}^\reg(k_1,k_2,k_3)  I^\reg_{32}(k_{1234},k_5,k_6) }{s_{123}s_{1234}} \; + \; \frac{ I_{21,31}^\reg(k_{12},k_3,k_4,k_{56} )}{s_{12}s_{56}} \; + \; \frac{ I^\reg_{21,43}(k_{123},k_4,k_5,k_6) }{s_{12}s_{123}} \notag \\
 &\; + \; \frac{ I^\reg_{21}(k_{12},k_3,k_{4})  I^\reg_{32}(k_{1234},k_5,k_6)  }{s_{12}s_{1234}}
\; + \;  \frac{ I^\reg_{21,31,41}(k_1,k_2,k_3,k_4,k_{56}) }{s_{56}} \; + \; \frac{I^\reg_{21,31,54}(k_{12},k_3,k_4,k_5,k_6)}{s_{12}}  \notag \\
 &\; + \; \frac{ I_{21}^\reg(k_1,k_2,k_3) I^\reg_{21,43}(k_{123},k_4,k_5,k_6) }{ s_{123}} \; + \; \frac{ I_{21,31}^\reg(k_1,k_2,k_3,k_{4}) I^\reg_{32}(k_{1234},k_5,k_6) }{s_{1234}} \; + \; I^\reg_{21,31,41,65}
\notag 
 \end{align}
None of the $I_{\ldots}^\reg$ shown depends on $k_{A_i}^2$. 

According to the discussion in subsection \ref{sec:block}, we construct the
$F^\si$ from the $\nu=3$ sector of the KK basis and think of its integrals
$Z(1,2_\rho,3_\rho,7,4_\rho,5_\rho,6)$ as forming six blocks labeled by
$\{2_\rho,3_\rho\}$. The $\{3,5\}$ functions are characterized by polynomials $
\in \{ z_{31} z_{51} , \, z_{51} z_{53} \} \times \{ z_{42} z_{64} , \, z_{62}
z_{64} \}$ and do not exhibit any poles. Let us display the residues necessary
to construct the $\ap$-expansion of the $\{2,5\}$ block (covering the $\{3,4\}$
block by parity):
\begin{align}
\prod_{i=2}^5 \int_0^{z_{i+1}} \dd z_i \ \frac{ \prod_{i<j}^6 |z_{ij}|^{s_{ij}} }{z_{21} z_{51} z_{34} z_{46}} \eq&  \frac{ I_{31,42}^\reg(k_{12},k_{34},k_5,k_6) }{s_{12} s_{34}}  \ + \ \frac{ I_{41,32,53}^\reg(k_{12},k_3,k_4,k_5,k_6) }{s_{12}} \notag \\
& \ \ + \ \frac{ I_{21,41,53}^\reg(k_1,k_2,k_{34},k_5,k_6) }{s_{34}} \ + \ I_{21,51,43,64}^\reg
\label{02,37}
\\
%
\prod_{i=2}^5 \int_0^{z_{i+1}} \dd z_i \ \frac{ \prod_{i<j}^6 |z_{ij}|^{s_{ij}} }{z_{25} z_{51} z_{34} z_{46}} \eq&- \, \frac{ I^\reg_{41,42,53}(k_1,k_2,k_{34},k_5,k_6) }{s_{34}} \ - \ I_{51,52,43,64}^\reg 
\label{02,38} \\
%
\prod_{i=2}^5 \int_0^{z_{i+1}} \dd z_i \ \frac{ \prod_{i<j}^6 |z_{ij}|^{s_{ij}} }{z_{21} z_{51} z_{36} z_{46}} \eq& \frac{ I^\reg_{41,52,53}(k_{12},k_3,k_4,k_5,k_6) }{s_{12}} \ + \ I_{21,51,63,64}^\reg
\label{02,39} \\
\prod_{i=2}^5 \int_0^{z_{i+1}} \dd z_i \ \frac{ \prod_{i<j}^6 |z_{ij}|^{s_{ij}} }{z_{25} z_{51} z_{36} z_{46}} \eq& -\,I_{51,52,63,64}^\reg
\label{02,40}
\end{align}
with mass dependence
\begin{align}
I_{31,42}^\reg(k_{12},k_{34},k_5,k_6) &\eq I_{31,42}^\reg[s_{1234}-s_{12}, s_{15}+s_{25},s_{35}+s_{45},s_{36}+s_{46},s_{56}] \label{ktos21} \\
I_{21,41,53}^\reg(k_1,k_2,k_{34},k_5,k_6) &\eq I_{21,41,53}^\reg[s_{12},s_{134},s_{15},s_{23}+s_{24} ,s_{25},s_{35}+s_{45} ,s_{26},s_{36}+s_{46},s_{56}]\notag \\
I^\reg_{41,42,53}(k_1,k_2,k_{34},k_5,k_6) &\eq I^\reg_{41,42,53}[s_{12},s_{134},s_{15},s_{23}+s_{24} ,s_{25},s_{35}+s_{45} ,s_{26},s_{36}+s_{46},s_{56}] \notag
\end{align}
The Mandelstam arguments of six-point contact vertices are displayed in the order
\beq
I^\reg_{ij,kl,mn}(k_1,k_2,k_3,k_{4},k_5) \eq I^\reg_{ij,kl,mn}[s_{12},s_{13},s_{14},s_{23},s_{24},s_{34},s_{25},s_{35},s_{45}]
\label{ktos23}
\eeq
which is reversed by $i \mapsto 6-i$ world-sheet parity. The $\{2,4\}$ and $\{
4,5\}$ blocks are analogously addressed in appendix \ref{poleblock}.  Let us
now turn to the remaining task of expanding the regular parts $I_{\ldots}^\reg$
in $\ap$.


\section{Polylogarithms - Calculation of the regulated integrals}
\label{secPolylogs}

This section is devoted to the evaluation of the regular part of $N$-point disk
integrals. As explained in the previous section, their singular parts are
recursively determined by lower-point disk amplitudes. Thus, we will assume
that the pole residues have already been accounted for by the methods of
subsection \ref{sec1,23}. In the following, we set the stage for extracting the
intrinsic $N$-point information required for the $(N-3)!$ basis functions
$F^\si$ from a convenient set of regulated parts $I^\reg$ defined in
(\ref{02,202}). More precisely, we will provide methods to expand integrals
\begin{eqnarray}
I^\reg&=&
\prod_{k=2}^{N-2} \int^{z_{k+1}}_0 \frac{ \dd z_k }{z_k - a_k} \prod_{i<j}^{N-1} |z_{ij}|^{s_ij} \co a_k \in \{ \, 0,z_{k+1},z_{k+2},\ldots,z_{N-2},1 \, \}\nnl
&=&
  \prod_{k=2}^{N-2} \int^{z_{k+1}}_0 \frac{ \dd z_k }{z_k - a_k} \prod_{i<j}^{N-1} \sum_{n_{ij}=0}^{\infty}(s_{ij})^{n_{ij}} \; \frac{(\ln |z_{ij}|)^{n_{ij}}}{n_{ij}!}
\label{constraint}
\end{eqnarray}
where the constraint on the coefficients $a_k$ guarantees that one can
successively integrate over $z_2,z_3,\ldots,z_{N-2}$ with only one differential
form $\frac{ \text{\sffamily{d}} z_k }{z_k - a_k}$ depending on the integration variable $z_k$
in each step\footnote{There are numerous situations in which techniques for solving iterated integrals have been applied and explored, for example \cite{BrownGM, Bogner:2012dn}. In particular, handling the criterion in \eqn{constraint} was discussed in \cite{Brown:2008um} and - in a physics context - in \cite{Anastasiou:2013srw}.}. This leaves us with $(N-2)!$ functions\footnote{On the one hand,
the ability to expand the regular part of the $(N-2)!$ functions in
(\ref{constraint}) appears to be fully sufficient for the construction of an
$(N-3)!$ basis of $F^\si$. On the other hand, several of these $(N-2)!$
functions exhibit incompatible pole channels such that we cannot properly
assemble the regular and singular parts with the methods from the previous section. Hence, the
applicability of the present techniques to $(N-2)!$ regular parts is crucial
for gathering the complete information on the $N$-point integral basis.} in
(\ref{constraint}) accessible to the techniques described below, and one can
always apply partial fraction to express a generic integrand in this form.

In the following, expressions of the form (\ref{constraint}) shall be thought
of as disk integrals (\ref{02,202}) whose poles have already been subtracted. That is why the Koba-Nielsen factor is expanded via (\ref{02,201}). The
various Mandelstam variables $s_{ij}$ which have been pulled out from the
integral will contribute to the $\ap$-expansion
of the matrix $F$ in \eqn{00,1a}, while the multiple integration yields the
MZVs defined in section \ref{secap}. While for lower transcendentalities those
types of integrals can be solved by standard methods, we will here formalize
the way to obtain a solution for any multiplicity and weight. Multiple
polylogarithms serve as a suitable language to achieve this formalization and
to convert the integrals within (\ref{02,202}) or (\ref{constraint}) into
MZVs. So before explicitly solving the integral, let us collect a couple of
facts about and relations among polylogarithms in the next subsection. 

\subsection{Multiple polylogarithms}
Multiple polylogarithms---sometimes referred to as Goncharov polylogarithms---are 
defined as\footnote{Our definitions for polylogarithms agree with
Goncharovs paper \cite{GoncharovTate} and references \cite{SSMZV} and
\cite{DGR}. Other aspects of mutiple polylogarithms are discussed for example in references \cite{Bluem1,Bluem2}.} 
\begin{equation}
  G(a_1,a_2,\ldots,a_n;z) \ \ := \ \ \int_0^z\frac{\dd t}{t-a_1}G(a_2,\ldots,a_n;t)
  \label{eqn:defG}
\end{equation}
where $G(z)=G(;z)=1$ except for $G(\vec{a};0)=G(;0)=0$. In order to keep the terminology clear, we will refer to $\vec{a}=(a_1,\ldots,a_n)$ as the
\textit{label} and to $z$ as \textit{argument} of a polylogarithm $G$. Multiple
polylogarithms constitute a graded Lie algebra with the shuffle product (cf.
subsection \ref{sec1,12})
\begin{equation}
  G(a_1,\ldots,a_r;z)G(a_{r+1},\ldots,a_{r+s};z)=\sum_{\s\in \Sigma(r,s)}G(a_{\s(1)},\ldots,a_{\s(r+s)};z)
  \label{eqn:shuffle}
\end{equation}
where $\Sigma(r,s)$ is the subset of the permutation group $S_{r+s}$ acting on
$\{a_1,\ldots,a_{r+s}\}$ which leaves the order of the elements of the
individual sets $\{a_1,
\ldots,a_r\}$ and $\{a_{r+1},\ldots,a_{r+s}\}$ unchanged. The unit element for shuffling is $G(;z)$=1.

Multiple polylogarithms of uniform labels are related to powers of ordinary logarithms via
\begin{eqnarray}
  G(\underbrace{0,0,\ldots,0}_w;z)&=&\frac{1}{w!}(\ln\ z)^w \mand G(\underbrace{1,1\ldots,1}_w;z) \ = \ \frac{1}{w!}\ln^w(1-z)\nnl
  G(\underbrace{a,a,\ldots,a}_w;z)&=&\frac{1}{w!}\ln\left(1-\frac{z}{a}\right)^w 
  \label{eqn:lntoHG}
\end{eqnarray}
and satisfy the scaling property
\begin{equation}
  G(k\vec{a};kz)=G(\vec{a};z)\co k\neq 0\,.
  \label{scaling}
\end{equation}
The scaling relation does not apply to labels $\vec{a}$ with $a_i=0\,\forall\,i$. 

MZVs as defined in \eqn{00,4} are special cases of multiple polylogarithms
with $a_i\in\{0,1\}$ evaluated at argument $z=1$ (where the numbers below the
underbraces denote the number of entries.\footnote{Note, that the order of the
$n_1\ldots n_r$ on the right hand side of the equation is reversed with respect
to the left hand side. Our convention for the MZVs agrees with references
\cite{SSMZV, GoncharovTate} and \cite{BrownTate}.}):
\beq
  \zeta_{n_1,\ldots,n_r} \eq (-1)^r G(\underbrace{0,0,\ldots,0,1}_{n_r},\ldots,\underbrace{0,0,\ldots,0,1}_{n_1};1),
  \label{eqn:defZeta}
\eeq
As an immediate benefit of the scaling property \eqn{scaling}, multiple polylogarithms $G(\vec{a},z)$ with $a_i\in\{0,z\}$ can be rescaled to yield
\begin{equation}
  G(\{0,z\}_w;z)=G(\{0,1\}_w;1)
  \label{eqn:secretZeta}
\end{equation}
which in turn can be translated into MZVs via \eqn{eqn:defZeta}. In the above
equation $\{a,b,\ldots\}_w$ refers to a word of length $w$ built from the
letters $a,b,\ldots\,\,\,$.

Multiple polylogarithms are divergent integrals in general. As can be seen from
the definition \eqn{eqn:defG}, the divergences occur, when either $a_1=z$ or
$a_n=0$. Regularization of those integrals is discussed in detail in references
\cite{GoncharovTate,GoncharovGalois}. The general idea is to slightly move the
endpoints of the integration by a small parameter $\ve$ and to expand in this
parameter afterwards. The regularized value is defined to be the piece in the
expansion, which does not depend on the parameter $\ve$. Extracting the
$\ve$-independent piece by using shuffle relations, one can show that for the
case where $a_1=z$ the regularized value can be obtained via 
\begin{eqnarray}
 G(z,a_2,\ldots,a_n;z)&=& G(z;z)\,G(a_2,\ldots,a_n;z)-G(a_2,z,a_3,\ldots,a_n;z)\nnl
 &&\qquad-G(a_2,a_3,z,a_4,\ldots,a_n;z)-\ldots-G(a_2,\ldots,a_n,z;z)
  \label{shufflereg1}
\end{eqnarray}
where one defines 
\begin{equation}
 G(z,\ldots,z;z)=0  \ .
  \label{shufflereg1a}
\end{equation}
The other case, where $a_n=0$ can be dealt with in the same way 
\begin{align}
  G(a_1,a_2,\ldots,&a_{n-1},0;z)\eq G(a_1,a_2,\ldots,a_{n-1};z)G(0;z)-G(a_1,a_2,\ldots,0,a_{n-1};z)\notag \\
  &- G(a_1,a_2,\ldots,0,a_{n-2},a_{n-1};z)-\ldots-G(0,a_2,\ldots,a_{n-1};z)\,.
  \label{shufflereg2}
\end{align}
Here, however, $G(0;z) = \ln (z) \neq 0$. The above rewriting in order to keep
the pure logarithms explicit will nevertheless prove convenient below in order
to rewrite the polylogarithms in a form where the identity \eqn{eqn:GIdentity}
can be readily used.

In the same way as multiple polylogarithms have to be regulated, this is true
for the MZVs. In particular, one defines all MZVs with $n_r=1$ by their
shuffled version using \eqn{shufflereg1}. From \eqn{shufflereg1a} one
immediately finds $G(1,\ldots,1;1)=0$ using \eqn{scaling}. 

Before continuing with the evaluation of the integral in \eqn{constraint}, a
couple of remarks are appropriate concerning the multiple zeta data mine
\cite{Datamine}. This collection of identities between different MZVs for
weights up to $w=22$ allows expressing any MZV in the basis spelled out in
table \ref{zetaBasis} for weights $w\leq 12$. The accompanying articles and the
formulae presented in \cite{Datamine, RV} make use of harmonic polylogarithms,
which are - up to a sign - multiple polylogarithms with entries
$a_i\in\{-1,0,1\}$: 
\begin{equation}
  H(\vec{a};z)=(-1)^kG(\vec{a};z) \qquad a_i\in\{-1,0,1\} 
\end{equation}
where $k$ is the number of elements equal to $(+1)$ in $\vec{a}$
\footnote{This definition agrees with Remiddi and Vermaseren \cite{RV} and
Maitre \cite{MaitreHPL}.}.

\subsection{Performing the integration using polylogarithm identities}
In order to employ the rules from \eqn{eqn:lntoHG}, one has to slightly rewrite
the terms in the integrand of \eqn{constraint}. For example, at $N=5$ one finds
the term 
\begin{align}
  I^\reg_{\text{Ex${}_1$}}& \ \ := \ \ \frac{1}{2} s_{23}^2 \int_0^1\frac{\dd z_3}{z_3-1}\int_0^{z_3}\frac{\dd z_2}{z_2} \, (\ln(z_3-z_2))^2
  \label{IregExample} \\
  &\eq \frac{1}{2} s_{23}^2 \int_0^1\frac{\dd z_3}{z_3-1}\int_0^{z_3}\frac{\dd z_2}{z_2} \, \left[ \,(\ln(z_3))^2+2\,\ln z_3\,\ln\left(1-\frac{z_2}{z_3}\right)+\left(\ln\left(1-\frac{z_2}{z_3}\right)\right)^2 \, \right]\notag
\end{align}
as part of the integral $I^\reg_{21,34}$, where the absolute value is
automatically taken care of by our parametrization $z_i <z_{i+1} \in \ZR$ of
the disk boundary. After having written all terms in the expansion of the
integrand in \eqn{constraint} in the above form, one can express them in terms
of multiple polylogarithms $G$ using \eqn{eqn:lntoHG}, which will remove the
combinatorical factors from expanding powers of logarithms.

By rewriting logarithms and employing the rules in \eqn{eqn:lntoHG}, one can
express the complete Koba-Nielsen factor in terms of multiple
polylogarithms\footnote{The $k=N-1$ term in the double product of (\ref{KN})
covers the factors of $\prod_{i=2}^{N-2} \sum_{m_{i}=0}^{\infty}
s_{i,N-1}^{m_{i}} G(\vec{1}_{m_{i}};z_i)$.
}:
\begin{align}
 \prod_{i<j}^{N-1} |z_{ij}|^{s_{ij}}\,=\,
 \left\lbrace\prod_{i=2}^{N-2}\sum_{n_i=0}^\infty\Bigg(\sum_{l=1}^{i-1}s_{il}\Bigg)^{n_i}G(\vec{0}_{n_i};z_i)\right\rbrace\times
 \left\lbrace\prod_{2\leq j<k\leq N-1}\sum_{n_{jk}=0}^\infty s_{jk}^{n_{jk}}G((\vec{z}_k)_{n_{jk}};z_j)\right\rbrace\,,
 \label{KN}
\end{align}
where $\vec{0}_{n_i}$ and $(\vec{z}_k)_{n_{jk}}$ denotes label vectors with
$n_i$ entries $0$ and $n_{jk}$ entries $z_k$, respectively. For the particular
example in \eqn{IregExample} one will obtain
\begin{equation}
  I^\reg_{\text{Ex${}_1$}}\eq s_{23}^2 \int_0^1\frac{\dd z_3}{z_3-1}\int_0^{z_3}\frac{\dd z_2}{z_2}\,\big(G(0,0;z_3)\,+\,G(0;z_3)\,G(z_3;z_2)\,+\,G(z_3,z_3;z_2)\big)\,.
  \label{GPLExplicit}
\end{equation}
After the integrand has been cast into polylogarithms, one can start with the
formal integration. As long as the label $\vec{a}$ does not contain the
argument $z$ one can readily apply \eqn{eqn:defG}. If, however, the integrand
should not contain the integration variable at all, it has to be replaced by
$G(;z)=1$, which is the case for the first term in \eqn{GPLExplicit}.
Performing the integral over $z_2$ promotes \eqn{GPLExplicit} to
\begin{equation}
  I^\reg_{\text{Ex${}_1$}}\eq s_{23}^2 \int_0^1\frac{\dd z_3}{z_3-1}\big(G(0,0;z_3)G(0;z_3)\,+\,G(0;z_3)\,G(0,z_3;z_3)\,+\,G(0,z_3,z_3;z_3)\big)\,.
  \label{}
\end{equation}
In the above equation, one can recognize several MZVs in the integrand after
rescaling with $k=1/z_3$ (cf. \eqn{eqn:secretZeta}). However, as already
mentioned above, scaling is not allowed for labels consisting of zeros
exclusively. 

After rewriting $G(0,z_3;z_3)= -\zeta_2$ and $G(0,z_3,z_3;z_3)=  \zeta_3$, only
the first term contains the product of two multiple polylogarithms with
argument $z_3$. Using the shuffle relation \eqn{eqn:shuffle}, one finds
$G(0,0;z_3)\,G(0;z_3)=3\,G(0,0,0;z_3)$, which allows integration by means of
\eqn{eqn:defG} and leads to 
\begin{eqnarray}
  I^\reg_{\text{Ex${}_1$}}&=& s_{23}^2 \int_0^1\frac{\dd z_3}{z_3-1}\big(3\,G(0,0,0;z_3)\,+\,G(0;z_3)\,G(0,1;1)\,+\,G(0,1,1;1)\big)\nnl
  &=& s_{23}^2 \big(3\,G(1,0,0,0;1)\,+\,G(1,0;1)\,G(0,1;1)\,+\,G(1;1)\,G(0,1,1;1)\big)\nnl
  &=& s_{23}^2 \big(-3\,(G(0,1,0,0;1)+G(0,0,1,0;1)+G(0,0,0,1;1))\,-\,G(0,1;1)\,G(0,1;1)\big)\nnl
  &=& s_{23}^2 \big(3\,\z_4\,-\,\z_2^2\big)\nnl
  &=& \frac{s_{23}^2}{5} \z_2^2\,,
  \label{05,03}
\end{eqnarray}
where we performed the shuffle regulation \eqn{shufflereg1} for $G(1,0;1)$ and
$G(1,0,0,0;1)$ in the third line as well as recognized $G(1;1)=0$.

Unfortunately not all integrals can be solved with the same ease as
\eqn{IregExample}. Consider for example
\begin{equation}
   I^\reg_{\text{Ex${}_2$}}\ \ := \ \ \int_0^1\frac{\dd z_3}{z_3-1}\int_0^{z_3}\frac{\dd z_2}{z_2}\,G(z_3;z_2)\,G(1;z_2)
  \label{05,04}
\end{equation}
which is part of the calculation for the coefficient of $s_{23}s_{24}$ in
$I^\reg_{21,43}$. After shuffling the product of polylogarithms and performing
the integration over $z_2$ one is faced with the following integral:
\begin{equation}
   I^\reg_{\text{Ex${}_2$}}\eq\int_0^1\frac{\dd z_3}{z_3-1}\big(G(0,z_3,1;z_3)\,+\,G(0,1,z_3;z_3)\big)\,.
\end{equation}
Obviously, the definition \eqn{eqn:defG} can not be applied here, because the
vector of labels, $\vec{a}$, still contains the argument $z_3$ in both
polylogarithms. Furthermore, the disturbing $z_3$ in $\vec{a}$ can not be
removed by means of \eqn{eqn:secretZeta}. The general problem is the following:
how can a multiple polylogarithm of the form
\begin{equation}
  G(\{0,a_1,a_2,\ldots,z,\ldots,a_n\}_w;z)
  \label{eqn:zArg}
\end{equation}
be rewritten as a linear combination of objects which do not contain the
argument $z$ in their labels any more and can thus be integrated using
\eqn{eqn:defG}? Naturally, the desired objects have to be of the same weight as
the original polylogarithms \eqn{eqn:zArg}. A canonical ansatz is of the form
\begin{align}
  G(\{0,&a_1,a_2,\ldots,z,\ldots,a_n\}_w;z)\eq \sum c_i\,G(\{0,a_1,a_2,\ldots,a_n\}_w;z)_i\notag \\
  &\,+\,\z_2\sum c_j\,G(\{0,a_1,a_2,\ldots,a_n\}_{w-2};z)_j \,+\,\z_3\sum c_k\,G(\{0,a_1,a_2,\ldots,a_n\}_{w-3};z)_k\,+ \notag \\
  &\,\vdots\,\notag \\
  &\,+\,\z_5\sum c_l\,G(\{0,a_1,a_2,\ldots,a_n\}_{w-5};z)_l\,+ \,\z_2\z_3\sum c_m\,G(\{0,a_1,a_2,\ldots,a_n\}_{w-5};z)_m\, + \notag \\
  &\,\vdots\,\notag \\
  &\,+\,c_{max} \zeta_w
  \label{05,06b}
\end{align}
with rational coefficients $c$. While the first line consists of the obvious
sum, the other lines contain terms in which the weight is partially carried by
MZVs. For weights $w\geq 5$ one has to consider all basis elements at
weight $w$ spelled out in table \ref{zetaBasis}, and the last term $c_{max}
\zeta_w$ is a shorthand for having one coefficient $c_{max} $ for each $\ZQ$
independent MZV product at weight $w$, see appendix \ref{sec:zeta1}. The
remaining sums run over a basis in the space of polylogarithms of the form
$G(\{0,a_1,a_2,\ldots,a_n\}_{w-k};z)$ with $0 \leq k \leq w$.

The canonical way to derive those relations, that is, to find all coefficients
$c$ in \eqn{05,06b}, is thoroughly described in \cite{DuhrHopf}: one uses the fact that the
multiple polylogarithms constitute a Hopf algebra endowed with a coproduct,
which iteratively provides - colloquially spoken - the opposite operation of
the shuffle product \eqn{eqn:shuffle}. Demanding agreement of the coproduct at
every iteration of its calculation on both sides of \eqn{05,06b} allows to fix
all coefficients but the one(s) in the last line. Considering the ansatz
numerically at a special value determines those coefficients as well. The
so-called symbol \cite{GoncharovOrig,Goncharov:2010jf}, which has been proven a
valuable tool to derive identities between (poly)logarithms is equivalent to
the maximally iterated coproduct and thus a special case of the method used
here.

While we initially used the method described in \cite{DuhrHopf}, it turned out
that the identities produced in this way can be cast into a closed formula. For
simplicity we will write the equality for the case of only one occurrence of
the argument $z$ in the label $\vec{a}$:
\begin{subequations}
  \label{eqn:GIdentity}
  \begin{align}
  G(a_1,
  \ldots,a_{i-1},z,a_{i+1},\ldots,a_n;z)&=G(a_{i-1},a_1,\ldots,a_{i-1},\hat{z},a_{i+1},\ldots,a_n;z)\label{line1}\\
  &\qquad-G(a_{i+1},a_1,\ldots,a_{i-1},\hat{z},a_{i+1},\ldots,a_n;z)\label{line2}\\
  &\qquad-\int_0^z\frac{\dd t}{t-a_{i-1}}G(a_1,\ldots,\hat{a}_{i-1},t,a_{i+1},\ldots,a_n;t)\label{line3}\\
  &\qquad+\int_0^z\frac{\dd t}{t-a_{i+1}}G(a_1,\ldots,a_{i-1},t,\hat{a}_{i+1},\ldots,a_n;t)\label{line4}\\
  &\qquad+\int_0^z\frac{\dd t}{t-a_1}G(a_2,\ldots,a_{i-1},t,a_{i+1},\ldots,a_n;t)\label{line5}\,.
  \end{align}
\end{subequations}

In the above equation a hat denotes omission of the corresponding element\footnote{For illustration, here is an example employing the above formula:
  \begin{align}
    G(0,z,1;z)&=G(0,0,1;z)-G(1,0,1;z)-\int_0^z\frac{\dd t}{t-0}G(t,1;t)+\int_0^z\frac{\dd t}{t-1}\underbrace{G(0,t;t)}_{-\z_2}+\int_0^z\frac{\dd t}{t-0}G(t,1;t)\nnl
    &=G(0,0,1;z)-G(1,0,1;z)-\z_2G(1;z)\nonumber
  \end{align}}.
Thus, one can rewrite the $G(a_1,\ldots,a_{i-1},z,a_{i+1},\ldots,a_n;z)$ in
terms of a sum of polylogarithms which are simpler in the following
sense:\\[-22pt]
 \begin{itemize}
   \item terms (\ref{line1}) and (\ref{line2}) do not contain the argument in their labels any more
   \item terms (\ref{line3}), (\ref{line4}) and (\ref{line5}) still contain the argument $z$ in their labels, but their weight is reduced by one compared to the initial polylogarithm.
 \end{itemize} Applying \eqn{eqn:GIdentity} recursively allows to completely remove the argument $z$ in the labels of the multiple polylogarithms on the right hand side. A couple of remarks are in order here:
\begin{itemize}
  \item As mentioned above, \eqn{eqn:GIdentity} deals with the situation, where $z$ appears only once in the label of the polylogarithm. If there are several $z$'s in $\vec{a}$, one has to write down the first four terms on the right hand side for each of the occurrences of $z$. The cancellations between neighboring terms ensure that the reduction still leads to an expression where the labels of the polylogarithms on the right hand side are independent of $z$ or shorter. 
  \item If the argument $z$ occurs at the last position in the label, that is $a_n=z$, the term (\ref{line4}) in \eqn{eqn:GIdentity} has to be dropped. In addition, the term (\ref{line2}) has to be modified to read
    \begin{equation}
     -G(0,a_1,\ldots,a_{i-1},\hat{z};z)\,.
      \label{}
    \end{equation} 
  \item The opposite situation, where the argument $z$ resides at first position in the label of the polylogarithm would require special attention as well. For simplicity we will assume here, that those polylogarithms will be dealt with applying the shuffle regulation rule \eqn{shufflereg1}.
  \item While $a_n=0$ does not require special attention in \eqn{eqn:GIdentity}, it is nevertheless convenient to apply \eqn{shufflereg2} first in order to keep the number of different identities small. Especially for higher weights this shortens the calculation significantly. 
\end{itemize}
Another property of the above identity is that it preserves the shuffle
regulation: If the initial polylogarithm to be expanded is shuffle-regulated,
so will be all expressions on the right hand side. 

The comments above will become clear in proving the identity: let us write
formula (\ref{eqn:GIdentity}) as an integral of its derivative: 
\begin{equation}
G(a_1,
  \ldots,a_{i-1},z,a_{i+1},\ldots,a_n;z) \eq
 \int_0^z \dd t\ \frac{\dd}{\dd t}G(a_1,\ldots,a_{i-1},t,a_{i+1},\ldots,a_n;t)\,.  \label{eqn:proof}
\end{equation}
The total $t$ derivative in (\ref{eqn:proof}) splits into partial derivatives
acting on the $t$ among the labels
\begin{eqnarray}
  \frac{\pd}{\pd a_i}G(\vec{a};z)&=&\frac{1}{a_{i-1}-a_i}G(\ldots,\hat{a}_{i-1},\ldots;z) \ + \ \frac{1}{a_i-a_{i+1}}G(\ldots,\hat{a}_{i+1},\ldots;z)\nnl
  &-&\frac{a_{i-1}-a_{i+1}}{(a_{i-1}-a_i)(a_i-a_{i+1})}G(\ldots,\hat{a}_i,\ldots;z)
  \label{eqn:derivativeG}
\end{eqnarray}
and the argument
\begin{equation}
  \frac{\pd}{\pd z}G(\vec{a};z)=\frac{1}{z-a_1}G(a_2,\ldots,a_n;z).
  \label{eqn:derivativeG1}
\end{equation}
Applying partial-fraction identities leads straightforwardly to \eqn{eqn:GIdentity}. 

If there are several occurrences of $z$ in the label of the polylogarithm,
several partial derivatives (\ref{eqn:derivativeG}) contribute to
(\ref{eqn:proof}) which explains the first point in the remarks above. If the
argument appears in the last position, the modified derivative rule reads
\begin{equation}
  \frac{\pd}{\pd a_n}G(\vec{a};z)=\frac{1}{a_{n-1}-a_n}G(\ldots,\hat{a}_{n-1},a_n;z)
  -\frac{a_{n-1}}{(a_{n-1}-a_n)a_n}G(\ldots,a_{n-1};z)
   \label{eqn:derivativeG2}
\end{equation}
which explains the necessity to drop the term (\ref{line4}) in
\eqn{eqn:GIdentity}. Finally, comparing the last term in \eqn{eqn:derivativeG2}
with the last term in \eqn{eqn:derivativeG1} shows that the term (\ref{line2})
in \eqn{eqn:GIdentity} should be modified as mentioned in the remarks
above.\hfill$\Box$

Applying \eqn{eqn:GIdentity} recursively one finds identities like
\begin{align}
 G(a_1,0,z;z) &= G(0,0,a_1;z)-G(0,a_1,a_1;z)-G(a_1;z)\z_2\notag \\
 G(a_1,z,a_2;z) &= G(a_1,a_1,a_2;z)-G(a_2,0,a_1;z)+G(a_2,a_1,a_1;z)-G(a_2,a_1,a_2;z)
\end{align}
where in the above formul\ae\,\,$a_1,a_2 \neq 0$. In addition, a collection of 
identities for weights two and three can be found in
appendix \ref{polyexpl}.  With the methods described above one can now solve
all regular integrals $I^\reg$. Let us now combine all previous information and
finally obtain the results for the functions $F^\si$ in the next section.

\section{Assembling the basis}
\label{sec3}

The techniques of sections \ref{secKinPoles} and \ref{secPolylogs} provide
access to the $\ap$-expansion of any integral whose rational function takes the
form (\ref{constraint}) provided that its field-theory limit does not involve
dual pole channels, that is, more than one cubic YM diagram. In some cases,
these two constraints are incompatible. For instance the seven-point world-sheet integrals
from the $\{2,3\}$-block -- see section \ref{sec:block} -- cannot be brought
into a basis of functions of the form (\ref{constraint}) with one field-theory
diagram each\footnote{A factor of $(z_{45} z_{46})^{-1}$ in the integrand
violates the criterion (\ref{constraint}) imposed by our polylogarithm
integration techniques whereas $(z_{45} z_{56})^{-1}$ introduces incompatible
poles in $s_{45}^{-1}$ and $s_{56}^{-1}$. 

To give a related example for the limitation of our method: It is not possible
to directly construct the six-point basis from the $\nu=4$ sector (\ref{02,76})
of the pole channel basis: On the one hand, the integrand $\sim
(z_{23}z_{24}z_{14})^{-1}$ of the $Z[s_{23}s_{234}s_{1234}]$ function violates
the criterion (\ref{constraint}). On the other hand, partial-fraction
elimination of the obstruction $(z_{23}z_{24})^{-1}$ introduces a rational
factor $(z_{23}z_{34})^{-1}$ which triggers singularities in incompatible pole
channels $s_{23}$ and $s_{34}$.}. We will now introduce methods to bypass this
subtlety.

In bypassing this subtlety, we pay particular attention to increase the
computational efficiency. As we will see, the number of disk integrals whose
$\ap$-expansions need to be performed independently can be pushed well below the number $(N-3)!$ of basis
functions $F^\si$, more precisely to $1,3$ and $11$ at multiplicities $N=5,6$
and $N=7$, respectively. We will present the basis construction for these
multiplicities and investigate possible shortcuts from exploiting world-sheet
parity and cyclic transformations along the lines of \cite{Stieberger:2006te}.

The action of world-sheet parity is implemented by $z_{i,j} \mapsto z_{N-i,N-j},
\ k_i \mapsto k_{N-i}$ on the integrand, accompanied by reflection of the
integration domain. Let $\si^t$ denote the image of the permutation $\si$ under
reversal of its elements, then one finds
\beq
F^{i_1 i_2 \ldots i_{N-3}} \, \Big|_{i \mapsto N-i} \eq F^{N-i_{N-3},N-i_{N-4},\ldots,N-i_{1}} \ .
\label{06,0a}
\eeq
Cyclic shifts have a more subtle action on the $F^\si$. As in section
\ref{secap}, $\BA_\YM$ denotes the vector of BCJ-independent subamplitudes, and
${\bf F}$ is the corresponding $(N-3)!$ component vector of  functions $F^\si$ entering the canonically ordered disk amplitude
(that is, the first row of the matrix $F$ defined in \eqn{00,1a}). Then, the
shift $i \mapsto i \pm 1$ (subject to cyclic identifications $i := i+N$)
can be implemented by $(N-3)! \times (N-3)!$ matrices
\beq
\BA_\YM \, \Big|_{i \mapsto i\pm 1} \ \ = \ \ \CU_{\pm} \BA_\YM
\eeq
Their entries are quotients of Mandelstam invariants determined by the BCJ
relations\footnote{The exact form of the matrices $\CU_{\pm}$ follows by
expanding the images $A_{\te{YM}}(1,2,\si(3,\ldots,N-1),N)$ and
$A_{\te{YM}}(\si(1,\ldots,N-3),N-2,N-1,N)$ under $i \mapsto i+1$ and $i \mapsto
i-1$, respectively, in the original basis of
$A_{\te{YM}}(1,\si(2,\ldots,N-2),N-1,N)$ used in (\ref{00,1}).} \cite{BCJ}.
Since the composition of opposite shifts $(i \mapsto i\pm 1) (i \mapsto i\mp
1)$ leaves $\BA_\YM$ invariant, the matrices $\CU_{\pm}$ satisfy
\beq
\CU_{\pm}^{-1} \eq \CU_{\mp} \, \Big|_{i \mapsto i\pm 1} \ .
\label{06,0b}
\eeq
Cyclic invariance of the open-string subamplitude $A_\open(1,2,\ldots,N)$
follows from total symmetry of the CFT correlator\footnote{We emphasize again
that before specifying the integration domain $D(\Pi)$ the open-string
amplitude (\ref{01,7}) is totally symmetric in its labels for the same reasons
as the gravity amplitude (\ref{01,4}), see subsection \ref{sec1,12}.} together
with cyclic invariance of the integration domain defined by $z_i<z_{i+1}$.
Hence, the cyclic $i \mapsto i\pm1$ action on the functions $F^\si$ must be
linear
\beq
{\bf F} \, \Big|_{i \mapsto i\pm 1} \eq  {\cal V}_{\pm} \, {\bf F}
\eeq
with $(N-3)! \times (N-3)!$ matrices $\CV_{\pm}$ determined by
\beq
\BA_\YM^t \, {\bf F} \eq \BA_\YM^t \, \CU_{\pm}^t \, \CV_{\pm} \, {\bf F} \ .
\label{06,0c}
\eeq
Together with (\ref{06,0b}), we obtain:
\beq
\CV_{\pm} \eq \CU_{\mp}^{t} \, \Big|_{i \mapsto i\pm 1}\,.
\label{06,0d}
\eeq
At five points, for instance, the $2\times 2$ matrices $\CU_+,\CV_+$
implementing a cyclic $i\mapsto i+1$ shift on $A_\YM(1,\si(2,3),4,5)$ and
$F^{\si(23)}$ are given by
\beq
\CU_+ \eq  \ccb 1 &0 \\ -1-s_{34}/s_{35} &s_{13}/s_{35} \cce   \co
\CV_+ \eq \ccb  1 &-1-s_{23}/s_{13} \\ 0 &s_{35}/s_{13} \cce  \ .
\label{06,0e}
\eeq
They can be checked to obey $\CU_+ \CV^t_+ =1$ using momentum conservation.

In order to finally assemble the functions $F^\si$, one has to perform the following steps
\begin{itemize}
  \item express the functions $F^\si$ in terms of $Z(1,2_\si,\ldots,\nu_\si,N,(\nu+1)_\si,\ldots,(N-2)_\si,N-1)$ choosing the most convenient representation $\nu=\lfloor N/2\rfloor$ as in \eqn{01,25}. 
  \item convert the functions $Z(\ldots)$ into the pole basis $Z[\ldots]$ using the partial-fraction manipulations described in subsection \ref{sec1,22}
  \item decompose the elements of the pole basis $Z[\ldots]$ into their residues as in section \ref{sec1,23} and rewrite the residues in terms of lower-point regular integrals as in \eqn{02,204}.
  \item evaluate all regular integrals using the techniques from section \ref{secPolylogs}.
\end{itemize}
In order to keep the dependence of the regular integrals on squares $k_{A_i}^2$ of off-shell momenta explicit, the arguments are given in terms of Mandelstam invariants (see
(\ref{ktos23}) for our conventions for regular six-point integrals). In
addition to the direct construction of $F^\si$ following the enumeration above,
we demonstrate how parity and cyclic invariance of the open-string amplitude
allows to bypass the expansion of some basis functions.

\subsection{Five-point}

Given the closed form expression (\ref{02,212}) for the four-point basis
function $F^{(2)}$, we shall start our series of examples by assembling the
five-point basis integrals $F^{\si(23)}$. The $\ap$-expansion of their $\nu=2$
representation (\ref{01,26}) follows from the pole structures of the underlying
disk integrals $Z(1,\si(2),5,\si(3),4)$ given in (\ref{02,21}) and
(\ref{02,22}),
\begin{align}
F^{(23)} \eq &s_{12} s_{34} \,Z[s_{12}s_{34}] 
\notag \\
\eq &1 \ + \ s_{12} \, I_{21}^\reg[s_{12},s_{23}+s_{24}] \ + \ s_{34} \, I_{21}^\reg[s_{34},s_{13}+s_{23}] \ + \ s_{12}s_{34} \, I_{21,43}^\reg \notag  \\ 
F^{(32)} \eq &s_{13} s_{24} \,Z[\zeta_2]  \eq  s_{13} s_{24} \, I_{31,42}^\reg \ . \label{06,01}
\end{align}  
It is instructive to compare with the equivalent $\nu=3$ representation (\ref{01,27})
\begin{align}
F^{(23)} \eq &1 \ + \ s_{34} \, I_{21}^\reg[s_{34},s_{13}+s_{23}] \ + \ s_{12} \, \big( \, I_{21}^\reg[s_{123},s_{24}+s_{34}] \ + \  I_{21}^\reg[s_{12},s_{23}] \, \big) \notag \\
& \ \ + \ s_{12} \, \big( (s_{13}+s_{23}) \, I_{21,31}^\reg \ + \ s_{23} \, I_{31,32}^\reg \, \big) \label{06,02}
\\
F^{(32)} \eq &s_{13} \, \big( \, I_{21}^\reg[s_{13}+s_{23},s_{34}] \ - \ I_{21}^\reg[s_{123},s_{24}+s_{34}] \ + \ s_{12} \, I_{21,31}^\reg \ - \ s_{23} \, I_{31,32}^\reg \, \big) \notag
\end{align}
following from (\ref{02,25}) and (\ref{02,26}) with $s_{12}
I_{21}^\reg[s_{12},s_{23}] = s_{23} I^\reg_{21}[s_{23},s_{12}]$. Due to the
larger number of four-point contact vertices, \eqn{06,02} is less appealing
for practical purposes. The two representations (\ref{06,01}) and (\ref{06,02})
are equal if the $I_{\ldots}^\reg$ satisfy
\begin{align}
I_{21}^\reg[s_{12},s_{23}+s_{24}] \ + \ s_{34} \, I_{21,43}^\reg \eq &
I_{21}^\reg[s_{123},s_{24}+s_{34}] \ + \  I_{21}^\reg[s_{12},s_{23}] \notag \\
& \ \   + \ (s_{13}+s_{23}) \, I_{21,31}^\reg \ + \ s_{23} \, I_{31,32}^\reg  \label{06,03}
\end{align}
as well as
\beq
s_{24} \, I_{31,42}^\reg \eq s_{12} \, I_{21,31}^\reg \ - \ s_{23} \, I_{31,32}^\reg \ + \ I_{21}^\reg[s_{13}+s_{23},s_{34}] \ - \ I_{21}^\reg[s_{123},s_{24}+s_{34}]  \ .
\label{06,04}
\eeq
Relations among the $I_{\ldots}^\reg$ which follow from comparing different
$F^{\si}$-representations (\ref{01,24}) mix regular parts at different
multiplicity. In other words, integration by parts identities (\ref{01,18}) and
(\ref{01,19}) only apply to the full integrals including all their singular
terms. The naive attempt to lift the vanishing of integrals over $\partial_2
\prod_{i<j}^4 |z_{ij}|^{s_{ij}} /z_{31}$ to regular parts $s_{12}
I_{21,31}^\reg - s_{23}  I_{31,32}^\reg-s_{24} I_{31,42}^\reg$ fails because of
the admixture of $I_{21}^\reg$ in (\ref{06,04}).

\subsubsection{The cyclicity shortcut}

Cyclic properties of the functions $F^{(23)}$ and $F^{(32)}$ allow to bypass
the evaluation of $I_{31,42}^\reg$ in \eqn{06,01}. Using the matrix $\CV_+$
given in (\ref{06,0e}) and encoding the cyclic action on $F^{\si(23)}$, one can
infer the second basis function from
\beq
F^{(32)} \eq \frac{ s_{13} }{s_{13}+s_{23}} \; \Big( \, F^{(23)} \ - \ F^{(23)} \, \Big|_{i \mapsto i+1} \, \Big) \ .
\label{virtue}
\eeq
Since $I_{21}^\reg$ and $I_{32}^\reg$ are available in closed form
(\ref{02,213}) and (\ref{02,215}), only $I_{21,43}^\reg$ is left to expand through the methods of section \ref{secPolylogs}, i.e. half of the effort at five points is saved by virtue of (\ref{virtue}).

\subsection{Six-point}

The six-point basis integrals $F^\si$ can be obtained from the $\nu=3$ sector
of the pole basis. The singularity structure (\ref{02,27}) and (\ref{02,28}) of
$Z[s_{12}s_{123}s_{45}]$ and $Z[s_{23}s_{123}s_{45}]$ yields the following
expressions:
\begin{align}
F^{(234)} \eq& s_{12}(s_{13}+s_{23}) s_{45}\, Z[s_{12}s_{123}s_{45}]  \ + \ s_{12} s_{23} s_{45} \, Z[s_{23}s_{123}s_{45}] 
\notag \\ 
\eq &(1 \ +\  s_{12} \, I^\reg_{21}[s_{12}, s_{23}]) \, (1 \ + \ s_{45}\, I^\reg_{21}[s_{45}, s_{14} + s_{24} + s_{34}] )\notag \\
& \ \ + \ (s_{13} + 
    s_{23})\, I^\reg_{21}[s_{13} + s_{23}, s_{34} + s_{35}]  \ + \ 
 s_{12}  \, I^\reg_{21}[ s_{123}, s_{24} + s_{25} + s_{34} + s_{35}] \notag \\
 & \  \ + \ s_{12} (s_{13} + 
    s_{23}) \, I_{21,31}^\reg [s_{12},  s_{13},  s_{23},  s_{24} + s_{25},   s_{34} + s_{35}] \notag \\
    & \ \ + \ s_{45} (s_{13} + 
    s_{23}) \, I^\reg_{21,43} [s_{13} + s_{23},  s_{14} + s_{24},  s_{34}, s_{35},  s_{45}]  \notag \\
    & \ \ + \ 
 s_{12} s_{23} \, I^\reg_{31,32}[s_{12},s_{13}, s_{23},  s_{24} + s_{25}, s_{34} + s_{35}]  \notag \\
 & \ \ + \ 
 s_{12} s_{45} \, I^\reg_{21,43} [s_{123}, s_{14},  s_{24} + s_{34},  s_{25} + s_{35}, s_{45}] \notag \\
 & \ \ + \  s_{12}(s_{13}+s_{23}) s_{45}\, I^\reg_{21,31,54} \ + \ s_{12} s_{23} s_{45} \, I^\reg_{31,32,54}
 \label{06,05}  \\
F^{(324)} \eq& s_{12} s_{13} s_{45} \, Z[s_{12}s_{123}s_{45}]   \ - \ s_{13} s_{23} s_{45} \, Z[s_{23}s_{123}s_{45}]  \notag  \\
\eq & s_{13} \, \Big( \, I^\reg_{21} [s_{13} + s_{23}, s_{34} + s_{35}]  \ - \ I^\reg_{21} [s_{123},  s_{24} + s_{25} + s_{34} + s_{35}] \notag \\
& \ \ +  \
 s_{12} \, I^\reg_{21,31} [s_{12}, s_{13}, s_{23},  s_{24} + s_{25}, s_{34} + s_{35}] \ + \ 
 s_{45}  \,I^\reg_{21,43} [s_{13} + s_{23}, s_{14} + s_{24},  s_{34}, 
  s_{35}, s_{45}] \notag \\
  & \ \ - \ s_{23} \, I^\reg_{31,32}[ s_{12}, s_{13}, s_{23}, s_{24} + s_{25},  s_{34} + s_{35}] \ - \ 
  s_{45} \,I^\reg_{21,43} [s_{123},s_{14},  s_{24} + s_{34},s_{25} + s_{35}, s_{45}]  \notag \\
  & \ \ + \ s_{12}  s_{45} \, I^\reg_{21,31,54}  \ - \ s_{23} s_{45} \, I^\reg_{31,32,54} \, \Big)
  \label{06,06}
\end{align}
We have used $s_{12} I_{21}^\reg[s_{12},s_{23}] = s_{23}
I^\reg_{21}[s_{23},s_{12}]$ in order to cast these two basis functions into a
manifestly local form. The remaining four functions are built from
significantly less $I^\reg_{\ldots}$ since the underlying $Z[\ldots]$ have at
most single poles as spelled out in \eqns{02,29}{02,30}:
\begin{align}
F^{(243)} \eq& s_{12}(s_{14}+s_{24}) s_{35} \, Z[\zeta_2 s_{12}]  \ + \ s_{12} s_{24} s_{35}  \, \tilde Z[\zeta_3]\notag \\
\eq& s_{12}(s_{14}+s_{24}) s_{35} \, I^\reg_{21,41,53} \ + \ s_{12} s_{24} s_{35}  \, I^\reg_{41,42,53} \notag \\
& \ \ + \ (s_{14}+s_{24}) s_{35} \, I_{31,42}^\reg[s_{13}+s_{23},s_{14}+s_{24},s_{34},s_{35},s_{45}]
\label{06,07} \\
F^{(423)} \eq& s_{12} s_{14}s_{35} \, Z[\zeta_2 s_{12}]\ - \ s_{14}s_{24}s_{35} \, \tilde Z[\zeta_3] \notag \\
\eq & s_{12} s_{14}s_{35} \, I^\reg_{21,41,53} \ - \ s_{14}s_{24}s_{35} \, I^\reg_{41,42,53} \notag \\
& \ \ + \ s_{14}s_{35} \, I_{31,42}^\reg[s_{13}+s_{23},s_{14}+s_{24},s_{34},s_{35},s_{45}] 
\label{06,08}\\
F^{(342)} \eq &s_{13}(s_{14}+s_{34}) s_{25} \, Z[\zeta_3] \ + \ s_{13} s_{34} s_{25} \,  Z[\zeta_2 s_{34}] \notag \\
\eq & s_{13}(s_{14}+s_{34}) s_{25} \, I^\reg_{31,41,52} \ + \ s_{13} s_{34} s_{25} \, I^\reg_{41,43,52}  \notag \\
& \ \ + \ s_{13}s_{25} \, I^\reg_{31,42}[s_{12},s_{134},s_{23}+s_{24},s_{25},s_{35}+s_{45}]
\label{06,09} \\
F^{(432)} \eq &s_{13} s_{14}s_{25} \,  Z[\zeta_3] \ - \ s_{14}s_{34} s_{25} \,  Z[\zeta_2 s_{34}] \notag \\
\eq& s_{13} s_{14}s_{25} \, I^\reg_{31,41,52} \ - \ s_{14}s_{34} s_{25} \, I^\reg_{41,43,52} \notag \\
& \ \ - \ s_{14} s_{25} \, I^\reg_{31,42}[s_{12},s_{134},s_{23}+s_{24},s_{25},s_{35}+s_{45}]
\label{06,10}
\end{align}
This pedestrian approach to the six-point $F^\si$ basis requires the expansion
of six integrals. Let us now demonstrate how parity and cyclicity reduce this
number.

\subsubsection{The parity shortcut}

As a first method to improve the efficiency of the basis construction,
we employ the parity transformation of the functions $F^\si$ in \eqn{06,0a}.
Under parity, one finds two singlets $F^{(234)}$ and $F^{(432)}$ and two
doublets
\beq
F^{(243)}  \eq  F^{(324)} \, \Big|_{i \mapsto 6-i} \co F^{(423)} \eq F^{(342)} \, \Big|_{i \mapsto 6-i} \ .
\label{06,11}
\eeq
This allows to bypass the evaluation of $I^\reg_{21,41,53} =Z[s_{12} \zeta_2]$
and $I^\reg_{41,42,53}=\tilde Z[\zeta_3]$. Parity has reduced the required
number of $I^\reg_{ij,kl,mn}$-expansions by one third, and we will next show
that cyclicity leads to a further reduction. 

\subsubsection{The cyclic shortcut}
Each of the pole basis integrals $Z[\ldots]$ contains partial information on the $F^\si$ basis such as
\begin{align}
Z[s_{12}s_{123}s_{45}] \eq &\frac{1}{s_{123}s_{45}} \, \Big( \, \frac{ F^{(234)} }{ s_{12} } \ + \ \frac{ F^{(324)} }{s_{13} } \, \Big)
\label{06,12} \\
Z[\zeta_2 s_{12}] \eq &
\frac{1}{s_{124}s_{35}} \, \Big( \, \frac{ F^{(243)} }{ s_{12} } \ + \ \frac{ F^{(423)} }{s_{14} } \, \Big)
\label{06,13} \\
Z[\zeta_3] \eq & \frac{1}{s_{134}s_{25}} \, \Big( \, \frac{ F^{(342)} }{ s_{13} } \ + \ \frac{ F^{(432)} }{s_{14} } \, \Big) \ ,
\label{06,14}
\end{align}
see (\ref{invmom2}). In order to obtain the missing information on the $F^\si$
beyond (\ref{06,12}) to (\ref{06,14}), we apply cyclic transformations to their
right hand sides. At six-point level, the shifts $i \mapsto i \pm 1$ act on the
$\BA_\YM$ and ${\bf F}$ through $6 \times 6$ matrices $\CU_{\pm}$ and
$\CV_{\pm}$. The entries of $\CU_{\pm}$ are again straightforwardly determined
by BCJ relations, and the $\CV_{\pm}$ then follow from (\ref{06,0d}). Instead
of displaying $\CV_+$ in matrix form, let us list the cyclic images of the
basis functions relevant for the $i \mapsto i+1$ shift of (\ref{06,14}):
\begin{align}
&F^{(342)}\, \Big|_{i \mapsto i+1} \eq - \, \frac{F^{(342)}
   \left(s_{134}+s_{24}\right)
   s_{36}}{s_{13}
  s_{134}} \ - \ \frac{F^{(432)} s_{24}s_{36}}{s_{14}
   s_{134}} \ + \ \frac{F^{(324)}
   s_{36}}{s_{13}}     \label{06,18}
   \\
 &F^{(432)}\, \Big|_{i \mapsto i+1} \eq - \, \frac{F^{(342)}
s_{36}
   \left(s_{13}-s_{46}\right)}{s_{13}
s_{134}}\ - \ \frac{F^{(432)} \left(s_{24}+s_{45}\right)
s_{36} }{s_{14}
  s_{134}} \ . \label{06,20}
\end{align}
We can then obtain the decomposition of $Z[\zeta_3] \, \big|_{i \mapsto i+1},
Z[\zeta_3] \, \big|_{i \mapsto i-1}$ and $Z[\zeta_3] \, \big|_{i \mapsto i-2}$
into a basis of $F^\si$ from (\ref{06,14}) together with $F^{\si(234)}$ transformations such as (\ref{06,18}) and
(\ref{06,20}). The $6 \times 6$ matrix of expansion coefficients for
$Z[s_{12}s_{123}s_{45}] , Z[\zeta_2 s_{12} ] , Z[\zeta_3]$ and cyclic images
thereof has an inverse with local entries: 
\begin{align}
&\vecb \! F^{(234)} \! \\ \!  F^{(243)}\!  \\ \! F^{(324)}\!  \\ \! F^{(342)}\!  \\ \!  F^{(423)}\!  \\ \! F^{(432)}\!  \vece \ = \ 
\left(
\begin{array}{ccccc}
\! s_{12} s_{123} s_{45} \!& -s_{12}^2 \left(s_{24}\! + \!s_{25}\right)
   &\! -s_{12} \left( s_{25} s_{134}\! + \!s_{24}(s_{245}\! + \!s_{134})\right) \!&
 \!  -s_{12} s_{24} s_{245} \!&\ldots
   \\
 0 &\! s_{12} \left(s_{14}\! + \!s_{24}\right) s_{35} \! & 0 & 0 &\ldots \\
 0 & s_{12} s_{13} \left(s_{24}\! + \!s_{25}\right) & s_{13}
 \left( s_{25} s_{134}\! + \!s_{24}(s_{245}\! + \!s_{134})\right)
  & s_{13} s_{24}
   s_{245} &\ldots
   \\
 0 & s_{12} s_{13} s_{25} & s_{13} s_{25} s_{134} & 0 &\ldots \\
 0 & s_{12} s_{14} s_{35} & 0 & 0 &\ldots\\
 0 & -s_{12} s_{14} s_{25} & 0 & 0 &\ldots
\end{array}
\right.
\notag \\
&  \ \ \ \ \  \left. \begin{array}{ccc} \ldots & s_{12} \left(s_{24}\! + \!s_{25}\right)
   \left(s_{23}\! + \!s_{24}\! + \!s_{25}\! + \!s_{35}\right) & s_{12} s_{24} \left(s_{24}\! + \!s_{25}\right)
   \\
   \ldots & s_{12} s_{24} s_{35} & s_{12}
   s_{24} s_{35} \\
 \ldots & -s_{13} \left(s_{24}\! + \!s_{25}\right)
   \left(s_{23}\! + \!s_{24}\! + \!s_{25}\! + \!s_{35}\right) & -s_{13} s_{24} \left(s_{24}\! + \!s_{25}\right)
   \\
   \ldots &
   -s_{13} s_{25} \left(s_{23}\! + \!s_{24}\! + \!s_{25}\! + \!s_{35}\right) & -s_{13} s_{24} s_{25}
   \\
   \ldots & -s_{14} s_{24} s_{35} & -s_{14} s_{24} s_{35} 
   \\
 \ldots  & s_{14} s_{25}
   \left(s_{23}\! + \!s_{24}\! + \!s_{25}\! + \!s_{35}\right) & s_{14} s_{24} s_{25}
 \end{array} \right) 
 \, \vecb Z[s_{12}s_{123}s_{45}] \\ Z[\zeta_2 s_{12}]\\ Z[\zeta_3] \\ Z[\zeta_3] \, \big|_{i \mapsto i+1} \\Z[\zeta_3] \, \big|_{i \mapsto i-1} \\Z[\zeta_3] \, \big|_{i \mapsto i-2}  \vece 
\end{align}
In this setting, the $\ap$-expansions of $Z[s_{12}s_{123}s_{45}],  \,Z[\zeta_2
s_{12}]$ and $Z[\zeta_3]$ obtained on the basis of (\ref{02,27}), (\ref{02,29})
and (\ref{02,31}), respectively, encode the low-energy behavior of the complete
six-point amplitude.

\subsection{Seven-point}
\label{7basis}

The pole expansions given in subsection \ref{7ptexpl} are already adapted to a
parity- and cyclicity-inspired construction of the seven-point basis for
$F^{\si}$. We start by computing the basis functions from the parity
independent $\{3,5\}, \, \{2,4\} ,\, \{2,5\}$ and $\{4,5\}$ blocks. The
$\{3,5\}$ block does not involve any poles and directly translates into $
I^\reg_{ij,kl,mn,pq}$ 
\beq
\vecb  \frac{F^{(3524)}}{s_{1 3} s_{4 6}}  \\ \frac{F^{(3542)}}{s_{13}  s_{2 6}}  \\  \frac{ F^{(5324)} }{s_{1 5} s_{4 6}} \\ \frac{F^{(5342)} }{s_{1 5} s_{2 6}} \vece 
\ = \ \ccccb (s_{1 5}+s_{35}) s_{2 4} & s_{3 5} s_{2 4} &(s_{1 5}+s_{35}) s_{2 6} & s_{3 5} s_{2 6} \\
-  (s_{1 5}+s_{35}) s_{2 4} &-  s_{3 5} s_{2 4}  &(s_{1 5}+s_{35}) (s_{24}+s_{4 6}) &s_{3 5} (s_{24}+s_{4 6}) \\
s_{1 3}  s_{2 4} &-   s_{3 5} s_{2 4} &s_{1 3} s_{2 6} &-  s_{3 5} s_{2 6} \\
  -  s_{1 3}  s_{2 4} & s_{3 5} s_{2 4} &s_{1 3}   (s_{2 4}+s_{46}) &- 
s_{3 5} (s_{2 4}+s_{46})
\cccce
\, \vecb I^\reg_{31,51,42,64} \\ I^\reg_{51,53,42,64} \\ I^\reg_{31,51,62,64} \\ I^\reg_{51,53,62,64} \vece 
\label{06,37a}
\eeq
As explained in \cite{Mafra:2011nw}, the $\ap$-expansions of
$F^{(3524)},F^{(3542)},F^{(5324)}$ and $F^{(5342)}$ start at order $\ap^4
\zeta_4$ thanks to the absence of poles and the properties of the
$I^\reg_{31,51,62,64}$.

All other blocks incorporate functions with two simultaneous poles. For the $\{2,4\}$ block of the $F^\si$-basis,
\begin{align}
 \vecb ( s_{1 2} s_{5 6})^{-1}\, F^{(2435)}  \\ (s_{12}  s_{3 6})^{-1} \, F^{(2453)}  \\ (s_{1 4} s_{5 6})^{-1} \, F^{(4235)}  \\ (s_{1 4} s_{3 6})^{-1} \, F^{(4253)}  \vece \eq &\ccccb (s_{1 4}+s_{24}) s_{3 5} &- s_{2 4} s_{3 5}  &(s_{1 4}+s_{24}) s_{3 6}  & -  s_{2 4} s_{3 6}
\\
- (s_{1 4}+s_{24}) s_{3 5} &s_{2 4} s_{3 5} &(s_{1 4}+s_{24}) (s_{35}+s_{5 6})  & - s_{2 4} (s_{35}+s_{5 6})
\\
  s_{1 2}  s_{3 5} & s_{2 4} s_{3 5} &s_{1 2} s_{3 6}  &s_{2 4} s_{3 6}
  \\
   - s_{1 2}  s_{3 5} &- s_{2 4} s_{3 5} &s_{1 2}   (s_{3 5}+s_{56}) &s_{2 4} (s_{3 5}+s_{56})
\cccce \notag
\\
& \ \ \ \ \ \times \
\prod_{i=2}^5 \int_0^{z_{i+1}} \dd z_i \  \prod_{i<j}^6 |z_{ij}|^{s_{ij}}  \, \vecb ( z_{21} z_{35} z_{41} z_{56} )^{-1} \\  ( z_{24} z_{35} z_{41} z_{56} )^{-1} \\  ( z_{21} z_{36} z_{41} z_{56} )^{-1} \\  ( z_{24} z_{36} z_{41} z_{56} )^{-1} \vece
\end{align}
the singularities for the integrals on the right hand side are given in (\ref{02,33}) to (\ref{02,36}). The Mandelstam variables in the $4 \times 4$
matrix then render the resulting expressions for the $F^\si$ manifestly local.
Similarly, locality of the following functions from the $\{2,5\}$ and $\{4,5\}$
block can be seen from interplay of the $4\times 4$ matrix entries with the
poles in (\ref{02,37}) to (\ref{02,40}) and (\ref{02,41}) to (\ref{02,44}):
\begin{align}
 \vecb ( s_{1 2} s_{4 6})^{-1}\, F^{(2534)}   \\ (s_{12}  s_{3 6})^{-1} \,  F^{(2543)}   \\ (s_{1 5} s_{4 6})^{-1} \, F^{(5234)}   \\ (s_{1 5} s_{3 6})^{-1} \, F^{(5243)}  \vece \eq &\ccccb (s_{1 5}+s_{25}) s_{3 4}
  &- s_{2 5} s_{3 4}  &(s_{1 5}+s_{25}) s_{3 6}  & -  s_{2 5} s_{3 6}
\\
- (s_{1 5}+s_{25}) s_{3 4} &s_{2 5} s_{3 4}  &(s_{1 5}+s_{25}) (s_{34}+s_{4 6})  & - s_{2 5} (s_{34}+s_{4 6})
\\
s_{1 2}  s_{3 4}&s_{2 5} s_{3 4} &s_{1 2} s_{3 6}  &s_{2 5} s_{3 6}
  \\
   - s_{1 2}  s_{3 4} &- s_{2 5} s_{3 4} &s_{1 2}   (s_{3 4}+s_{46}) &s_{2 5} (s_{3 4}+s_{46})\cccce \notag
\\
& \ \ \ \ \ \times \
\prod_{i=2}^5 \int_0^{z_{i+1}} \dd z_i \  \prod_{i<j}^6 |z_{ij}|^{s_{ij}}  \, \vecb ( z_{21} z_{51} z_{34} z_{46} )^{-1} \\  ( z_{25} z_{51} z_{34} z_{46} )^{-1} \\  ( z_{21} z_{51} z_{36} z_{46} )^{-1} \\  ( z_{25} z_{51} z_{36} z_{46} )^{-1} \vece
\label{something1}
\\
 \vecb (  s_{1 4} s_{3 6} )^{-1}\, F^{(4523)}  \\ (s_{14}  s_{2 6} )^{-1} \,  F^{(4532)}   \\ (s_{1 5} s_{3 6})^{-1} \, F^{(5423)}  \\ (s_{1 5} s_{2 6})^{-1} \, F^{(5432)}  \vece \eq &\ccccb (s_{1 5}+s_{45}) s_{2 3}
  &s_{4 5} s_{2 3}  &(s_{1 5}+s_{45}) s_{2 6}   & s_{4 5} s_{2 6}
\\
- (s_{1 5}+s_{45}) s_{2 3} &-s_{4 5} s_{2 3}  &(s_{1 5}+s_{45}) (s_{23}+s_{3 6}) & s_{4 5} (s_{23}+s_{3 6})
\\
s_{1 4}  s_{2 3} &- s_{4 5} s_{2 3} &s_{1 4} s_{2 6}  &-s_{4 5} s_{2 6}
  \\
   - s_{1 4}  s_{2 3} &s_{4 5} s_{2 3} &s_{1 4}   (s_{2 3}+s_{36}) &- s_{4 5} (s_{2 3}+s_{36})  \cccce \notag
\\
& \ \ \ \ \ \times \
\prod_{i=2}^5 \int_0^{z_{i+1}} \dd z_i \  \prod_{i<j}^6 |z_{ij}|^{s_{ij}}  \, \vecb ( z_{32} z_{63} z_{41} z_{51}  )^{-1} \\  ( z_{32} z_{63} z_{51} z_{54}  )^{-1} \\  ( z_{62} z_{63} z_{41} z_{51}  )^{-1} \\  ( z_{62} z_{63} z_{51} z_{54} )^{-1} \vece
\label{something2}
\end{align} 

\subsubsection{The parity shortcut}

Four out of six seven-point blocks of basis functions contain a doublet and two
fixed points under world-sheet parity. On the contrary, the functions from the
$\{3,4\}$ block transform into those of the $\{ 2,5\}$ block; one can therefore
infer
\begin{align}
F^{(3425)} &\eq F^{(2534)}  \, \Big|_{i \mapsto 7-i} \co F^{(3452)} \eq F^{(5234)}  \, \Big|_{i \mapsto 7-i}\label{06,53} \\
F^{(4325)} &\eq F^{(2543)}\, \Big|_{i \mapsto 7-i}  \co F^{(4352)} \eq F^{(5243)} \, \Big|_{i \mapsto 7-i}  \ .\label{06,54}
\end{align}
Within the $\{2,4\}, \{3,5\}$ and $\{4,5\}$ blocks, on the other hand, one can
take advantage of parity to bypass one out of the four integral expansions
each, e.g.
\begin{align}
\prod_{i=2}^5 \int_0^{z_{i+1}} \! \! \dd z_i \ \frac{ \prod_{i<j}^6 |z_{ij}|^{s_{ij}} }{ z_{35} z_{51} z_{26} z_{46} } &\ \,= \, \ \prod_{i=2}^5 \int_0^{z_{i+1}} \! \! \dd z_i \ \frac{ \prod_{i<j}^6 |z_{ij}|^{s_{ij}} }{ z_{31} z_{51} z_{26} z_{46} } \ - \ \left( \, \prod_{i=2}^5 \int_0^{z_{i+1}}\! \! \dd z_i \ \frac{ \prod_{i<j}^6 |z_{ij}|^{s_{ij}} }{ z_{31} z_{51} z_{24} z_{46} }  \, \right) \, \Big|_{ i \mapsto 7-i } \ .
\label{06par3} 
\end{align}
The analogous relations for the $\{ 2,4\}$ and $\{ 4,5\}$ blocks follow by
relabelling $(3,5) \leftrightarrow (2,4)$ and $3\leftrightarrow4$,
respectively. Note that the form (\ref{constraint}) of the integrand required
for applicability of the polylogarithm techniques of section \ref{secPolylogs}
is not invariant under parity, this explains the asymmetry in equation
(\ref{06par3}).
 

\subsubsection{The cyclicity shortcut}
\label{cyc7}

So far we have addressed the $\ap$-expansion of 20 basis functions, the results
are given in (\ref{06,37a}) to (\ref{something2}) together with their parity
images (\ref{06,53}) and (\ref{06,54}). As explained in the previous
subsection, parity can be used to obtain the $\{2,4\}, \{3,5\}$ and $\{4,5\}$
blocks from three independent integrals each. In addition, the cyclic transformation
$Z(1,5,3,7,2,4,6) = -Z(1,4,2,7,5,3,6) \big|_{i \mapsto i+4}$ allows to infer
another $\{3,5\}$ integral from the $\{2,4\}$ block:
\beq
\prod_{i=2}^5 \int_0^{z_{i+1}} \dd z_i \ \frac{ \prod_{i<j}^6 |z_{ij}|^{s_{ij}} }{ z_{51} z_{53} z_{42} z_{64} } \eq  \prod_{i=2}^5 \int_0^{z_{i+1}} \dd z_i \ \frac{ \prod_{i<j}^6 |z_{ij}|^{s_{ij}} }{ z_{41} z_{42} z_{53} z_{63} } \, \Big|_{ i \mapsto i+4 } \,. 
\label{06par5}
\eeq
Similarly, the $i \mapsto i+1$ shift of the disk integrals $Z(1,2,5,7,3,4,6)$ and $Z(1,2,5,7,4,3,6)$ from the $\{2,5\}$ block yields two elements in the $\{4,5\}$ block:
\begin{align}
\prod_{i=2}^5 \int_0^{z_{i+1}} \dd z_i \ \frac{ \prod_{i<j}^6 |z_{ij}|^{s_{ij}} }{ z_{54} z_{41} z_{63} z_{32} } \eq  \prod_{i=2}^5 \int_0^{z_{i+1}} \dd z_i \ \frac{ \prod_{i<j}^6 |z_{ij}|^{s_{ij}} }{ z_{52} z_{21} z_{64} z_{43} } \, \Big|_{ i \mapsto i+1 }
\label{06par17}
\\
\prod_{i=2}^5 \int_0^{z_{i+1}} \dd z_i \ \frac{ \prod_{i<j}^6 |z_{ij}|^{s_{ij}} }{ z_{54} z_{51} z_{63} z_{32} } \eq  \prod_{i=2}^5 \int_0^{z_{i+1}} \dd z_i \ \frac{ \prod_{i<j}^6 |z_{ij}|^{s_{ij}} }{ z_{52} z_{21} z_{63} z_{43} } \, \Big|_{ i \mapsto i+1 }\,.
\label{06par18}
\end{align} 
In summary, parity relations such as (\ref{06par3}) as well as cyclicity
relations (\ref{06par5}) to (\ref{06par18}) reduce 20 basis functions outside
the $\{2,3\}$ block to ten cyclicity- and parity-independent computations. Now,
the $\{2,3\}$ block with basis functions $F^{(2345)},F^{(2354)},F^{(3245)}$ and
$F^{(3254)}$ is left to determine, and we shall again make use of cyclicity
methods.

 \medskip
The cyclic transformation of the seven-point $F^\si$ can be understood along the
lines of the six-point integrals, see the discussion around (\ref{06,18}) and
(\ref{06,20}). We can infer $F^{(3245)}$ and $F^{(3254)}$ by solving the cyclic
transformation of two known functions $F^{(3425)}$ and $F^{(4235)}$ shown in (\ref{06,55}) and (\ref{06,56}).
These identities relate the unknowns $F^{(3245)}$ and $F^{(3254)}$ to the 20
basis functions from above. Knowledge of $F^{(3245)}$ by virtue of
(\ref{06,55}) allows to determine $F^{(2354)}$ via parity,
\beq
F^{(2354)} \eq F^{(3245)} \, \Big|_{i \mapsto 7-i} \ .
\label{06,57}
\eeq
The final basis function $F^{(2345)}$ with non-vanishing field-theory limit
cannot be obtained from parity and cyclicity (relations of that type would lead
to contradictions in the $\ap \rightarrow 0$ limit). Hence, we make use of the
basis expansion (\ref{invmom2}) and the pole decomposition (\ref{02,32}) for
the integral
\begin{align}
Z[s_{12}s_{123}s_{1234}s_{56}] \eq& \frac{1}{s_{1234} s_{56}} \sum_{\si \in S_3} \frac{ F^{(\si(234)5)} }{s_{1\si(2)} s_{1\si(23)} }
\label{06,57a}
\end{align}
and obtain the last basis function by solving (\ref{06,57a}) for $F^{(2345)}$.


\section{Conclusions}

This paper aims to deepen the conceptual and computational understanding of
superstring disk amplitudes and their $\ap$-corrections. On the conceptual
side, the world-sheet integrand for the open-string tree amplitude has been
presented in close analogy to field-theory trees in supergravity in
subsection \ref{KLTsec}. The underlying dictionary between YM subamplitudes and
disk integrals at various permutations is supported by world-sheet analogues of
KK and BCJ relations in subsection \ref{sec1,12}.

On the computational side, we provide the tools to calculate the
$\ap$-expansion  for any multiplicity and to any order in $\ap$ in principle.
The poles of world-sheet disk integrals are classified by the pole criterion
(\ref{pole}) and the recursive structure of their residues is accounted for by
the diagrammatic methods of subsection \ref{sec1,23}. The regular part carrying
the intrinsic $N$-point information on contact terms can be analytically
evaluated by the polylogarithm method introduced in section \ref{secPolylogs}.
The polylogarithm identity (\ref{05,06b}) renders any world-sheet integration
elementary.  The limitation is set by calculational power as the number of
terms to consider and the number of steps necessary to solve the world-sheet
integrals iteratively grows immensely at higher multiplicities and orders in
$\ap$. Nevertheless, the methods have allowed us to determine the
$\ap$-corrections to the open-string tree-level $N$-point amplitude up to the
order\footnote{Order $\ap^{22}$ will be made available soon.} $\ap^{21}$ (weight $w=21$) for $N=5$, up to $\ap^9$ (weight $w=9$) for
$N=6$ and $\ap^7$ (weight $w=7$) for $N=7$. For these cases we could prove the explicit form (\ref{00,13}) of the image of the motivic superstring amplitude.
We refer the reader to Ref. \cite{BSST} and the data at \cite{WWW} for more details.

These results provide further testing grounds for the algebraic structure of
the open-string amplitude explored in \cite{SSMZV}. Further mathematical
investigation towards expressing all tree-level amplitudes in open superstring
theory using the Drinfeld associator as started for $N=4$ in
\cite{Drummond:2013vz} will be left for another publication \cite{WIP}.

\section*{Acknowledgments}

We are grateful to Claude Duhr for sharing his expertise on multiple
polylogarithms. Furthermore we would like to thank Carlos Mafra for numerous discussions, in particular for pointing
out similarities in the pole treatment of one-loop and tree level integrals. In
addition, we would like to express our gratitude towards Herbert Gangl,
Johannes Henn, Oliver Schnetz and Don Zagier for helpful discussions. 
Moreover, we are very thankful to Jos Vermaseren for streamlining 
and improving our FORM code to be more efficient for obtaining the 
$\alpha'$-expansion of the five--point amplitude at the weights $w=17$ 
through $w=22$. A portion of the computations have been performed on the SGE cluster
of the Arnold~Sommerfeld~Center for Theoretical Physics at M\"unchen.

JB would like to thank Michael Green and the DAMTP for hospitality. OS is supported by Michael Green and the European Research Council Advanced Grant No. 247252. Moreover, OS is grateful to CERN and CQUeST for generous hospitality during preparation of this work and to Dimitri Polyakov for useful discussions.

\section*{Appendix}
\appendix


\section{Introduction to (motivic) multiple zeta values}
\label{sec:zeta}

This appendix gathers some background information on MZVs, their motivic
versions, their isomorphic images $f_2^k f_{i_1}\ldots f_{i_p}$ in
(\ref{00,13}) and the underlying Hopf algebra structure \cite{BrownDecomp,
SSMZV}. The choice of material is adapted to the needs of the present article.

\subsection{Multiple zeta values}
\label{sec:zeta1}

MZVs have been a very active field of research during the last years. While
there is a vast amount of articles available on the subject, we will here
briefly collect the necessary information. MZVs $\zeta_{n_1,\ldots,n_r}$ of
depth $r$ and weight $w=\sum_{i=1}^r n_i$ are defined in (\ref{00,4}). There
are numerous relations between different MZVs such as for example the
quasi-shuffle (or stuffle) relation 
\begin{equation}
  \z_m\,\z_n\eq\z_{m,n}+\z_{n,m}+\z_{m+n}.
  \label{qshuffle}
\end{equation}
The collection of all relations allows to define the commutative graded algebra
$\CZ$ as the span of all MZVs over the rational numbers $\ZQ$, where the
algebra is conjectured to be graded by the weights of the MZVs 
\begin{equation}
  \CZ=\bigoplus \CZ_w\,.
  \label{Zalgebra}
\end{equation}
The dimension $d_w$ of $\CZ_w$ has been conjectured to be $d_w=d_{w-2}+d_{w-3}$
where $d_0=1$, $d_1=0$ and $d_2=1$ \cite{Zagier23}. A possible choice of basis
elements at each weight $w$ is given in table \ref{zetaBasis}, cf. also 
\cite{Datamine}.

\begin{table}[h]
  \[
\begin{array}{|l|l|l|l|l|l|l|l|l|l|ll|ll|}
\noalign{\hrule}
 w &2 &3 & 4  & 5 & 6& 7 & 8 & 9 & 10 & 11 &&12&\nnl
\noalign{\hrule}
\CZ_w & \zeta_2  &\zeta_3  & \zeta_2^2 &\zeta_5 & \zeta_3^2 &\zeta_7  &\zeta_{3,5} &\zeta_9 &\zeta_{3,7} & \zeta_{3,3,5}&\zeta_2\  \zeta_3^3 & \zeta_{1,1,4,6}&\zeta_2\ \zeta_{3,7} \nnl
\ &\  &\  & \  &\zeta_2\ \zeta_3 & \zeta_2 ^3 &\zeta_2\ \zeta_5  &\zeta_3\ \zeta_5 &\zeta_3^3&\zeta_3\ \zeta_7 &\zeta_{3,5}\ \zeta_3&\zeta_2\ \zeta_9  
&\zeta_{3,9}&\zeta_2^2\ \zeta_{3,5}\nnl
\ &\  &\  & \  &\ &\ & \zeta_2^2\ \zeta_3  &\zeta_2\ \zeta_3 ^2&\zeta_2\ \zeta_7 &\zeta_5^2&\zeta_{11}& \zeta_2^2\ \zeta_7  &\zeta_3\ \zeta_9&
\zeta_2\ \zeta_5^2\nnl
\ &\  &\  & \  &\ &\ &\  &\zeta_2^4&\zeta_2^2\ \zeta_5 &\zeta_2\ \zeta_{3,5} &\zeta_3^2\ \zeta_5&\zeta_2^3\ \zeta_5   &\zeta_5\ \zeta_7&\zeta_2\ \zeta_3\ \zeta_7\nnl
\ &\  &\  & \  &\ &\ &\  &\ &\zeta_2^3\ \zeta_3 &\zeta_2\ \zeta_3\ \zeta_5  &   &\zeta_2^4\ \zeta_3   &\zeta_3^4&\zeta_2^2\ \zeta_3\ \zeta_5\nnl
\ &\  &\  & \  &\ &\ &\  &\ &\ &\zeta_2^2\ \zeta_3^2 & &  & &\zeta_2^3\ \zeta_3^2 \nnl
\ &\  &\  & \  &\ &\ &\  &\ &\ &\zeta_2^5 & &  & & \zeta_2^6 \nnl
\noalign{\hrule}
d_w &1 &1  &1 &2 &2 &3  &4 &5 &7 & 9& &12& \nnl
\noalign{\hrule}
\end{array}
\]
\caption{A possible choice for the basis elements of $\CZ_w$ for $2\leq w\leq 12$.}
\label{zetaBasis}
\end{table}

There is a profound difference between single $\z$-functions of even and odd
weight. All single zeta values of even weight can be expressed as rational
multiples of even powers of $\pi$ and are thus expressable as powers of $\z_2$,
e.g. $\z_2^2=\frac{5}{2}\z_4$. This immediately renders them transcendental
numbers. There is no analogue of this property for the remaining single zeta
values: there are no known relations of this type relating two single zeta
values of different odd weight. Another difficulty arising for $\z$-values of
odd weight is their conjectured transcendentality: the only fact which have
been proven are the irrationality of $\z_3$ and that there is an infinite
number of odd irrational $\z$'s \cite{Apery,BallRivoal}.

Once the entries $F_\Pi{}^\s$ are expanded in terms of basis MZVs shown in
table \ref{zetaBasis}, the $\ap$-expansion of the matrix $F$ in \eqn{00,1}
takes the following form
\begin{align}
  F=\,\,&1_{(N-3)!\times(N-3)!}+\z_2 P_2+\zeta_3 M_3+\z_2^2 P_4+\z_2\z_3 P_2M_3 + \z_5 M_5\nnl
  &+ \z_2^3 P_6 + \frac{1}{2} \z_3^2 M_3 M_3 + \z_7 M_7 + \z_2\z_5 P_2M_5 + \z_2^2\z_3 P_4 M_3\nnl
  &+\z_2^4 P_8 + \z_3\z_5 M_5 M_3 + \frac{1}{2} \zeta_2 \zeta_3^2 P_2 M_3 M_3 + \frac{1}{5}\z_{3,5}[M_5,M_3]\nnl
  &+\ldots+\left(9\z_2\z_9+\frac{6}{25}\z_2^2\z_7-\frac{4}{35}\z_2^3\z_5+\frac{1}{5}\z_{3,3,5}\right)[M_3,[M_5,M_3]]\nnl
  &+\ldots+\z_{3,5}\z_{3,7}\frac{208926}{894845}\ [M_3[M_3[M_7,M_5]]]+\ldots\;,
  \label{00,5}
\end{align}
see \cite{SSMZV} for the explicit results up to weight $w=16$. In (\ref{00,5}),
$P_w$ and $M_w$ denote $(N-3)!\,\times\,(N-3)!$-matrices of homogeneity degree
$w$ in Mandelstam variables (\ref{01,2}) and are defined in \eqn{MsPs}. The form of the
expansion in \eqn{00,5} is far from obvious. In particular the fact that any
product of single $\z$'s comes with the appropriate product of matrices and in
addition the appearance of MZVs as prefactors of commutators containing the
corresponding matrices has not been expected at all.

In spite of its beautiful structure, the expansion \eqn{00,5} has three 
drawbacks: the coefficients in front of higher order terms related to
commutators of matrices turn into unwieldy rational numbers soon, as can be
seen by the examples at weight 11 and 18 noted in the last two lines of
\eqn{00,5}. Furthermore, the form of \eqn{00,5} depends on the particular
choice of a basis for $\CZ_w$. Moreover, the graded algebra spanned by 
the matrices $M_w$ with the Lie bracket $[\ldots,\ldots]$ depends on $N$. E.g. for $N=5$ at weight $w=18$ the commutator structure becomes ambiguous \cite{BSST}.
So it would be preferable to find a language, in
which the rational coefficients disappear, which simultaneously does not
depend on a particular choice of basis and is not sensitive to the dependence of the underlying algebra on $N$.

\subsection{Hopf algebra structure of MZVs} 

The language providing the desired simplification is the graded noncommutative
Hopf algebra comodule $\CU$ composed from words 
\begin{equation}
f_{2i_1+1}\ldots f_{2i_r+1}\ f_2^k\ \ \ ,\ \ \ r,k\geq 0,i_1,\ldots,i_r\geq 1
  \label{UBasis}
\end{equation}
of weight $w=2 (i_1+\ldots+i_r)+r+2k$. The objects $f_{2i+1}$ alone constitute
a noncommutative Hopf algebra. Upon adding powers of $f_2$, which commute with
all $f_{2i+1}$, the resulting object is a Hopf algebra comodule
\cite{BrownTate}. It is not difficult to convince oneself that the bases of the
gradings $\CU_w$ do indeed have the same dimension as $\CZ_w$: writing down all
noncommutative words of the form in \eqn{UBasis} yields the correct number
$d_w$. 

The missing piece is the link between the $\ZQ$-algebra $\CZ$ of MZVs
\eqn{Zalgebra} and $\CU$. Before constructing a map, one first needs to promote
the algebra $\CZ$ to a Hopf algebra. A Hopf algebra is an algebra which is its
own coalgebra and encompasses - besides the usual product of the algebra - a
coproduct in a way that product and coproduct are compatible \cite{DuhrHopf}.
While there is an obvious candidate for the coproduct of MZVs
\cite{BrownTate}, it is however not easy to promote $\CZ$ to a Hopf algebra: a
consistent coproduct needs to be $\ZQ$-linear. This would not pose a problem at
all if one could prove, that all MZVs are transcendental. However, as the
transcendental nature of many MZVs is undetermined, one needs to a employ a
mathematical tool, which ensures correct treatment of this uncertainty:
\textit{motives}. An introduction to the theory of motivic zeta values in the
context of string theory amplitudes can be found in \cite{SSMZV}; the original
papers are \cite{BrownDecomp, BrownTate}. 

Lifting the regular $\z$'s to their motivic versions $\z^\fm$ allows to promote
the commutative graded $\ZQ$-algebra $\CZ$ to the commutative graded Hopf
algebra of motivic $\z$-values $\CH$, which is defined over a finite extension
of $\ZQ$. What remains is the construction of an isomorphism $\phi$ from the
commutative, but non-cocommutative Hopf algebra $\CH$ to the noncommutative,
but co-commutative algebra comodule $\CU$:
\begin{equation}
  \phi:\CH\rightarrow\CU\,.
  \label{Phimap}
\end{equation}
The construction of the map $\phi$ assigning a linear combination of
noncommutative words (\ref{UBasis}) to each motivic MZV is described in
\cite{BrownDecomp}, where a complete set of examples for $w \leq 16$ can be
found in \cite{SSMZV}. In order to fix the normalization of the map $\phi$, we
choose single $\z$'s to be directly translated into one-letter words
\begin{equation}
 \phi(\z_w^\fm)\eq f_w \co f_{2k} \ \ := \ \ \frac{ \zeta_{2k} }{(\zeta_2)^k} \, f_2^k  \,.
 \label{00,10}
\end{equation} 
The map $\phi$ preserves all the different relations between the MZVs, e.g.
(cf. \eqn{qshuffle}):
\begin{equation}
  \phi(\z^\fm_k\,\z^\fm_l)\eq\phi(\z^\fm_{k,l})+\phi(\z^\fm_{l,k})+\phi(\z^\fm_{k+l})\,.
  \label{nettesBeispiel}
\end{equation}
In addition one finds for example:
\begin{eqnarray}
  \phi(\z^\fm_{3,9}) &\eq& -6 f_5 f_7 - 15 f_7 f_5 - 27 f_9 f_3 \nnl
  \phi(\z^\fm_{3,3,5}) &\eq&-5 f_5f_3f_3 + \frac{4}{7}f_5f_2^3-\frac{6}{5}f_7f_2^2-45f_9f_2\,.
  \label{phiExamples}
\end{eqnarray}
Employing the second example simplifies the coefficient of $[M_3,[M_5,M_3]]$ in
\eqn{00,5} to $-f_5 f_3 f_3$.

Applying the map $\phi$ to $F^\fm$, the motivic version of the matrix $F$
defined in \eqn{00,1a}, yields 
\begin{eqnarray}
\phi (F^\fm) &=& (\   1  +  f_2 P_2  +  f_2^2 P_4   +   f_2^3 P_6  +  f_2^4 P_8  +   f_2^5 P_{10}  +   f_2^6 P_{12}  +\ldots  \  )\nnl
&&\times  (\  1  +  f_3\ M_3  +  f_5\ M_5  +  f_3^2\ M_3^2  +  f_7\ M_7  +  f_3 f_5\ M_5 M_3  +  f_5 f_3\ M_3 M_5 \nnl 
&&\qquad +  f_9\ M_9+  f_3^3\ M_3^3  +  f_5^2\ M_5^2  +  f_3 f_7\ M_7 M_3  +  f_7 f_3\ M_3 M_7+  f_{11}\ M_{11}   \nnl
&&\qquad  +  f_3^2 f_5\ M_5 M_3^2  +  f_3 f_5 f_3\ M_3 M_5 M_3  +  f_5 f_3^2\ M_3^2 M_5 +  f_3^4\ M_3^4  +  f_3 f_9\ M_9 M_3    \nnl
&&\qquad  +  f_9 f_3\ M_3 M_9  +  f_5 f_7\ M_7 M_5   +  f_7 f_5\ M_5 M_7  +  \ldots)\,.
  \label{00,12}
\end{eqnarray}
Thus, by means of the map $\phi$ all rational coefficients in \eqn{00,5} are
converted to $1$. The result is a sum over all words in \eqn{UBasis}. In the
odd part of \eqn{00,12}, words $f_{i_1} f_{i_2}\ldots f_{i_p}$ are accompanied
by matrix products $M_{i_p} \ldots M_{i_2} M_{i_1}$ with reversed ordering of
indices. Writing \eqn{00,12} in a closed form yields the formula (\ref{00,13})
which completely covers the structure of the open superstring amplitude. All
the $s_{i_1\ldots i_p}$ content is expressed in terms of matrices, which
already appeared as coefficients of single $\z$'s in \eqn{00,5}. In order to
use the same matrices in both expressions, one has to fix the
freedom\footnote{The $\phi$ action on MZVs
$\zeta^\fm_{n_1,\ldots,n_r}$ of depth $r \geq 2$ is fixed by the
coproduct, except for the coefficient of $f_{n_1+\ldots +n_r}$. In
the conventions of \cite{SSMZV}, the $\phi$ images of the basis MZVs in table
\ref{zetaBasis} do not involve $f_{n_1+\ldots +n_r}$, with the exception of
single zeta values.} in the construction of the map $\phi$ \cite{BrownTate} such
that the only elements of weight $w$ in table \ref{zetaBasis} whose images
contain $f_w$ or $(f_2)^{w/2}$ are $\z_w$ and $(\z_2)^{w/2}$, respectively.
However, for any choice of basis for the MZVs the amplitude can be brought
into the form \eqn{00,13} by appropriately constructing the map and possibly
redefining the matrices $P_w$ and $M_w$. 

\section{Additional material on field-theory patterns in world-sheet integrals}
\label{appA}

This appendix provides supplementing material for section \ref{ftpatterns}. In
particular, it supports the correspondence between field-theory amplitudes
$A_{\te{YM}}(\si)$ and world-sheet disk integrals $Z_\Pi(\si)$ by connecting KK
and BCJ relations between $A_{\te{YM}}$ with partial-fraction and integration
by parts manipulations on the world-sheet, see subsection \ref{sec1,12}.

\subsection{KK relations between disk integrals}

KK relations among the $A_\YM(\si)$ correspond to partial-fraction relations between various permutations of the Green's functions $(z_{12} z_{23} \ldots z_{N-1,N} z_{N,1})^{-1}$ in the integrand of $Z_\Pi(\si)$. Let us demonstrate this through the four-point example, where the photon decoupling identity 
  \begin{equation}
   A_\YM(1,2,3,4)+A_\YM(1,2,4,3)+A_\YM(1,3,2,4)=0 
    \label{4ptphotondec}
  \end{equation}
  corresponds to a partial-fraction relation at the level of the integrand,
\begin{align}
Z_\Pi(1,2,3,4)  \ +& \ Z_\Pi(1,2,4,3)  \ + \ Z_\Pi(1,3,2,4) \ \ \sim \ \ \frac{1}{\CV_{\te{CKG}}} \prod_{i=1}^4 \int_{D(\Pi)} \dd z_i \  \prod_{i<j}^4 |z_{ij}|^{s_{ij}} \notag \\
\times \ &\left( \,
\frac{1}{  z_{12} z_{23} z_{34} z_{41} } \ + \ \frac{1}{  z_{12} z_{24} z_{43} z_{31} } \ + \ \frac{1}{  z_{13} z_{32} z_{24} z_{41} } \, \right) \eq  0 \ .
\label{01,17}
\end{align}
By iterating the manipulation shown in \eqn{01,17} one can verify the general KK relation (\ref{KKrelations}). As mentioned before, the choice of the KK basis $\big\{ Z_\Pi(1,\rho(2,3,\ldots,N),N-1) , \ \rho \in S_{N-2} \big\}$ in (\ref{KKrelations}) is convenient in view of the $SL(2,\ZR)$ fixing (\ref{01,fix}). Our methods for constructing the $F_\Pi{}^\si$ basis of functions take this $(N-2)!$ family of disk integrals as a starting point.

\subsection{BCJ relations between disk integrals}

BCJ relations correspond to the vanishing of a total derivative under the
world-sheet integral $\int_{D(\Pi)} \dd z_i$, regardless of the integration
domain $D(\Pi)$ defined by $z_{\Pi(i)}<z_{\Pi(i+1)}$. The linear Mandelstam
dependence enters through $z_i$ derivatives acting on the Koba-Nielsen factor:
\beq
\frac{\partial}{\partial z_k} \, \prod^{N-1}_{i<j} |z_{ij}|^{s_{ij}} \eq \prod_{i<j}^{N-1} |z_{ij}|^{s_{ij}} \, \sum^{N-1}_{l \neq k} \frac{ s_{kl} }{z_{kl}} 
\label{01,16}
\eeq
The world-sheet analogue of BCJ relations is most conveniently derived after $SL(2,\ZR)$ fixing:
\begin{align}
0 \eq &\int_{D(\Pi)} \dd z_2 \, \dd z_3 \ \frac{ \partial }{\partial z_3} \; \frac{ \prod_{i<j}^{4} |z_{ij}|^{s_{ij}}}{z_{12}} \eq \int_{D(\Pi)} \dd z_2 \, \dd z_3 \  \frac{ \prod_{i<j}^{4} |z_{ij}|^{s_{ij}}}{z_{12}} \; \left( \, \frac{s_{13}}{z_{31}} + \frac{ s_{23} }{z_{32}} + \frac{ s_{34}}{z_{34}} \, \right) \notag \\
\eq &-\, s_{13}\, Z_\Pi(1,3,2,5,4) \ - \ (s_{13}+
\label{01,18}
s_{23}) \,Z_\Pi(1,2,3,5,4) \ + \ s_{34} \, Z_\Pi(1,2,5,3,4)
\end{align}
Partial fraction $\frac{1}{z_{12} z_{31}} = \frac{1}{z_{12}
z_{32}}+\frac{1}{z_{13} z_{23}}$ in the first line of (\ref{01,18}) leads to a
literal copy of the five-point BCJ relation 
\begin{equation}
s_{13} \, A_\YM(1,3,2,5,4) \ + \ (s_{13}+s_{23})\, A_\YM(1,2,3,5,4) \eq s_{34} \,A_\YM(1,2,5,3,4)\,.  
  \label{5ptBCJ}
\end{equation}
More generally, we can obtain the $Z_\Pi$ analogue of the $N$-point BCJ
relation \cite{BCJ, BjerrumBohr:2009rd, Stieberger:2009hq} generalizing
(\ref{5ptBCJ}) by means of the following total derivative:
\begin{align}
0 \eq &- \ \prod_{i=2}^{N-2} \int_{D(\Pi)} \dd z_i \ \frac{ \partial }{\partial z_{N-2}} \;  \frac{ \prod_{i<j}^{N-1}  |z_{ij}|^{s_{ij}}}{z_{12} z_{23}\ldots z_{N-4,N-3}} \notag \\
\eq & s_{1,N-2} \, Z_\Pi(1,N-2,2,3,\ldots,N,N-1) \ + \ (s_{1,N-2}+s_{2,N-2}) \, Z_\Pi(1,2,N-2,3\ldots,N,N-1) \notag \\
& \ \ \ + \ \ldots \ + \ (s_{1,N-2}+\ldots+s_{N-3,N-2}) \, Z_{\Pi}(1,2,\ldots,N-3,N-2,N,N-1) \notag \\
&\ \ \ - \ s_{N-2,N-1} \, Z_{\Pi}(1,2,\ldots,N-3,N,N-2,N-1)
\label{01,19}
\end{align}
The proof of (\ref{01,19}) is shown in appendix \ref{secA2}.

\subsection{The momentum-kernel representation of the basis functions}
\label{secA1}

This subsection is devoted to proving equivalence between the momentum-kernel
representation $F_\Pi{}^\si =\sum_{\rho \in S_{N-3}} S[\rho|\si]_1
Z_{\Pi}(1,\rho(2,3,\ldots,N-2),N,N-1)$ of the basis functions and their
original form (\ref{01,1}) found in \cite{Mafra:2011nv,Mafra:2011nw}. Instead
of performing algebraic manipulations between (\ref{01,1}) and (\ref{01,5}), we
shall start from from the $(N-2)!$ representation
\begin{align}
A_\open(\Pi(1,\ldots,N) ) \ \, = \, & \sum_{\si \in N-2} n[1| \si(2,\ldots,N-2,N) | N-1]\,Z_\Pi(1,\si(2,3,\ldots,N-2,N),N-1) 
\label{appa1}
\end{align}
of the disk amplitude \cite{Mafra:2011nv}. The numerators $n[\ldots]$ therein
have been identified with BRST building blocks $T_{12\ldots p}$
\cite{Mafra:2010ir, Mafra:2010jq}
\beq
n[1| 2,\ldots ,\nu ,N,\nu+1,\ldots,N-2 | N-1] \eq \langle \, T_{12\ldots \nu} \, T_{N-1,N-2,\ldots,\nu+1} \, T_N \, \rangle \ ,
\label{appa1a}
\eeq
and the bracket $\langle \ldots \rangle$ denotes the pure-spinor zero-mode integration to extract superspace components \cite{Berkovits:2000fe}.

Without loss of generality, we restrict our discussion to the $(N-3)!$ terms of
(\ref{appa1}) with $\si(N)=N$ since the remaining terms where $\si(N)\neq N$
simply provide the BRST invariant completion towards $A_\YM$, see
\cite{Mafra:2010jq}. The BRST building blocks $T_{1\si(2\ldots N-2)}$ are
related to their corresponding Berends-Giele currents $M_{1\rho(2\ldots N-2)}$
\cite{Mafra:2010jq} through the momentum kernel,
\beq
T_{1\si(2\ldots N-2)}  \eq \sum_{\rho \in S_{N-3}}  S[ \, \si(2,\ldots,N-2) \, | \, \rho(2,\ldots,N-2) \, ]_1 \, M_{1\rho(2\ldots N-2)} \ .
\label{appa2}
\eeq
The latter build up field-theory subamplitudes via $A_{\te{YM}}(1,2,\ldots,N)=
\langle M_{12\ldots N-2} T_{N-1} T_N \rangle + \te{gauge invariant completion}$
\cite{Mafra:2010jq}. The claim (\ref{01,5}) then follows by substituting
(\ref{appa2}) into the $(N-2)!$ representation (\ref{appa1}) of the open-string
correlator:
\begin{align}
A_\open&(\Pi(1,\ldots,N)) \eq \sum_{\si \in S_{N-3}} \Big( \, \langle \, T_{1\si(23\ldots N-2)} T_{N-1} T_N \, \rangle\, Z_\Pi(1,\si(2,3,\ldots,N-2),N,N-1) \notag \\
& \ \ \ \ \ \ \ \ \ \ \ \ \ \ \ \ \ \ \ \ \ \ \ \ \ \ \ \ \ \ \ \ \ \ \ \ \ \ \ \ \ \ \ + \ \te{gauge invariant completion} \, \Big) \notag \\
\eq& \sum_{\si \in S_{N-3}} \Big( \, \sum_{\rho \in S_{N-3}} S[ \, \si(2,\ldots,N-2) \, | \, \rho(2,\ldots,N-2) \, ]_1  \, \langle \, M_{1\rho(23\ldots N-2)} T_{N-1} T_N \, \rangle\notag \\
& \ \ \ \ \ \  \times \  Z_\Pi(1,\si(2,3,\ldots,N-2),N,N-1) \ + \ \te{gauge invariant completion} \, \Big) \notag \\
\eq& \sum_{\si \in S_{N-3}}  \sum_{\rho \in S_{N-3}} S[ \, \si(2,\ldots,N-2) \, | \, \rho(2,\ldots,N-2) \, ]_1 \notag \\
& \ \ \ \ \ \ \times \ A_\YM(1,\rho(2,\ldots ,N-2),N-1,N) \, Z_\Pi(1,\si(2,3,\ldots,N-2),N,N-1)  
\label{appa3}
\end{align}
Matching the last line of (\ref{appa3}) with (\ref{00,1}) completes the proof
of (\ref{01,5}).

\subsection{Deriving the $N$-point BCJ relation for disk integrals}
\label{secA2}

We shall carry out the intermediate steps of (\ref{01,19}) here. The claim can
be reexpressed as
\begin{align}
- \ &\prod_{i=2}^{N-2} \int_{D(\Pi)} \dd z_i \ \frac{ \partial }{\partial z_{N-2}} \;  \frac{ \prod_{i<j}^{N-1}  |z_{ij}|^{s_{ij}}}{z_{12} z_{23}\ldots z_{N-4,N-3}} \eq - \ s_{N-2,N-1} \, Z_{\Pi}(1,2,\ldots,N\!-\!3,N,N\!-\!2,N\!-\!1) \notag \\
&+ \ \sum_{ j=1 }^{N-3} s_{N-2-j,N-2} \sum_{i=1}^j Z_\Pi(1,2,\ldots,N\!-\!2\!-\!i,N\!-\!2,N\!-\!1\!-\!i,\ldots,N\!-\!3,N,N\!-\!1) \ .
\label{01,90}
\end{align}
The left hand side contains $N-2$ terms from
\beq
 \frac{ \partial }{\partial z_{N-2}} \prod_{i<j}^{N-1}  |z_{ij}|^{s_{ij}} \eq \left( \, \sum_{k=1}^{N-3}  \frac{ s_{N-2,k} }{z_{N-2,k}} \ + \ \frac{ s_{N-2,N-1}}{z_{N-2,N-1}} \, \right) \, \prod_{i<j}^{N-1}  |z_{ij}|^{s_{ij}}
 \label{01,91}
 \eeq
So we have to check the $s_{N-2,N-1}$ coefficients to match,
\beq
 \prod_{i=2}^{N-2} \int_{D(\Pi)} \dd z_i \ \frac{\prod_{i<j}^{N-1}  |z_{ij}|^{s_{ij}}}{z_{12} z_{23}\ldots z_{N-4,N-3} \cdot z_{N-2,N-1}} \eq Z_{\Pi}(1,2,\ldots,N-3,N,N-2,N-1) \ ,
\label{01,92}
\eeq
and, after relabeling $k=N-2-j$,
\begin{align}
 \prod_{i=2}^{N-2} \int_{D(\Pi)} &\dd z_i \ \frac{\prod_{i<j}^{N-1}  |z_{ij}|^{s_{ij}}}{z_{12} z_{23}\ldots z_{N-4,N-3} \cdot z_{N-2,N-2-j}} \notag \\
 & = \ \ - \, \sum_{i=1}^j Z_\Pi(1,2,\ldots,N-2-i,N-2,N-1-i,\ldots,N-3,N,N-1)
\label{01,93}
\end{align}
for the range $j=1,2,\ldots,N-3$. As usual, we fix $z_1=0, \ z_{N-1} = 1$ and
$z_{N} \rightarrow \infty$ such that any ratio $\frac{z_{i,N}}{z_{j,N}}
\rightarrow 1$ for $i,j=1,2,\ldots,N-1$. Then, (\ref{01,92}) easily follows
from 
\begin{align}
Z_{\Pi}(1,2,\ldots,&N-3,N,N-2,N-1)
\eq \prod_{i=2}^{N-2} \int_{D(\Pi)} \dd z_i \notag \\
&\times \ \frac{\prod_{i<j}^{N-1}  |z_{ij}|^{s_{ij}}}{z_{12} z_{23}\ldots z_{N-4,N-3}z_{N-2,N-1}} \, \times \, \underbrace{ \frac{ z_{1,N-1} z_{1,N} z_{N-1,N}}{z_{N-3,N} z_{N,N-2} z_{N-1,1}} }_{\rightarrow \ 1 \ \te{as} \ z_N \rightarrow \infty}
\label{01,94}
\end{align}
and (\ref{01,93}) can be shown inductively: At $j=1$, it holds by virtue of
\begin{align}
Z_\Pi(1,2,\ldots&,N-3,N-2,N,N-1) \notag \\
\eq& \prod_{i=2}^{N-2} \int_{D(\Pi)} \dd z_i \ \frac{\prod_{i<j}^{N-1}  |z_{ij}|^{s_{ij}}}{z_{12} z_{23}\ldots z_{N-4,N-3}z_{N-3,N-2}} \, \times \, \underbrace{ \frac{ z_{1,N-1} z_{1,N} z_{N-1,N}}{z_{N-2,N} z_{N,N-1} z_{N-1,1}} }_{\rightarrow \ 1  \ \te{as} \ z_N \rightarrow \infty} \notag \\
\eq&  - \, \prod_{i=2}^{N-2} \int_{D(\Pi)} \dd z_i \ \frac{ \prod_{i<j}^{N-1}  |z_{ij}|^{s_{ij}}}{z_{12} z_{23}\ldots z_{N-4,N-3}\cdot z_{N-2,N-3}} \ .\label{01,95}
\end{align}
Assuming validity of (\ref{01,93}) at $j=l-1$, we conclude
\begin{align}
\sum_{i=1}^l Z_\Pi&(1,2,\ldots,N-2-i,N-2,N-1-i,\ldots,N-3,N,N-1) \notag \\
\eq&  Z_\Pi(1,2,\ldots,N-2-l,N-2,N-1-l,\ldots,N-3,N,N-1) \notag \\
& \ \ - \ \prod_{i=2}^{N-2} \int_{D(\Pi)} \dd z_i \ \frac{\prod_{i<j}^{N-1}  |z_{ij}|^{s_{ij}}}{z_{12} z_{23}\ldots z_{N-4,N-3} \cdot z_{N-2,N-1-l}} \notag \\
\eq& \prod_{i=2}^{N-2} \int_{D(\Pi)} \dd z_i \ \frac{\prod_{i<j}^{N-1}  |z_{ij}|^{s_{ij}}}{z_{12} z_{23}\ldots z_{N-4,N-3} \cdot z_{N-2,N-1-l}} \, \left( \, \frac{ z_{N-2-l,N-1-l} }{ z_{N-2-l,N-2} } \ - \ 1 \, \right) \notag \\
\eq& \prod_{i=2}^{N-2} \int_{D(\Pi)} \dd z_i \ \frac{\prod_{i<j}^{N-1}  |z_{ij}|^{s_{ij}}}{z_{12} z_{23}\ldots z_{N-4,N-3} \cdot z_{N-2-l,N-2}}  
\label{01,96}
\end{align}
Hence, the coefficients of all the $s_{N-2,k}$ match in (\ref{01,90}).


\section{Application of the pole criterion for disk integrals}
\label{exppole}

In this appendix, we gather examples for the rule (\ref{pole}) determining the
pole content of disk integrals from their KK basis (\ref{02,2}). All of the
results shown are in agreement with the closed formula (\ref{consFT}) which is
equivalent to having a consistent field-theory limit $F_\Pi{}^\si |_{\ap
\rightarrow 0} = \de_\Pi^\si$ for the disk amplitude (\ref{00,1}).

\subsection{Four-point examples}

The first kinematic poles occur at four points:
\begin{align}
Z(1,2,4,3) \ = \ \int^1_0 \dd z_2 \ \frac{ |z_{12}|^{s_{12}} |z_{23}|^{s_{23}}}{z_{12}} \ \ \Leftrightarrow \ \ R^{\nu=2}(z_{ij}) = z_{12} \ \ \Leftrightarrow  \ \ Z(1,2,4,3) \ \sim \ -\, \frac{1}{s_{12}}
\label{z4pt1}
 \\
Z(1,4,2,3) \ = \ \int^1_0 \dd z_2 \ \frac{ |z_{12}|^{s_{12}} |z_{23}|^{s_{23}}}{z_{23}} \ \ \Leftrightarrow \ \ R^{\nu=1}(z_{ij}) = z_{23} \ \ \Leftrightarrow \ \ Z(1,4,2,3) \ \sim \ -\, \frac{1}{s_{23}}
\label{z4pt2}
\end{align}
In the four-point KK basis $\{ Z(1,2,4,3), \, Z(1,4,2,3) \}$, the $R^{\nu}$ are
inevitably of the form $z_{i,i+1}$, so they directly translate into
two-particle channels $s_{i,i+1}$ according to (\ref{pole}). The cubic diagrams
arising from the low-energy limits of $Z(1,2,4,3)$ and $Z(1,4,2,3)$ are shown
in figure \ref{fig:ft4pt}:

\begin{figure}[htbp]
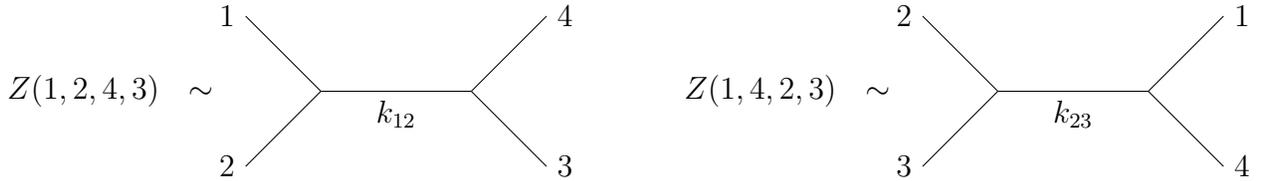
 
\begin{center}
\tikzpicture
\draw (-2.8,0) node{$  Z(1,2,4,3) \ \ \sim$};
\draw (0,0) -- (-1,1) node[left]{$1$};
\draw (0,0) -- (-1,-1) node[left]{$2$};
\draw (0,0) -- (2,0);
\draw (1,-0.3) node{$k_{12}$};
\draw (2,0) -- (3,1) node[right]{$4$};
\draw (2,0) -- (3,-1) node[right]{$3$};
\scope[xshift=9cm]
\draw (-2.8,0) node{$ Z(1,4,2,3) \ \ \sim$};
\draw (0,0) -- (-1,1) node[left]{$2$};
\draw (0,0) -- (-1,-1) node[left]{$3$};
\draw (0,0) -- (2,0);
\draw (1,-0.3) node{$k_{23}$};
\draw (2,0) -- (3,1) node[right]{$1$};
\draw (2,0) -- (3,-1) node[right]{$4$};
\endscope
\endtikzpicture
\caption{Cubic diagrams generated by four-point integrals.}
\label{fig:ft4pt}
\end{center}
\end{figure}

\subsection{Five-point examples}
\label{sec1,212}

Five points provide the first cases of incompatible pole channels within the
same function. The six KK basis elements $\{Z(1,2_\rho,3_\rho,5_\rho,4), \ \rho
\in S_3\}$ exhibit the following poles (cf. \ref{pole}):

\begin{table}[htdp]
\[
\begin{array}{|c|c||c|c|c|c|c||c|c||c|c|c|c|c|} \hline
Z(\ldots) &R^{\nu,\rho} &s_{12} &s_{23} &s_{34} &s_{123} &s_{234} &Z(\ldots) &R^{\nu,\rho} &s_{12} &s_{23} &s_{34} &s_{123} &s_{234}  \\ \hline \hline
Z(1,2,3,5,4)&z_{12} z_{23} &\checkmark &\checkmark &\times &\checkmark &\times 
&Z(1,5,2,3,4) &z_{23} z_{34} &\times &\checkmark &\checkmark &\times &\checkmark \\ \hline 
Z(1,3,2,5,4)&z_{13} z_{32} &\times &\checkmark &\times &\checkmark &\times 
&Z(1,5,3,2,4) &z_{32} z_{24} &\times &\checkmark &\times &\times &\checkmark \\ \hline 
Z(1,2,5,3,4)&z_{12} z_{34} &\checkmark &\times &\checkmark &\times &\times 
&Z(1,3,5,2,4) &z_{13} z_{24} &\times &\times &\times &\times &\times \\ \hline 
\end{array}
\]
\caption{Pole channels present in the five-point KK basis $Z(1,\rho(2,3,5),4)$.}
\label{pole5}
\end{table}

\begin{figure}[htbp]
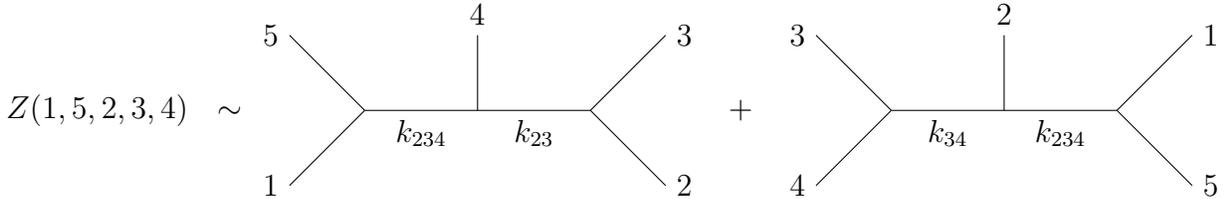
 
\begin{center}
\tikzpicture
\draw (-3.2,0) node{$ Z(1,5,2,3,4)  \ \ \sim$};
\draw (0,0) -- (-1,1) node[left]{$5$};
\draw (0,0) -- (-1,-1) node[left]{$1$};
\draw (0,0) -- (3,0);
\draw (0.75,-0.3) node{$k_{234}$};
\draw (2.25,-0.3) node{$k_{23}$};
\draw (1.5,0) -- (1.5,1) node[above]{$4$};
\draw (3,0) -- (4,1) node[right]{$3$};
\draw (3,0) -- (4,-1) node[right]{$2$};
\draw (5,0) node{$+$};
\scope[xshift=7cm]
\draw (0,0) -- (-1,1) node[left]{$3$};
\draw (0,0) -- (-1,-1) node[left]{$4$};
\draw (0,0) -- (3,0);
\draw (0.75,-0.3) node{$k_{34}$};
\draw (2.25,-0.3) node{$k_{234}$};
\draw (1.5,0) -- (1.5,1) node[above]{$2$};
\draw (3,0) -- (4,1) node[right]{$1$};
\draw (3,0) -- (4,-1) node[right]{$5$};
\endscope
\endtikzpicture
\caption{Cubic diagrams generated by the five-point integral $Z(1,5,2,3,4)$.}
\label{fig:ft5pt}
\end{center}
\end{figure}

The disk integrals $Z(1,3,2,5,4), \ Z(1,5,3,2,4)$ have the right number of poles to
describe a cubic five-point diagram of YM field theory:
\beq
Z(1,3,2,5,4) \ \ \sim \ \ \frac{-\,1}{s_{23} s_{123}} \co
Z(1,5,3,2,4) \ \ \sim \ \ \frac{-\,1}{s_{23} s_{234}} \co
Z(1,2,5,3,4)  \ \ \sim \ \ \frac{1}{s_{12} s_{34}}
\label{02,11}
\eeq
Two functions $Z(1,2,3,5,4)$ and $Z(1,5,2,3,4)$ are singular in three
Mandelstam variables with incompatible pairs therein: Neither $s_{12},s_{23}$
nor $s_{23},s_{34}$ can appear simultaneously. Hence, their field-theory limits
give rise to two cubic diagrams each\footnote{In the parametrization
(\ref{02,3}), the poles of $Z(1,2,3,5,4)$ can be traced back to the $x_{2}
\rightarrow 0$ limit of
\begin{align}
Z&(1,2,3,5,4) \eq \int^1_0 \dd z_3 \int^{z_3}_0 \dd z_2 \ \frac{ \prod_{i<j}^4 |z_{ij}|^{s_{ij}} }{z_{12} z_{23}} \notag \\
&\eq  \int^1_0 \dd x_1  \int^1_0 \dd x_2 \ x_1^{s_{12}-1} \, (1-x_1)^{s_{23}-1} \, x_2^{s_{123}-1} \, (1-x_2)^{s_{34}} \, (1-x_1 x_2)^{s_{24}} \ .
\label{02,6c}
\end{align}
The boundaries of the $x_1$ integration range give rise to the dual two-particle channel in either $s_{12}$ (from $x_1 \rightarrow 0$) or $s_{23}$ (from $x_1 \rightarrow 1$). }, as depicted in figure \ref{fig:ft5pt}:
\beq
Z(1,2,3,5,4) \ \ \sim \ \ \frac{1}{s_{123}} \, \left( \, \frac{1}{s_{12}}+\frac{1}{s_{23}} \, \right) \co
Z(1,5,2,3,4) \ \ \sim \ \ \frac{1}{s_{234}} \, \left( \, \frac{1}{s_{23}}+\frac{1}{s_{34}} \, \right)
\label{02,12}
\eeq
The sixth function $Z(1,3,5,2,4)$ is regular: As we shall see below, its
$\ap$-expansion starts at $\zeta_2$. According to table \ref{pole5} the
five-point KK basis exhausts all the five cubic diagrams compatible with the
color ordering $(1,2,3,4,5)$.

\subsection{Six-point examples}
\label{sec1,213}

At six points, the 24 elements $Z(1,2_\rho,3_\rho,4_\rho,6_\rho,5)$ of the KK basis can be split into four $S_3$ subsets according to $\nu=4,3,2,1$ in the notation of (\ref{02,2}). 
World-Sheet parity $(z_{i},k_i) \mapsto (z_{6-i},k_{6-i})$ connects the $\nu=4$ integrals $Z(1,2_\rho,3_\rho,4_\rho,6,5)$ with their $\nu =1$ counterparts $Z(1,6,2_\rho,3_\rho,4_\rho,5)$ and likewise the $Z(1,2_\rho,3_\rho,6,4_\rho,5)$ at $\nu =3$ with $Z(1,2_\rho,6,3_\rho,4_\rho,5)$ at $\nu=2$. Hence, it is sufficient to discuss the pole structure of the parity-independent $\nu=3,4$ sectors. Let us start with the $\nu=4$-sector:

\begin{table}[htdp]
\[
\begin{array}{|c|c||c|c|c|c|c|c|c|c|c|} \hline
Z(\ldots) &R^{\nu,\rho} &s_{12} &s_{23} &s_{34} &s_{45} &s_{1234} &s_{2345} &s_{123} &s_{234} &s_{345}  \\ \hline \hline
Z(1,2,3,4,6,5)&z_{12} z_{23} z_{34} &\checkmark &\checkmark &\checkmark &\times &\checkmark &\times &\checkmark &\checkmark &\times 
 \\ \hline 
Z(1,2,4,3,6,5)&z_{12} z_{24} z_{43} &\checkmark &\times &\checkmark &\times &\checkmark &\times &\times &\checkmark &\times \\ \hline 
Z(1,3,2,4,6,5)&z_{13} z_{32} z_{24} &\times &\checkmark &\times &\times &\checkmark &\times &\checkmark &\checkmark &\times \\ \hline 
Z(1,3,4,2,6,5)&z_{13} z_{34} z_{42} &\times &\times &\checkmark &\times &\checkmark &\times &\times &\checkmark &\times \\ \hline 
Z(1,4,2,3,6,5)&z_{14} z_{42} z_{23} &\times &\checkmark &\times &\times &\checkmark &\times &\times &\checkmark &\times \\ \hline 
Z(1,4,3,2,6,5)&z_{14} z_{43} z_{32} &\times &\checkmark &\checkmark &\times &\checkmark &\times &\times &\checkmark &\times \\ \hline 
\end{array}
\]
\caption{Pole channels present in the six-point integrals $Z(1,2_\rho,3_\rho,4_\rho,6,5)$ at $\nu=4$.}
\label{pole6a}
\end{table}

Two of the six KK integrals $Z(1,\ldots,6,5)$ have exactly three pole channels and thus corresponds to a single cubic six-point diagram,
\beq
Z(1,3,4,2,6,5) \ \ \sim \ \ \frac{1}{s_{34} s_{234} s_{1234}} \co Z(1,4,2,3,6,5) \ \ \sim \ \ \frac{1}{s_{23} s_{234} s_{1234}} \ .
\label{02,13}
\eeq
Three functions involve pairs of incompatible pole channels $(s_{12},s_{234}), (s_{123},s_{234})$ and $(s_{23},s_{34})$, respectively, so their field-theory limits involve two diagrams
\begin{align}
Z(1,2,4,3,6,5) \ \ &\sim \ \ \frac{1}{s_{1234}s_{34} } \, \left( \, \frac{1}{s_{12}} + \frac{1}{s_{234}} \, \right) \notag \\
Z(1,3,2,4,6,5) \ \ &\sim \ \ \frac{1}{s_{1234}s_{23} } \, \left( \, \frac{1}{s_{123}} + \frac{1}{s_{234}} \, \right)
\label{02,14} \\
Z(1,4,3,2,6,5) \ \ &\sim \ \ - \, \frac{1}{s_{1234}s_{234} } \, \left( \, \frac{1}{s_{23}} + \frac{1}{s_{34}} \, \right) \ . \notag 
\end{align}
The $Z(1,2,3,4,6,5)$ integral has six pole channels, five of which form a cycle $s_{12},s_{23},s_{34},s_{123},s_{234}$ with incompatible neighbors. The only way of exhausting all possible singularities is a sum of five diagrams
\begin{align}
Z(1,2,3,4,6,5) \ \ &\sim \ \ - \, \frac{1}{s_{1234}} \, \left( \, \frac{1}{s_{12}s_{34}} + \frac{1}{s_{23}s_{123}} +\frac{1}{s_{34}s_{234}} +\frac{1}{s_{123}s_{12}} +\frac{1}{s_{234}s_{23}} \, \right) \ .
\label{02,15}
\end{align}

\begin{table}[htdp]
\[
\begin{array}{|c|c||c|c|c|c|c|c|c|c|c|} \hline
Z(\ldots) &R^{\nu,\rho} &s_{12} &s_{23} &s_{34} &s_{45} &s_{1234} &s_{2345} &s_{123} &s_{234} &s_{345}  \\ \hline \hline
Z(1,2,3,6,4,5)&z_{12} z_{23} z_{45} &\checkmark &\checkmark &\times &\checkmark &\times &\times &\checkmark &\times &\times \\ \hline 
Z(1,3,2,6,4,5)&z_{13} z_{32} z_{45} &\times &\checkmark &\times &\checkmark &\times &\times &\checkmark &\times &\times \\ \hline 
Z(1,2,4,6,3,5)&z_{12} z_{24} z_{35} &\checkmark &\times &\times &\times &\times &\times &\times &\times &\times \\ \hline 
Z(1,4,2,6,3,5)&z_{14} z_{42} z_{35} &\times &\times &\times &\times &\times &\times &\times &\times &\times \\ \hline 
Z(1,3,4,6,2,5)&z_{13} z_{34} z_{25} &\times &\times &\checkmark &\times &\times &\times &\times &\times &\times \\ \hline 
Z(1,4,3,6,2,5)&z_{14} z_{43} z_{25} &\times &\times &\checkmark &\times &\times &\times &\times &\times &\times \\ \hline 
\end{array}
\]
\caption{Pole channels present in the six-point integrals $Z(1,2_\rho,3_\rho,6,4_\rho,5)$ at $\nu=3$.}
\label{pole6b}
\end{table}

The $\nu=3$ sector incorporates two integrals with a non-vanishing field-theory limit which resembles the five-point $Z(1,2_\rho,3_\rho,5,4)$ functions (up to an additional $s_{45}$-propagator):
\beq
Z(1,2,3,6,4,5) \ \ \sim \ \ -\, \frac{1}{s_{123} s_{45}} \, \left( \, \frac{1}{s_{12}}+\frac{1}{s_{23}} \, \right) \co Z(1,3,2,6,4,5) \ \ \sim \ \  \frac{1}{s_{123}s_{23} s_{45}} 
\label{02,16}
\eeq
Table \ref{pole6b} provides further examples of a general feature of $(N\geq 5$)-point integrals: some of them contribute to fewer pole channels than present in a cubic YM diagrams. Functions with $k=0,1,2,\ldots,N-5$ pole channels decouple from the field-theory limit, and their leading low-energy contribution occurs at transcendentality $\zeta_{N-3-k}$ at the residue of $k$ simultaneous poles\footnote{Note that the non-existence of (a shuffle-regularized version of) $\zeta_1$ ties in with the absence of $N-4$ simultaneous poles in the leading low energy behavior of disk integrals.}. From table \ref{pole6b} one can extract three six-point examples
\beq
Z(1,2,4,6,3,5) \ \ \sim \ \ -\, \frac{ \zeta_2}{s_{12}} \co Z(1,3,4,6,2,5) \ \ \sim \ \  \frac{ \zeta_2}{s_{34}}\co Z(1,4,3,6,2,5) \ \ \sim \ \ -\,\frac{ \zeta_2}{s_{34}} 
\label{02,17}
\eeq
with a single pole channel $k=1$ and leading transcendentality $\zeta_2$. The integral $Z(1,4,2,6,3,5)$ is free of poles and furnishes a $k=0$ example: the leading terms of its $\ap$-expansion is accompanied by $\zeta_3$.

\subsection{Seven-point examples}

Instead of showing the lengthy list of 120 seven-point KK integrals
$Z(1,2_\rho,3_\rho,4_\rho,5_\rho,7_\rho,6)$ including all permutations $\rho \in S_5$, we
shall present some examples here, others can be found in \cite{Mafra:2011nw}.
In the $\nu=5$ sector $Z(1,2_\rho,3_\rho,4_\rho,5_\rho,7,6)$ of (\ref{02,2}),
the pole prescription (\ref{pole}) applies to polynomials $R^{\nu=5,\rho} =
z_{12_\rho} z_{2_\rho 3_\rho} z_{3_\rho 4_\rho} z_{4_\rho 5_\rho}$.
\begin{itemize}
\item $Z(1,2,3,4,5,7,6)$ has ten pole channels
  $s_{12},s_{23},s_{34},s_{45},s_{123},s_{234},s_{345},s_{1234},s_{2345}$ and
  $s_{12345}$ where the former nine are subject to various incompatibilities.
  The complete set of singularities is captured by summing fourteen diagrams
  resembling a six-point YM subamplitude:
\begin{align}
Z(1,2,3,&4,5,7,6)  \ \ \sim \ \ \frac{1}{s_{12345}} \, \bigg( \, \frac{ 1}{s_{12} s_{123} s_{1234} } + \frac{ 1}{s_{23} s_{123} s_{1234} } +\frac{ 1}{s_{23} s_{234} s_{1234} } +\frac{ 1}{s_{34} s_{234} s_{1234} } \notag \\
& +\frac{ 1}{s_{12} s_{34} s_{1234} } +\frac{ 1}{s_{23} s_{234} s_{2345} } +\frac{ 1}{s_{34} s_{234} s_{2345} } +\frac{ 1}{s_{34} s_{345} s_{2345} } +\frac{ 1}{s_{45} s_{345} s_{2345} } \notag \\
&+\frac{ 1}{s_{23} s_{45} s_{2345} } +\frac{ 1}{s_{12} s_{123} s_{45} } +\frac{ 1}{s_{23} s_{123} s_{45} } +\frac{ 1}{s_{12} s_{34} s_{345} } +\frac{ 1}{s_{12} s_{45} s_{345} } \, \bigg)
\end{align}
\item $Z(1,2,3,5,4,7,6)$ has seven pole channels, two universally compatible
  ones $s_{12345},s_{45}$ and a cycle of five pole channels
  $s_{12},s_{23},s_{345},s_{123},s_{2345}$ with incompatibility between
  neighbors
\beq
Z(1,2,3,5,4,7,6)  \ \ \sim \ \ - \, \frac{ 1}{s_{12345} s_{45}} \, \bigg( \, \frac{1}{s_{12} s_{345}} + \frac{1}{s_{23} s_{123}} +\frac{1}{s_{345} s_{2345}} + \frac{1}{s_{12} s_{123}} + \frac{1}{s_{23} s_{2345}} \, \bigg)
\eeq
\item $Z(1,2,5,4,3,7,6)$ has six pole channels, two universally compatible ones
  $s_{12345} ,s_{345}$, and two incompatible pairs $s_{12},s_{2345}$ as well as
  $s_{34},s_{45}$.
\beq
Z(1,2,5,4,3,7,6)  \ \ \sim \ \  \frac{1}{s_{12345} s_{345}} \, \left( \, \frac{1}{s_{12}} + \frac{1}{s_{2345}} \, \right)\, \left( \, \frac{1}{s_{34}} + \frac{1}{s_{45}} \, \right)
\eeq
\item $Z(1,2,4,5,3,7,6)$ has five pole channels, $s_{12345},s_{345},s_{45}$
 being universally compatible and $s_{12},s_{2345}$ as an incompatible pair:
\beq
Z(1,2,4,5,3,7,6) \ \ \sim \ \ - \, \frac{1}{s_{12345} s_{345} s_{45}} \, \left( \, \frac{1}{s_{12}} + \frac{1}{s_{2345}} \, \right)
\eeq
\item $Z(1,3,5,4,2,7,6)$ has four pole channels $s_{45},s_{345},s_{2345},
  s_{12345}$ leading to a unique diagram
\beq
Z(1,3,5,4,2,7,6) \ \ \sim \ \  \frac{1}{s_{45}s_{345}s_{2345} s_{12345}}
\eeq
\item $Z(1,3,5,2,4,7,6)$ has two pole channels $s_{2345}, s_{12345}$ such that its leading low-energy contribution occurs at $\zeta_2$ order
\beq
Z(1,3,5,2,4,7,6) \ \ \sim \ \  \frac{ \zeta_2 }{s_{2345} s_{12345}}
\eeq
\end{itemize}
The $\nu \leq 4$ sectors of the seven-point KK basis additionally involve
integrals with leading transcendentality $\zeta_3$ and $\zeta_4$, respectively,
e.g.
\begin{align}
Z(1,2,5,3,7,4,6) \ \ \sim \ \ \frac{ 2 \zeta_3 }{s_{12}} \co
Z(1,5,3,7,2,4,6) \ \ \sim \ \ - \, \frac{17}{4} \, \zeta_4\ .
\end{align}
Further examples of integrals with leading low-energy behavior $\sim
\zeta_{N-3-k}\prod_{i=1}^k (s_{a_i ,a_i+1,\ldots,b_i})^{-1}$ can be found in
subsection \ref{sec1,22}.


\newpage
\section{Singularity structure of further six-point integrals}
\label{app6pt}

This appendix supplements the discussion of subsection \ref{subsec:pole6pt} on
residues in six-point integrals. We investigate pole structures present in the
$\nu =4$ sector (\ref{02,76}) of the pole basis which we can relate to the
$F^\si$ via (\ref{invmom2}):
\begin{align}
Z[s_{12}s_{123}s_{1234}] \eq &\frac{1}{s_{1234}} \, \Big( \, \frac{ F^{(234)} }{s_{12} s_{123}} \ + \ \frac{ F^{(243)} }{s_{12} s_{124}} \ + \ \frac{ F^{(324)} }{s_{13} s_{123}} \ + \ \frac{ F^{(342)} }{s_{13} s_{134}} \ + \ \frac{ F^{(423)} }{s_{14} s_{124}} \ + \ \frac{ F^{(432)} }{s_{14} s_{134}} \, \Big) \notag \\
Z[s_{12}s_{34}s_{1234}] \eq &\frac{1}{s_{1234}} \, \Big( \, \frac{ F^{(234)} }{s_{12} s_{34}} \ +\ \frac{  F^{(342)} }{s_{34} s_{134} } \ - \ \frac{ F^{(432)}}{s_{134}}  \, \Big(\frac{1}{s_{14}} +\frac{1}{s_{34}} \Big) \notag \\
& - \ \frac{ F^{(423)} }{s_{14} s_{124}} \ - \
 \frac{ F^{(243)} }{s_{12}} \, \Big(\frac{1}{s_{124}} + \frac{1}{s_{34}}\Big) \, \Big) \notag
\\
Z[s_{23}s_{123}s_{1234}] \eq &\frac{1}{s_{1234}} \, \Big( \, \frac{ F^{(234)} }{s_{23} s_{123}} \ - \ \frac{ F^{(324)} }{s_{123}} \, \Big( \frac{1}{s_{13}}+\frac{1}{s_{23}} \Big) \ - \ \frac{ F^{(342)} }{s_{13} s_{134}} \notag \\
& +\ \frac{ F^{(423)} }{s_{14} s_{23}}  \ - \ \frac{ F^{(432)} }{s_{14} } \, \Big( \frac{1}{s_{23}}+\frac{1}{s_{134}} \Big) \, \Big) \notag 
\\
Z[s_{34}s_{234}s_{1234}] \eq &\frac{1}{s_{1234}} \, \Big( \, \frac{ F^{(234)} }{s_{34} s_{234}}  \ - \ F^{(243)} \, \Big( \frac{1}{s_{34}s_{234}}+\frac{1}{s_{24}s_{124}}+\frac{1}{s_{24}s_{234}} \Big) \ + \ \frac{ F^{(324)} }{s_{24} s_{234}}  \label{expa6pt}  \\
& - \ F^{(342)} \, \Big( \frac{1}{s_{34}s_{134}}+\frac{1}{s_{24}s_{234}} + \frac{1}{s_{34}s_{234}} \Big) \ + \ F^{(423)} \, \Big( \frac{1}{s_{14}s_{124}}+\frac{1}{s_{24}s_{124}} +\frac{1}{s_{24}s_{234}} \Big) \notag \\
& + \ F^{(432)} \, \Big( \frac{1}{s_{14}s_{134}}+\frac{1}{s_{34}s_{134}} + \frac{1}{s_{34}s_{234}} \Big) \, \Big) \notag 
\\
Z[s_{23}s_{234}s_{1234}] \eq &\frac{1}{s_{1234}} \, \Big( \,
\frac{ F^{(234)} }{s_{23} s_{234}} \ + \ \frac{ F^{(243)} }{s_{24}} \, \Big( \frac{1}{s_{124}}+\frac{1}{s_{234}} \Big)  \ - \ \frac{ F^{(324)} }{s_{234}} \, \Big( \frac{1}{s_{23}}+\frac{1}{s_{24}} \Big)  \ + \ \frac{ F^{(342)} }{s_{24} s_{234}} \notag \\
& - \ F^{(423)} \, \Big( \frac{1}{s_{14} s_{23}}+\frac{ 1}{s_{14}s_{124}} + \frac{1}{s_{23}s_{234}} +\frac{1}{s_{124} s_{24}}+\frac{1}{s_{24}s_{234}} \Big) \notag \\
& + \ \frac{ F^{(432)} }{s_{23}} \, \Big( \frac{1}{s_{14}}+\frac{1}{s_{234}} \Big) \, \Big)  \notag 
\\
Z[\zeta_2 s_{1234}] \eq &\frac{1}{s_{1234}} \, \Big( \, \frac{ F^{(243)} }{s_{24}s_{124}} \ + \ \frac{ F^{(324)} }{s_{13}s_{24}} \ - \ \frac{ F^{(342)} }{s_{13}} \, \Big( \frac{1}{s_{134}}+\frac{1}{s_{24}}\Big) \notag \\
&- \ \frac{ F^{(423)} }{s_{124}} \, \Big( \frac{1}{s_{14}}+\frac{1}{s_{24}} \Big) \ - \ \frac{ F^{(432)} }{s_{14} s_{134}} \, \Big) \notag 
\end{align}
Figures \ref{fig:s12s123s1234} and \ref{fig:s12s34s1234} summarize the diagrams
contributing to the singular part of $Z[s_{12}s_{123}s_{1234}]$ and
$Z[s_{12}s_{34}s_{1234}]$. They cover the two topologies possible for cubic six
point graphs. The methods of subsection \ref{sec1,23} yield the following
residues for the six functions in (\ref{expa6pt}):
\begin{align}
Z[s_{12}s_{123}s_{1234}] \eq &
\frac{1}{s_{12} s_{123}s_{1234} } \ + \ \frac{I_{21}^\reg(k_1,k_2,k_3)}{s_{123} s_{1234}} \ + \ \frac{I_{21}^\reg(k_{12},k_3,k_4)}{s_{12} s_{1234}} \ + \ \frac{I_{21}^\reg(k_{123},k_4,k_5)}{s_{12} s_{123} } \notag \\
& + \ \frac{I_{21,31}^\reg( k_1,k_2,k_3,k_4)}{s_{1234} } \ + \ \frac{I_{21}^\reg(k_1,k_2,k_3) I_{21}^\reg(k_{123},k_4,k_5)}{ s_{123}} \notag \\
& + \ \frac{I_{21,31}^\reg( k_{12},k_3,k_4,k_5 )}{s_{12}  } \ + \ I^\reg_{21,31,41} \label{expa6pt1}
\\
Z[s_{12}s_{34}s_{1234}] \eq&
\frac{1}{s_{12} s_{34}s_{1234} } \ + \ \frac{I_{21}^\reg(k_1,k_2,k_{34})}{ s_{34}s_{1234} } \ + \ \frac{I_{32}^\reg(k_{12},k_3,k_4)}{s_{12} s_{1234} } \ + \ \frac{I_{21}^\reg(k_{12},k_{34},k_5) }{s_{12} s_{34}}  \notag \\
&+ \ \frac{I_{21,43}^\reg(k_1,k_2,k_3,k_4)}{ s_{1234} } \ + \ \frac{I_{21,31}^\reg(k_1,k_2,k_{34},k_5)}{ s_{34}  } \notag \\
& + \ \frac{I_{31,32}^\reg(k_{12},k_3,k_4,k_5)}{s_{12}  } \ + \ I^\reg_{21,43,41} \label{expa6pt2}
\\
Z[s_{23}s_{123}s_{1234}] \eq &
\frac{1}{s_{23} s_{123} s_{1234}} \ + \ \frac{  I_{21}^\reg(k_{123},k_4,k_5) }{s_{23} s_{123}} \ + \ \frac{ I_{32}^\reg(k_1,k_2,k_3) }{s_{123} s_{1234}} \ + \ \frac{ I_{21}^\reg(k_1,k_{23},k_4) }{s_{23}s_{1234}}   \notag \\
&+ \ \frac{ I_{21,31}^\reg(k_1,k_{23},k_4,k_5)}{s_{23}} \ + \ \frac{ I_{21}^\reg(k_{123},k_4,k_5) I_{32}^\reg(k_1,k_2,k_3) }{s_{123}} \notag \\
& + \ \frac{ I_{31,32}^\reg(k_1,k_2,k_3,k_4) }{s_{1234}} \ + \ I_{31,32,41}^\reg \label{expa6pt3}
\\
Z[s_{23}s_{234}s_{1234}] \eq &
\frac{1}{s_{23} s_{234} s_{1234}} \ + \ \frac{ I_{21}^\reg(k_1,k_{234},k_5) }{s_{23} s_{234}} \ + \ \frac{ I_{32}^\reg(k_{1},k_{23},k_4) }{s_{23} s_{1234}} \ + \ \frac{ I_{21}^\reg(k_2,k_3,k_4) }{s_{234} s_{1234}} \notag \\
&+ \ \frac{ I_{23,24}^\reg(k_1,k_2,k_3,k_4)}{s_{1234}}  \ + \ \frac{  I_{21}^\reg(k_1,k_{234},k_5) I_{21}^\reg(k_2,k_3,k_4) }{s_{234}} \notag \\
& + \ \frac{ I_{31,32}^\reg(k_1,k_{23},k_4,k_5) }{s_{23}} \ + \ I_{32,41,42}^\reg \label{expa6pt4}
\\
Z[s_{34}s_{234}s_{1234}]  \eq &
\frac{1}{s_{34} s_{234} s_{1234}} \ + \ \frac{ I_{21}^\reg(k_{1},k_{234},k_5) }{s_{34} s_{234}} \ + \ \frac{ I_{32}^\reg(k_2,k_3,k_4) }{s_{1234}s_{234}} \ + \ \frac{ I_{32}^\reg(k_1,k_2,k_{34}) }{s_{34} s_{1234}} \notag \\
& + \ \frac{ I_{31,32}^\reg(k_1,k_2,k_{34},k_5) }{s_{34}} \ + \ \frac{ I_{21}^\reg(k_{1},k_{234},k_5) I_{32}^\reg(k_2,k_3,k_4) }{s_{234}} \notag \\
& + \ \frac{ I_{42,43}^\reg(k_1,k_2,k_3,k_4)}{s_{1234}} \ + \ I_{41,42,43}^\reg \label{expa6pt5}
\\
Z[\zeta_2 s_{1234}] \eq &
\frac{ I^\reg_{31,42}(k_1,k_2,k_3,k_4)}{s_{1234}} \ + \ I^\reg_{31,41,42} \label{expa6pt6}
\end{align}

\begin{figure}[h] 
\begin{center}
\tikzpicture[scale=0.65]
\draw (0,0) -- (-1,1) node[left]{$1$};
\draw (0,0) -- (-1,-1) node[left]{$2$};
\draw (0,0) -- (4.5,0);
\draw (0.75,-0.3) node{$k_{12}$};
\draw (2.25,0.3) node{$k_{123}$};
\draw (3.75,-0.3) node{$k_{1234}$};
\draw (1.5,0) -- (1.5,-1) node[below]{$3$};
\draw (3,0) -- (3,-1) node[below]{$4$};
\draw (4.5,0) -- (5.5,1) node[right]{$6$};
\draw (4.5,0) -- (5.5,-1) node[right]{$5$};
\draw (6.5,0) node{$+$};
\draw (6.5,-3) node{$+$};
\draw (6.5,-6) node{$+$};
\draw (15,0) node{$+$};
\draw (-2,-3) node{$+$};
\draw (-2,-6) node{$+$};
\draw (15,-3) node{$+$};
\scope[xshift=8.5cm]
\draw (0,0) -- (-1,1) node[left]{$1$};
\draw (0,0) -- (-1,-1) node[left]{$2$};
\draw (1.5,0) -- (4.5,0);
%
\draw (2.25,0.3) node{$k_{123}$};
\draw (3.75,-0.3) node{$k_{1234}$};
\draw (1.5,0) -- (1.5,-1) node[below]{$3$};
\draw (3,0) -- (3,-1) node[below]{$4$};
\draw (4.5,0) -- (5.5,1) node[right]{$6$};
\draw (4.5,0) -- (5.5,-1) node[right]{$5$};
\draw (0.75,0) ellipse (0.75cm and 0.5cm);
\draw (0.75,0) node{$I_{21}^\reg$};
\endscope
\scope[xshift=17cm]
\draw (0,0) -- (-1,1) node[left]{$1$};
\draw (0,0) -- (-1,-1) node[left]{$2$};
\draw (0,0) -- (1.5,0);
\draw (4.5,0) -- (3,0);
\draw (0.75,-0.3) node{$k_{12}$};
\draw (3.75,-0.3) node{$k_{1234}$};
\draw (1.5,0) -- (1.5,-1) node[below]{$3$};
\draw (3,0) -- (3,-1) node[below]{$4$};
\draw (4.5,0) -- (5.5,1) node[right]{$6$};
\draw (4.5,0) -- (5.5,-1) node[right]{$5$};
\draw (2.25,0) ellipse (0.75cm and 0.5cm);
\draw (2.25,0) node{$I_{21}^\reg$};
\endscope
\scope[xshift=0cm,yshift=-3cm]
\draw (0,0) -- (-1,1) node[left]{$1$};
\draw (0,0) -- (-1,-1) node[left]{$2$};
\draw (0,0) -- (3,0);
\draw (0.75,-0.3) node{$k_{12}$};
\draw (2.25,0.3) node{$k_{123}$};
\draw (1.5,0) -- (1.5,-1) node[below]{$3$};
\draw (3,0) -- (3,-1) node[below]{$4$};
\draw (4.5,0) -- (5.5,1) node[right]{$6$};
\draw (4.5,0) -- (5.5,-1) node[right]{$5$};
\draw (3.75,0) ellipse (0.75cm and 0.5cm);
\draw (3.75,0) node{$I_{21}^\reg$};
\endscope
\scope[xshift=8.5cm,yshift=-3cm]
\draw (0,0) -- (-1,1) node[left]{$1$};
\draw (0,0) -- (-1,-1) node[left]{$2$};
\draw (3,0) -- (4.5,0);
%
\draw (3.75,-0.3) node{$k_{1234}$};
\draw (1.5,-0.5) -- (1.5,-1) node[below]{$3$};
\draw (3,0) -- (3,-1) node[below]{$4$};
\draw (4.5,0) -- (5.5,1) node[right]{$6$};
\draw (4.5,0) -- (5.5,-1) node[right]{$5$};
\draw (1.5,0) ellipse (1.5cm and 0.5cm);
\draw (1.5,0) node{$I_{21,31}^\reg$};
\endscope
\scope[xshift=17cm,yshift=-3cm]
\draw (0,0) -- (-1,1) node[left]{$1$};
\draw (0,0) -- (-1,-1) node[left]{$2$};
\draw (1.5,0) -- (3,0);
%
\draw (2.25,0.3) node{$k_{123}$};
\draw (1.5,0) -- (1.5,-1) node[below]{$3$};
\draw (3,0) -- (3,-1) node[below]{$4$};
\draw (4.5,0) -- (5.5,1) node[right]{$6$};
\draw (4.5,0) -- (5.5,-1) node[right]{$5$};
\draw (0.75,0) ellipse (0.75cm and 0.5cm);
\draw (0.75,0) node{$I_{21}^\reg$};
\draw (3.75,0) ellipse (0.75cm and 0.5cm);
\draw (3.75,0) node{$I_{21}^\reg$};
\endscope
\scope[xshift=0cm,yshift=-6cm]
\draw (0,0) -- (-1,1) node[left]{$1$};
\draw (0,0) -- (-1,-1) node[left]{$2$};
\draw (0,0) -- (1.5,0);
\draw (0.75,-0.3) node{$k_{12}$};
\draw (1.5,0) -- (1.5,-1) node[below]{$3$};
\draw (3,-0.5) -- (3,-1) node[below]{$4$};
\draw (4.5,0) -- (5.5,1) node[right]{$6$};
\draw (4.5,0) -- (5.5,-1) node[right]{$5$};
\draw (3,0) ellipse (1.5cm and 0.5cm);
\draw (3,0) node{$I_{31,32}^\reg$};
\endscope
\scope[xshift=8.5cm,yshift=-6cm]
\draw (0,0) -- (-1,1) node[left]{$1$};
\draw (0,0) -- (-1,-1) node[left]{$2$};
%
\draw (1.5,-0.5) -- (1.5,-1) node[below]{$3$};
\draw (3,-0.5) -- (3,-1) node[below]{$4$};
\draw (4.5,0) -- (5.5,1) node[right]{$6$};
\draw (4.5,0) -- (5.5,-1) node[right]{$5$};
\draw (2.25,0) ellipse (2.25cm and 0.5cm);
\draw (2.25,0) node{$I_{21,31,41}^\reg$};
\endscope
\endtikzpicture
\caption{Pole structure of the function $Z[s_{12}s_{123}s_{1234}]$.}
\label{fig:s12s123s1234}
\end{center}
\end{figure}


\begin{figure}[h] 
\begin{center}
\tikzpicture[scale=0.8]
\draw (5,0) node{$+$};
\draw (5,-3.5) node{$+$};
\draw (5,-7) node{$+$};
\draw (12,0) node{$+$};
\draw (-2,-3.5) node{$+$};
\draw (-2,-7) node{$+$};
\draw (12,-3.5) node{$+$};
\draw (0,0) -- (-1,1) node[left]{$1$};
\draw (0,0) -- (-1,-1) node[left]{$2$};
\draw (0,0) -- (3,0);
\draw (0.75,0.3) node{$k_{12}$};
\draw (2.25,0.3) node{$k_{1234}$};
\draw (1.9,-0.5) node{$k_{34}$};
\draw (1.5,0) -- (1.5,-1);
\draw (1.5,-1) -- (0.5,-2) node[left]{$3$};
\draw (1.5,-1) -- (2.5,-2) node[right]{$4$};
\draw (3,0) -- (4,1) node[right]{$6$};
\draw (3,0) -- (4,-1) node[right]{$5$};
\scope[xshift=7cm]
\draw (0,0) -- (-1,1) node[left]{$1$};
\draw (0,0) -- (-1,-1) node[left]{$2$};
\draw (1.5,0) -- (3,0);
\draw (0.75,0) ellipse (0.75cm and 0.5cm);
\draw (0.75,0) node{$I_{21}^\reg$};
\draw (2.25,0.3) node{$k_{1234}$};
\draw (1.9,-0.5) node{$k_{34}$};
\draw (1.5,0) -- (1.5,-1);
\draw (1.5,-1) -- (0.5,-2) node[left]{$3$};
\draw (1.5,-1) -- (2.5,-2) node[right]{$4$};
\draw (3,0) -- (4,1) node[right]{$6$};
\draw (3,0) -- (4,-1) node[right]{$5$};
\endscope
\scope[xshift=14cm,yshift=0cm]
\draw (0,0) -- (-1,1) node[left]{$1$};
\draw (0,0) -- (-1,-1) node[left]{$2$};
\draw (0,0) -- (3,0);
\draw (1.5,-0.5) ellipse (0.5cm and 0.5cm);
\draw (1.5,-0.5) node{$I_{32}^\reg$};
\draw (0.75,0.3) node{$k_{12}$};
\draw (2.25,0.3) node{$k_{1234}$};
%
\draw (1.5,-1) -- (0.5,-2) node[left]{$3$};
\draw (1.5,-1) -- (2.5,-2) node[right]{$4$};
\draw (3,0) -- (4,1) node[right]{$6$};
\draw (3,0) -- (4,-1) node[right]{$5$};
\endscope
\scope[xshift=0cm,yshift=-3.5cm]
\draw (0,0) -- (-1,1) node[left]{$1$};
\draw (0,0) -- (-1,-1) node[left]{$2$};
\draw (0,0) -- (1.5,0);
\draw (2.25,0) ellipse (0.75cm and 0.5cm);
\draw (2.25,0) node{$I_{21}^\reg$};
\draw (0.75,0.3) node{$k_{12}$};
\draw (1.1,-0.5) node{$k_{34}$};
\draw (1.5,0) -- (1.5,-1);
\draw (1.5,-1) -- (0.5,-2) node[left]{$3$};
\draw (1.5,-1) -- (2.5,-2) node[right]{$4$};
\draw (3,0) -- (4,1) node[right]{$6$};
\draw (3,0) -- (4,-1) node[right]{$5$};
\endscope
\scope[xshift=7cm,yshift=-3.5cm]
\draw (0,0) -- (-1,1) node[left]{$1$};
\draw (0,0) -- (-1,-1) node[left]{$2$};
\draw (1,0) -- (3,0);
%
\draw (2.25,0.3) node{$k_{1234}$};
%
\draw (0.9,0)[rotate=330] ellipse (0.9cm and 0.5cm);
\draw (0.8,-0.5) node{$I_{21,43}^\reg$};
\draw (1.5,-1) -- (0.5,-2) node[left]{$3$};
\draw (1.5,-1) -- (2.5,-2) node[right]{$4$};
\draw (3,0) -- (4,1) node[right]{$6$};
\draw (3,0) -- (4,-1) node[right]{$5$};
\endscope
\scope[xshift=14cm,yshift=-3.5cm]
\draw (0,0) -- (-1,1) node[left]{$1$};
\draw (0,0) -- (-1,-1) node[left]{$2$};
%
\draw (1.9,-0.8) node{$k_{34}$};
\draw (1.5,-0.5) -- (1.5,-1);
\draw (1.5,0) ellipse (1.5cm and 0.5cm);
\draw (1.5,0) node{$I_{21,31}^\reg$};
\draw (1.5,-1) -- (0.5,-2) node[left]{$3$};
\draw (1.5,-1) -- (2.5,-2) node[right]{$4$};
\draw (3,0) -- (4,1) node[right]{$6$};
\draw (3,0) -- (4,-1) node[right]{$5$};
\endscope
\scope[xshift=0cm,yshift=-7cm]
\draw (0,0) -- (-1,1) node[left]{$1$};
\draw (0,0) -- (-1,-1) node[left]{$2$};
\draw (0,0) -- (2,0);
\draw (0.75,0.3) node{$k_{12}$};
%
\scope[xshift=3cm]
\draw (0.9,0)[rotate=210] ellipse (0.9cm and 0.5cm);
\endscope
\draw (2.15,-0.5) node{$I_{31,32}^\reg$};
\draw (1.5,-1) -- (0.5,-2) node[left]{$3$};
\draw (1.5,-1) -- (2.5,-2) node[right]{$4$};
\draw (3,0) -- (4,1) node[right]{$6$};
\draw (3,0) -- (4,-1) node[right]{$5$};
\endscope
\scope[xshift=7cm,yshift=-7cm]
\draw (0,0) -- (-1,1) node[left]{$1$};
\draw (0,0) -- (-1,-1) node[left]{$2$};
%
%
\draw (1.5,0) ellipse (1.5cm and 1cm);
\draw (1.5,0) node{$I_{21,43,41}^\reg$};
\draw (1.5,-1) -- (0.5,-2) node[left]{$3$};
\draw (1.5,-1) -- (2.5,-2) node[right]{$4$};
\draw (3,0) -- (4,1) node[right]{$6$};
\draw (3,0) -- (4,-1) node[right]{$5$};
\endscope
\endtikzpicture
\caption{Pole structure of the function $Z[s_{12}s_{34}s_{1234}]$.}
\label{fig:s12s34s1234}
\end{center}
\end{figure}

Following the off-shell completion (\ref{21mass}) of the quartic contact
vertices, the non-vanishing mass contributions to $I_{21}^\reg$ and
$I_{23}^\reg$ are given by
\begin{align}
I_{21}^\reg(k_1,k_{23},k_4) \eq &I_{21}^\reg[s_{123},s_{24}+s_{34}] \co I_{21}^\reg(k_{1},k_{234},k_5) \eq &I_{21}^\reg[s_{1234},s_{25}+s_{35}+s_{45}] \notag \\
I_{21}^\reg(k_{12},k_{34},k_5) \eq &I_{21}^\reg[s_{13}+s_{14}+s_{23}+s_{24}+s_{34},s_{35}+s_{45}]  \label{ktos6} \\
I_{32}^\reg(k_{1},k_{23},k_4) \eq &I_{21}^\reg[s_{234},s_{12}+s_{13}]
\notag
\end{align}
Also, we discover additional off-shell completions for five-point contact vertices
\begin{align}
I_{21,31}^\reg(k_{1},k_2,k_{34},k_5) \eq &I_{21,31}^\reg[s_{12},s_{134},s_{23}+s_{24},s_{25},s_{35}+s_{45}] \label{ktos7}
\\
I_{21,31}^\reg(k_1,k_{23},k_4,k_5)\eq &I_{21,31}^\reg[s_{123},s_{14},s_{24}+s_{34},s_{25}+s_{35},s_{45}] \label{ktos8} \\
 I_{31,32}^\reg(k_1,k_2,k_{34},k_5) \eq &I_{31,32}^\reg[s_{12},s_{134},s_{23}+s_{24},s_{25},s_{35}+s_{45}] \ . \label{ktos9}
 %
\end{align}
Unfortunately, the regular part $I^\reg_{32,41,42}$ is incompatible with the
prerequisite (\ref{constraint}) for the polylogarithm methods of section
\ref{secPolylogs}, hence, we cannot determine the dependence of
$I_{31,32}^\reg(k_1,k_{23},k_4,k_5)$ on $k_{23}^2$. The remaining six-point vertices
$I^\reg_{21,31,41},I^\reg_{21,43,41},I^\reg_{31,32,41},I^\reg_{41,42,43}$ and $I^\reg_{31,41,42}$,
on the other hand, can be inferred by equating the pole structures
(\ref{expa6pt1}) to (\ref{expa6pt6}) with the $F^{\si(234)}$ basis expansions
(\ref{expa6pt}) of the $Z[\ldots]$ functions. The low-energy expansion of these
$I^\reg_{\ldots}$ is relevant for seven point pole residues, so we can increase
the efficiency of our setup by extracting it from the six point integral basis.


\section{Polylogarithm material}
\label{polyexpl}

In this part of the appendix, we collect some examples of the identities derivable from \eqn{eqn:GIdentity} at weights two and three. Note that the idenities containing $0's$ are treated separately and thus $a_k\neq0$ below:
\begin{eqnarray}
  G(a_1, z; z) &=& -G(0, a_1; z) + G(a_1, a_1; z)\\[8pt]
G(0, z, a_1; z) &=& 
  G(0, 0, a_1; z) - G(a_1, 0, a_1; z) - 
   G(a_1; z) \,\z_2\nnl 
 G(0, a_1, z; z) &=& -2\,G(0, 0, a_1; z) + 
   G(0, a_1, a_1; z) + G(a_1, 0, a_1; z) + 
   G(a_1; z) \,\z_2\nnl 
 G(a_1, 0, z; z) &=& 
  G(0, 0, a_1; z) - G(0, a_1, a_1; z) - 
   G(a_1; z) \,\z_2\nnl 
 G(a_1, z, z; z) &=& 
  G(0, 0, a_1; z) - G(0, a_1, a_1; z) - 
   G(a_1, 0, a_1; z) + G(a_1, a_1, a_1; z)\nnl 
 G(a_1, z, a_2; z) &=& 
  G(a_1, a_1, a_2; z) - G(a_2, 0, a_1; z) + 
   G(a_2, a_1, a_1; z) - G(a_2, a_1, a_2; z)\nnl 
 G(a_1, a_2, z; z) &=& -G(0, a_1, a_2; z) - 
   G(a_1, 0, a_2; z) + G(a_1, a_2, a_2; z) \nnl
   && + G(a_2, 0, a_1; z) - G(a_2, a_1, a_1; z) + 
   G(a_2, a_1, a_2; z)\,.
\end{eqnarray}
These identities are given only for polylogs with the first entry not equal to the argument and, at the same time, nonzero last entry. The remaining polylogs have to be dealt with using the methods of \eqn{shufflereg1} and \eqn{shufflereg2} beforehand.

\section{Seven-point material}
\label{app7pt}

This appendix closes some of the gaps in the presentation of the seven-point $\ap$-expansion.


\subsection{The pole structure of integrals from the $\{2,4\}$ and $\{4,5\}$ blocks}
\label{poleblock}

In this appendix we show the singularity structure of seven point integrals from the $\{2,4\}$ and $\{4,5\}$ blocks which were omitted in subsection \ref{7ptexpl}.

The $\{2,4\}$ functions are characterized by polynomials $ \in \{ z_{21} z_{41} , \, z_{41} z_{42} \} \times \{ z_{53} z_{65} , \, z_{63} z_{65} \}$
\begin{align}
\prod_{i=2}^5 \int_0^{z_{i+1}} \dd z_i \ \frac{ \prod_{i<j}^6 |z_{ij}|^{s_{ij}} }{z_{21} z_{35} z_{41}  z_{56}}
 \eq &\frac{ I_{31,42}^\reg(k_{12},k_3,k_4,k_{56}) }{s_{12}s_{56}} \ + \ \frac{ I^\reg_{31,42,54}(k_{12},k_3,k_4,k_5,k_6) }{s_{12}}  \notag \\
& \  \ + \ \frac{I^\reg_{21,41,53}(k_1,k_2,k_3,k_4,k_{56}) }{s_{56}} \ + \ I_{21,41,53,65}^\reg
\label{02,33}
\\
\prod_{i=2}^5 \int_0^{z_{i+1}} \dd z_i \ \frac{ \prod_{i<j}^6 |z_{ij}|^{s_{ij}} }{z_{24} z_{35} z_{41} z_{56}} \eq& -\,\frac{ I^\reg_{41,42,53}(k_1,k_2,k_3,k_4,k_{56})}{s_{56} } \ - \ I_{41,42,53,65}^\reg
\label{02,35}
\\
\prod_{i=2}^5 \int_0^{z_{i+1}} \dd z_i \ \frac{ \prod_{i<j}^6 |z_{ij}|^{s_{ij}} }{z_{21} z_{36} z_{41} z_{56}} \eq& \frac{ I_{31,42}^\reg(k_{12},k_3,k_4,k_{56}) }{s_{12}s_{56}} \ + \ \frac{ I^\reg_{31,52,54}(k_{12},k_3,k_4,k_5,k_6) }{s_{12}} \notag \\
& \ \ + \ \frac{I^\reg_{21,41,53}(k_1,k_2,k_3,k_4,k_{56}) }{s_{56}} \ + \ I_{21,41,63,65}^\reg
\label{02,34}
\\
%
\prod_{i=2}^5 \int_0^{z_{i+1}} \dd z_i \ \frac{ \prod_{i<j}^6 |z_{ij}|^{s_{ij}} }{z_{24} z_{36} z_{41} z_{56}} \eq&- \, \frac{ I^\reg_{41,42,53}(k_1,k_2,k_3,k_4,k_{56})}{s_{56} } \ - \ I_{41,42,63,65}^\reg
\label{02,36}
\end{align}
without any mass dependence.


The $\{4,5\}$ functions are characterized by polynomials $ \in \{ z_{41} z_{51} , \, z_{51} z_{54} \} \times \{ z_{32} z_{63} , \, z_{62} z_{63} \}$
\begin{align}
\prod_{i=2}^5 \int_0^{z_{i+1}} \dd z_i \ \frac{ \prod_{i<j}^6 |z_{ij}|^{s_{ij}} }{z_{32} z_{63} z_{51} z_{54}} &\eq \frac{ I^\reg_{31,42}(k_1,k_{23},k_{45},k_6) }{s_{23} s_{45}}  \ + \ \frac{ I^\reg_{41,43,52}(k_1,k_{23},k_4,k_5,k_6) }{s_{23}} \notag \\
& \ \ + \ \frac{ I^\reg_{32,41,53}(k_1,k_2,k_3,k_{45},k_6)  }{s_{45}} \ + \ I_{32,51,54,63}^\reg
\label{02,41}  \\
%
\prod_{i=2}^5 \int_0^{z_{i+1}} \dd z_i \ \frac{ \prod_{i<j}^6 |z_{ij}|^{s_{ij}} }{z_{32} z_{63} z_{41} z_{51}} &\eq \frac{ I^\reg_{31,41,52}(k_1,k_{23},k_4,k_5,k_6) }{s_{23}} \ + \ I_{32,41,51,63}^\reg
\label{02,42} \\
%
\prod_{i=2}^5 \int_0^{z_{i+1}} \dd z_i \ \frac{ \prod_{i<j}^6 |z_{ij}|^{s_{ij}} }{z_{62} z_{63} z_{51} z_{54}} &\eq \frac{ I^\reg_{41,52,53}(k_1,k_2,k_3,k_{45},k_6)  }{s_{45}} \ + \ I_{51,54,62,63}^\reg 
\label{02,43}\\
%
\prod_{i=2}^5 \int_0^{z_{i+1}} \dd z_i \ \frac{ \prod_{i<j}^6 |z_{ij}|^{s_{ij}} }{z_{62} z_{63} z_{41} z_{51}} &\eq I_{41,51,62,63}^\reg
\label{02,44}
\end{align}
with mass dependence
\begin{align}
I^\reg_{31,42}(k_1,k_{23},k_{45},k_6)&\eq I^\reg_{31,42}[s_{123}, s_{145},s_{2345}-s_{23}-s_{45},s_{26}+s_{36},s_{46}+s_{56}]
\label{ktos22} \\
I^\reg_{41,43,52}(k_1,k_{23},k_4,k_5,k_6) &\eq I^\reg_{41,43,52}[s_{123},s_{14},s_{15},s_{24}+s_{34},s_{25}+s_{35},s_{45},s_{26}+s_{36},s_{46},s_{56}] \notag \\
I^\reg_{32,41,53}(k_1,k_2,k_3,k_{45},k_6) &\eq I^\reg_{32,41,53}[s_{12},s_{13},s_{145},s_{23},s_{24}+s_{25},s_{34}+s_{35},s_{26},s_{36},s_{46}+s_{56}]  \notag \\
I^\reg_{31,41,52}(k_1,k_{23},k_4,k_5,k_6) &\eq I^\reg_{31,41,52}[s_{123},s_{14},s_{15},s_{24}+s_{34},s_{25}+s_{35},s_{45},s_{26}+s_{36},s_{46},s_{56}] \notag \\
I^\reg_{41,52,53}(k_1,k_2,k_3,k_{45},k_6) &\eq I^\reg_{41,52,53}[s_{12},s_{13},s_{145},s_{23},s_{24}+s_{25},s_{34}+s_{35},s_{26},s_{36},s_{46}+s_{56}] \notag
\end{align}


\subsection{Cyclic transformations towards functions from the $\{2,3\}$ block}
\label{1more}

According to subsection \ref{cyc7}, the seven point basis functions $F^{(3245)}$ and $F^{(3254)}$ can be inferred from cyclicity. The required identities are
\begin{align}
F^{(3425)} &\, \Big|_{i\mapsto i+1} \eq \frac{F^{(3245)}
   \left(s_{36}+s_{37}\right)}{s_{13}} \ + \ F^{(2453)}   \ - \ \frac{F^{(4325) }s_{24}
   \left(s_{36}+s_{37}\right)}{s_{14}
     s_{134}} \ - \ \frac{ F^{(4253)}
(s_{14}+s_{24})}{s_{14}} \  \notag \\
  &- \ \frac{F^{(3425)}
   \left(  s_{134}+s_{24}\right)
   \left(s_{36}+s_{37}\right)}{s_{13}
  s_{134}}  \ - \ \frac{F^{(5324)}
   \left(s_{36}+s_{37}\right) \left(s_{25}+s_{45}\right)}{s_{15}
    s_{135}} \notag \\
    & + \ \frac{F^{(5423)}
   \left(s_{145}+s_{24}\right) \left(s_{1245}-s_{124}\right)}{s_{15}
   s_{145}} \ - \ \frac{F^{(5243)}
  \left(s_{1245}-s_{124}\right)}{s_{15}} \ + \ \frac{F^{(4523)} s_{24}
   \left(s_{1245}-s_{124}\right)}{s_{14}
   s_{145} } \notag \\
   &- \ \frac{F^{(3524)}
  \left(s_{36}+s_{37}\right)
   \left(s_{135}+s_{25}+s_{45}\right)}{s_{13}
 s_{135} } \ + \ \frac{F^{(5342)}
   \left(s_{36}+s_{37}\right) \left(s_{25}+s_{45}\right)
   \left(s_{1345}+s_{24}\right)}{s_{15}
     s_{135} s_{13 45} } \notag \\
     &+ \ \frac{F^{(3542)}
   \left(s_{36}+s_{37}\right)
   \left(s_{135}+s_{25}+s_{45}\right)
   \left(s_{1345}+s_{24}\right)}{s_{13}
    s_{135} s_{13 45} } \ + \ \frac{F^{(4352) }s_{24}
\left(s_{36}+s_{37}\right)
   \left(s_{1345}+s_{25} \right)}{s_{14}
 s_{134} 
 s_{1345} } \notag \\
 &+ \ \frac{F^{(5432)} s_{24}
\left(s_{36}+s_{37}\right) \left(s_{1245}-s_{124}\right) }{s_{15}
 s_{145}  s_{1345} } \ - \ \frac{F^{(3452) }s_{24}
 \left(s_{36}+s_{37}\right)
   \left(s_{56}+s_{57}\right)}{s_{13} s_{134} s_{1345} } \notag \\
   &- \ \frac{F^{(4532)} s_{24}
   \left(s_{14}-s_{25}\right) \left(s_{36}+s_{37}\right)}{s_{14}
   s_{145}  s_{1345} } \ . 
   \label{06,55}
 \end{align}
\begin{align}
F^{(4235)} &\, \Big|_{i\mapsto i+1} \eq - \ \frac{F^{(3254)}
   \left(s_{13}+s_{23}+s_{35}\right)}{s_{13}} \ + \ 
 F^{(2534)}  \ + \ \frac{F^{(3524)}
   \left(s_{13}+s_{23}+s_{35}\right) (s_{135}+s_{25})}{s_{13}
   s_{135}}  \notag \\
   &-\  \frac{F^{(5234)}(s_{15}+s_{25})}{s_{15}}  \ + \ \frac{F^{(5324)} s_{25}
   \left(s_{13}+s_{23}+s_{35}\right)}{s_{15}
   s_{135}}  \ - \ \frac{F^{(3425) }\left(s_{23}+s_{35}\right)
   (s_{46}+s_{47})}{s_{13}
   s_{134}} \notag \\
   & - \ \frac{F^{(5423)} s_{25}
   (s_{46}+s_{47})}{s_{15}
   s_{145}}  \ - \ \frac{F^{(4523)}
   \left(s_{14 5}+s_{25}\right)
   (s_{46}+s_{47})}{s_{14}
   s_{145}} \ + \ \frac{F^{(4253)}
   (s_{46}+s_{47})}{s_{14}} \notag \\
   &- \ \frac{F^{(4325)}
   \left(s_{134}+s_{23}+s_{35}\right)
   (s_{46}+s_{47})}{s_{14}
   s_{134}} \ + \ \frac{F^{(5432)} s_{25}
   (s_{46}+s_{47})
   \left(s_{1345}+s_{23}\right)}{s_{15}
   s_{145}
   s_{1345}} \notag \\
   & + \ \frac{F^{(3452)}
   \left(s_{23}+s_{35}\right) (s_{46}+s_{47})
   \left(s_{1345}+s_{25}\right)}{s_{13}
   s_{134}
   s_{1345}}  \ + \ \frac{F^{(3542)} s_{25}
   \left(s_{13}+s_{23}+s_{35}\right) (s_{46}+s_{47})}{s_{13}
   s_{135}    s_{1345}} \notag \\
   & + \ \frac{F^{(4352)}
   \left(s_{134}+s_{23}+s_{35}\right)
   (s_{46}+s_{47})
   \left(s_{1345}+s_{25}\right)}{s_{14}
   s_{134}
   s_{1345}}\
   - \ \frac{F^{(5342)}
   \left(s_{15}-s_{23}\right) s_{25} (s_{46}+s_{47})}{s_{15}
   s_{135}
   s_{1345}} \notag \\
   & - \ \frac{F^{(4532)} s_{25}
   \left(s_{36}+s_{37}\right)
   (s_{46}+s_{47})}{s_{14} s_{145}
   s_{1345}}   \ .
   \label{06,56}
\end{align}

\end{document}